\documentclass[final]{ws-ijmpa_my}

\usepackage{bm}
\usepackage{youngtab}
\usepackage{color}
\usepackage{amsmath,amsthm,amssymb}
\usepackage{amscd}
\usepackage{graphicx}
\usepackage{alphalph}

\makeatletter
\renewcommand{\theequation}{%
\thesection.\arabic{equation}}
\@addtoreset{equation}{section}
\makeatother

\newcommand{\bC}{\ensuremath{\mathbb{C}}}

\newcommand{\bN}{\ensuremath{\mathbb{N}}}

\newcommand{\bP}{\ensuremath{\mathbb{P}}}

\newcommand{\bR}{\ensuremath{\mathbb{R}}}

\newcommand{\bT}{\ensuremath{\mathbb{T}}}

\newcommand{\bZ}{\ensuremath{\mathbb{Z}}}

\newcommand{\scC}{\ensuremath{\mathcal{C}}}

\newcommand{\scF}{\ensuremath{\mathcal{F}}}

\newcommand{\scH}{\ensuremath{\mathcal{H}}}
\newcommand{\scI}{\ensuremath{\mathcal{I}}}

\newcommand{\scM}{\ensuremath{\mathcal{M}}}
\newcommand{\scN}{\ensuremath{\mathcal{N}}}
\newcommand{\scO}{\ensuremath{\mathcal{O}}}

\newcommand{\scV}{\ensuremath{\mathcal{V}}}

\def\beq#1\eeq{\begin{align}#1\end{align}}

\newcommand{\mr}[1]{{\mathrm{#1}}}

\newcommand{\bb}[1]{{\mathbb{#1}}}
\newcommand{\mca}[1]{{\mathcal{#1}}}

\newcommand{\Z}{\bb{Z}}
\newcommand{\Zh}{\bb{Z}_{\mathrm{h}}}

\newcommand{\V}{\mca{V}}

\newcommand{\pg}{\overset{+}{\succ}}
\newcommand{\pl}{\overset{+}{\prec}}
\newcommand{\mg}{\overset{-}{\succ}}
\newcommand{\ml}{\overset{-}{\prec}}

\newcommand{\tenchi}{{}^{\mathrm{t}}}

\newcommand{\Vmin}{\V_{\mathrm{min}}}

\newcommand{\type}{(\sigma,\theta\sss;\sss \mu,\nu)}
\newcommand{\sss}{\hspace{0.5pt}}

\newcommand{\coh}{\operatorname{coh}}
\newcommand{\module}{\operatorname{mod}}

\newcommand{\vs}[1]{\vspace{#1 mm}}
\newcommand{\hs}[1]{\hspace{#1 mm}}

\begin{document}

\title{
CRYSTAL MELTING AND WALL CROSSING PHENOMENA
}

\author{
MASAHITO YAMAZAKI
}

\address{
IPMU, University of Tokyo, Chiba 277-8586, Japan \\
and Department of Physics, University of Tokyo, Tokyo 113-0033, Japan\\
and California Institute of Technology, Pasadena, CA 91125, USA\\
yamazaki@hep-th.phys.s.u-tokyo.ac.jp
}

\maketitle


\begin{abstract}
This paper summarizes recent developments in the theory of Bogomol'nyi-Prasad-Sommerfield (BPS) state counting and the wall crossing phenomena, emphasizing in particular the role of the statistical mechanical model of crystal melting.

This paper is divided into two parts, which are closely related to each other. In the first part, we discuss the statistical mechanical model of crystal melting counting BPS states. Each of the BPS state contributing to the BPS index is in one-to-one correspondence with a configuration of a molten crystal, and the statistical partition function of the melting crystal gives the BPS partition function. We also show that smooth geometry of the Calabi-Yau manifold emerges in the thermodynamic limit of the crystal. This suggests a remarkable interpretation that an atom in the crystal is a discretization of the classical geometry, giving an important clue as to the geometry at the Planck scale.

In the second part we discuss the wall crossing phenomena. Wall crossing phenomena states that the BPS index depends on the value of the moduli of the Calabi-Yau manifold, and jumps along real codimension one subspaces in the moduli space. We show that by using type IIA/M-theory duality, we can provide a simple and an intuitive derivation of the wall crossing phenomena, furthermore clarifying the connection with the topological string theory. This derivation is consistent with another derivation from the wall crossing formula, motivated by multi-centered BPS extremal black holes.
We also explain the representation of the wall crossing phenomena in terms of crystal melting, and the generalization of the counting problem and the wall crossing to the open BPS invariants.

\end{abstract}

\ccode{PACS numbers: 11.25.Uv,11.25.Yb,11.25.Mj}


\section{Introduction and Overview} \label{chap.intro}
One of the most fundamental problems in theoretical physics in the 21st century is to construct a theory of quantum gravity. General relativity and quantum mechanics, which are the cornerstones of the 20th century physics, are mutually inconsistent. For example, general relativity is not renormalizable, and conventional perturbative techniques in quantum field theory does not apply. Not only are these two theories inconsistent, but there are indications that theoretical consistency requires unification of these two theories.
In general relativity, there is a celebrated singularity theorem \cite{PenroseSingularity}, which predicts the existence of singularities where general relativity breaks down. In this sense general relativity is not a consistent theory in itself, and singularity occurs precisely when quantum effects are expected to play crucial roles. The existence of singularities is classically not a problem if cosmic censorship conjecture \cite{PenroseCensorship} holds  and singularity is surrounded by the horizon. However, quantum mechanically black holes emit particles \cite{Hawking74,Hawking75}. This then leads us to the paradox of information loss \cite{Hawking76} (see Refs.~\refcite{Preskill,Page} for review), since we only have a thermal radiation coming out of the event horizon regardless of what goes inside the horizon. One should always keep in mind, however, that these arguments are based on semiclassical quantization of general relativity, and proper treatment of these problems require a theory of quantum gravity which unifies the general relativity and quantum mechanics in a consistent framework.

Over decades string theory has been the most promising candidate for quantum gravity. Just as in general relativity, string theory is severely constrained by 
various consistency conditions, and it is often impossible to tune the parameters by hand. This means that if we can extract quantitative predictions from string theory, we can perform pass/fail test of string theory as a theory of quantum gravity.
Computation of black hole entropy is an excellent example for such a stringent test.
One of the most successful predictions of string theory, as shown by Strominger and Vafa in 1996 \cite{StromingerVafa}, is that string theory correctly reproduces the semiclassical Bekenstein-Hawking entropy \cite{Bekenstein1,Bekenstein2,Bekenstein3,BardeenCH,Hawking76-2} 
of a class of supersymmetric extremal black holes (see e.g. Refs.~\refcite{MaldacenaReview,PeetTASI} for review). In statistical mechanics, entropy is given by the logarithm of total degeneracy of states, and Strominger and Vafa showed that string theory reproduces the correct number of states in the limit of large charges. Their analysis has subsequently been generalized to many other black holes, including four-dimensional black holes and non-extremal black holes. Most remarkably, string theory now reproduces not only the semi-classical Bekenstein-Hawking entropy of general relativity, but also the subleading entropy contributions from the higher curvature corrections to the Einstein-Hilbert action \cite{Cardoso1,Cardoso2,Cardoso3,OSV} (see Ref.~\refcite{Pioline} for review), as dictated by the Wald's formula \cite{Wald1,JacobsonKM,Wald2}. This gives a rather remarkable quantitative check of string theory as a theory of quantum gravity.

However, there are many issues that remain to be solved. For example, 
the entropy is typically determined only by the asymptotic growth of the microstate degeneracies (given by the Cardy formula \cite{Cardy}), in the limit of large charges.
However, we hope that string theory gives a more complete and detailed theory 
of the microstates, not just the asymptotic growth of their numbers. This will hopefully lead to rich and yet unknown aspects of quantum gravity.

This is closely related to the question: what is the geometry at the Planck scale? One of the key ideas in general relativity is the ``geometrization of physics'', where the physics notion (e.g. mass) are translated and reformulated in terms of geometry (e.g. curvature). If we follow a similar path, the central question in quantum gravity should be to identify the ``quantum geometry'', geometry at the Planck scale where spacetime itself fluctuates quantum mechanically.

\bigskip

In this paper, we will make small steps towards these ambitious goals. Unfortunately, solving string theory in gravity backgrounds is a notoriously difficult problem.
The strategy we take is to simplify the problem --- to replace the problem of gravity with a problem of gauge theory. In string theory compactifications, this corresponds to taking the Calabi-Yau manifold to be non-compact. Of course, the notion of a black hole is subtle in this limit since the gravity decouples and the Newton constant becomes zero. However, part of the important data in gravity theory still remain. For example, we can still discuss entropy of black holes since we can take a scaling limit where the mass of the black hole goes to infinity, while the entropy is kept finite. Moreover, it is generally believed that microstate degeneracies of black holes are captured by the BPS index, which is deformation invariant and stays constant in the decoupling process.
The counting of black hole microstates is now turned into a counting problem of BPS states in supersymmetric gauge theories. 

The counting problem of BPS states in string theories and in supersymmetric gauge theories is an important problem in itself, even if we forget about the motivation from black hole physics. For example, they provide primary tools to test various dualities in supersymmetric gauge theories and in string theory (e.g. non-perturbative checks \cite{SenDyon} of Montonen-Olive duality \cite{MontonenO} for $\scN=4$ supersymmetric Yang-Mills theory). 
BPS solitons in supersymmetric gauge theories have a rich structure, and provides a classic example of fruitful collaboration between physics and mathematics, as exemplified in the ADHM \cite{ADHM,DrinfeldM} and Nahm \cite{Nahm} construction. Furthermore, as we will see in later sections BPS state counting has an intimate connection with another counting problem in string theory, the topological string theory (mathematically Gromov-Witten theory).

\bigskip

In the first part of this paper, we show that when $X$ is a toric Calabi-Yau manifold (this roughly means that $X$ has an action of the three-dimensional complex torus; see Refs.~\refcite{Fulton,Oda} for review), we can give explicit answers to the BPS counting problem. More precisely, each of the BPS states contributing to the BPS index (defined in Section\,\ref{sec.index}) is in one-to-one correspondence with a configuration of a molten crystal, and the BPS partition function $Z_{\rm BPS}$ (defined in Section\,\ref{sec.index}) is the same as the statistical partition function of a crystal melting model \cite{OY1}:
\beq
Z_{\rm BPS}=Z_{\rm crystal}.
\eeq

Section\,\ref{chap.crystal} is devoted to the explanation of this result. Remarkably, the derivation of the formula above depends on the newly developed mathematical theory, the non-commutative Donaldson-Thomas theory \cite{Szendroi,MR}. This theory gives a new invariant for Calabi-Yau manifolds, which exactly coincides with the BPS index we are interested in. This means that BPS counting problem is important not only to physicists but also to the mathematicians alike.

In the next section (Section\,\ref{chap.thermodynamic}), we discuss the implication of these results to quantum gravity. We show that the thermodynamic limit of the crystal gives a projection of the shape of the mirror Calabi-Yau manifold \cite{OY2}. This in particular suggests that if we start from a classical smooth geometry and goes to smaller and smaller scales, the geometry gets discretized into a set of atoms when we finally arrive at the Planck scale. In this sense an atom in the crystal melting model is ``an atom/quark of internal geometry'', a discretized version of the geometry at the Planck scale. We therefore see that the two problems posed earlier are now related. Each of the microstate, which is an atom of the crystal, is the discretized version of the geometry; thus the problem of identifying black hole microstates is solved by the quantum structure of geometry!

\bigskip

In the second half of the paper, we move on to the discussion of the wall crossing phenomena (see Section\,\ref{sec.WC} for an introduction). Wall crossing phenomena states that the BPS degeneracy jumps as we change the value of the moduli of the Calabi-Yau manifold. Wall crossing phenomena, first discussed in Ref.~\refcite{CecottiVold} in the context of supersymmetric $\scN=(2,2)$ theories in two dimensions, have a long history of nearly two decades. 
They also play important roles in the Seiberg-Witten theory \cite{SW1,SW2,FerrariB,BilalF} and its stringy realization \cite{KLMVW,ShapereV} (see also Refs.~\refcite{BergmanF,MikhailovNS} as well as Ref.~\refcite{Bergman}). In these old days, it was possible to derive the jump of BPS states in simple cases, but generalization seemed to be difficult.

The recent breakthrough was partly triggered by the paper of Kontsevich and Soibelman \cite{KontsevichS}, who proposed a rather general formula for the jumps of BPS degeneracies (or mathematically ``generalized Donaldson-Thomas invariants''), generalizing the ``semi-primitive'' wall crossing formula previously proposed by Denef and Moore \cite{DenefM}. Physical interpretations of the formulas were subsequently given in Refs.~\refcite{GMN1,GMN2,CecottiV,GMN3,Galaxies}. We discuss these formulas in Section\,\ref{chap.WCF} and in \ref{app.KS}.

The wall crossing formulas are rather general, and can be applied to our setup, namely compactification on the toric Calabi-Yau manifold. In particular, the example of the resolved conifold is analyzed in Ref.~\refcite{JM} and independently in Ref.~\refcite{NN}. There it was shown that the non-commutative Donaldson-Thomas invariants (which are just generalized Donaldson-Thomas invariants in a certain chamber) discussed in Section\,\ref{chap.crystal} is related by wall crossing to the commutative (i.e. ordinary) Donaldson-Thomas invariants. In physics language, this means that the crystal melting partition function is related by wall crossings to the topological string partition function. This clarifies the connection of our crystal melting model and another crystal melting model proposed in Ref.~\refcite{ORV}, which describes the topological vertex \cite{AKMV}.

This is not the end of the story, however.
 First, it was observed in the literature that the BPS partition function computed by the non-commutative Donaldson-Thomas theory takes a beautiful infinite product form, and there should be an intuitive explanation of this result. From the viewpoint of non-commutative Donaldson-Thomas theory, this seems miraculous: we first compute the BPS indices separately by intricate mathematics, and only after summing up all of them and going through combinatorics can we see that the partition function takes such a simple form. Second, the partition function obtained in Section\,\ref{chap.crystal} is very similar to the topological string partition function, and it is necessary to understand the precise relation and to explain why the topological string partition function enters into the story.

In Section\,\ref{chap.M-theory} we give a simple, yet another derivation of the wall crossing phenomena from the viewpoint of M-theory \cite{AOVY}. This answers the questions raised in the previous paragraph. By lifting type IIA brane configurations to M-theory and by using the 4d/5d correspondence \cite{GSY,DVVafa}, the problem of counting BPS states is mapped (under certain conditions explained in Section\,\ref{chap.M-theory}) to a counting problem of free M2-brane particles in five dimensions, which span the free particle Fock space. This naturally explains the infinite product form of the BPS partition function. Also, the counting problem of M2-brane particles is a generalization of the Gopakumar-Vafa (GV) argument \cite{GV1,GV2}, which explains the appearance of the topological string partition function. More precisely, we prove a formula which schematically takes a form
\beq
Z_{\rm BPS}=Z_{\rm top}^2 \Big| _{\rm chamber}. \label{Z=Z2}
\eeq 
This derivation is consistent with another derivation from the wall crossing formula. As a bonus, we have new mathematical predictions for non-toric examples, which can be tested by future mathematical developments.

\bigskip

There is a generalization of the above-mentioned story to the case of open BPS invariants. Closed BPS invariants discussed up to this point are defined by counting D2-branes wrapping 2-cycles of the Calabi-Yau manifold. Open BPS invariants are defined by counting D2-branes wrapping on disks ending on another D-branes (D4-branes). Open BPS invariants are natural generalizations of closed invariants. Moreover, they give a useful computational tool to study closed invariants for complicated geometries, as shown in the topological vertex formalism \cite{AKMV}. 

In Section\,\ref{chap.open} we first give a definition of the ``non-commutative topological vertex'', which gives a basic building block for computing open BPS invariants. The definition uses the crystal melting model, and we perform several consistency checks of the proposal. We also discuss the wall crossing phenomena for the open BPS invariants both with respect to the open and closed string moduli, by again using the viewpoint of M-theory.

In the final section (Section\,\ref{chap.discussion}), we close this paper by pointing out several interesting problems which suggest directions for future research. We also collect slightly technical results in the appendices.

\bigskip

This paper is essentially the author's Ph.D thesis (submitted to University of Tokyo on November 2009), which in turn is mostly based on his papers Refs.~\refcite{OY1,OY2,AOVY,NY,AY}. We tried our best to make the presentation consistent and to present a unified perspective on the problem. However, BPS state counting and wall crossing phenomena are very active areas of research with long history, and is still growing very rapidly as of this writing. This inevitably means that there are many important omissions. For example, in this review there is almost no discussion of the wall crossing phenomena from 4d $\scN=2$ gauge theory viewpoint (see e.g. Refs~\refcite{GMN1,GMN2,GMN3} in this direction). and our discussion is mostly on the 6d Calabi-Yau side.
Moreover, in order to keep this paper to a reasonable length we could not include enough introductory material in this review. The complete introduction will require another review, but the good starting point is the lecture note Ref.~\refcite{Moorepitp}.

\bigskip

\begin{figure}[htbp]
\includegraphics[scale=0.7]{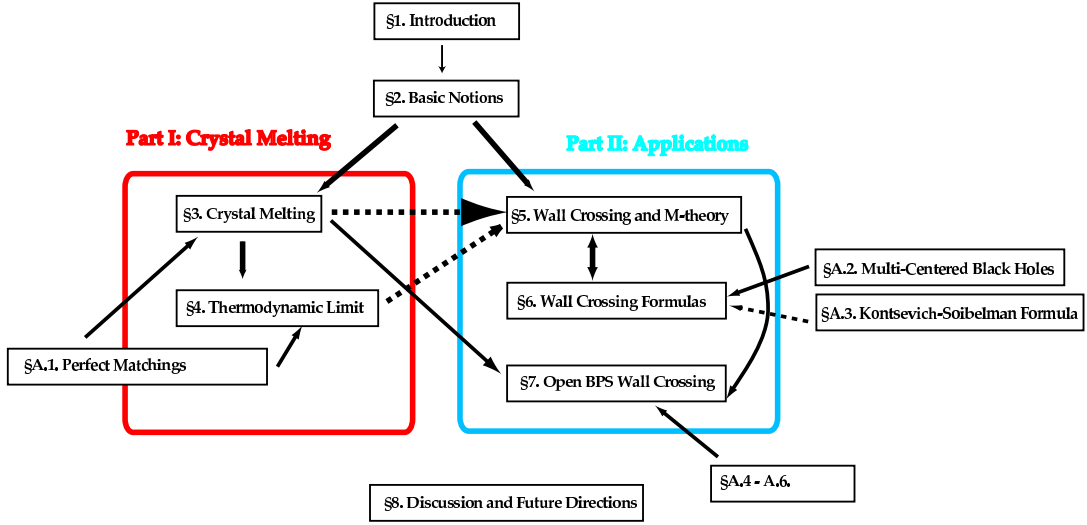}
  \caption{The interdependence of the sections of this paper.}
\label{fig.dependence}
\end{figure}

The interdependence of the sections are shown in Fig.\,\ref{fig.dependence}.

For the convenience of the reader, each section begins with a short summary
 of the section. The connections between different sections are emphasized throughout this review in order to organize the whole material into a unified framework. However, some sections are mostly independent, and do not require prior reading of earlier sections. For examples, most of section 5 does not require the prior reading of part I.

\bigskip

\section[Basic Notions]{Basic Notions}
\label{chap.basic}

Before venturing into the study of crystal melting and wall crossing phenomena beginning Section\,\ref{chap.crystal}, we here give a brief summary of the BPS states, BPS indices and their wall crossing phenomena. The material in this section is standard, and can safely be skipped by readers familiar with these topics.

\subsection{BPS States and BPS Index}\label{sec.index}

In theories with extended ($\scN>1$) supersymmetry in four dimensions, the supersymmetry algebra has a central extension $Z_{MN} (M,N=1,\ldots, \scN)$, which appears in the commutation relation of two supercharges $Q_{\alpha}^M$ (in the two-component notation):
\beq
\{Q_{\alpha}^M,Q_{\beta}^N \}=\epsilon_{\alpha \beta} Z_{MN}.
\eeq
From the above equation it follows that $Z_{MN}$ is an antisymmetric matrix. In the case of $\scN=2$ supersymmetry treated in this paper, $Z_{MN}$ has just one component (a complex number) $Z\equiv Z_{12}$. We can then show, purely from symmetry arguments, that the mass $M$ of a particle always satisfies the BPS inequality \footnote{This inequality was originally discovered by Bogomol'nyi, Prasad and Sommerfield \cite{Bogomolny,PrasadS} in the context of topological solitons, and was later shown to arise from the supersymmetry algebra \cite{WittenO}.}
\beq
M\ge \left| Z \right|.
\label{eq.BPS}
\eeq
States which saturates this inequality are called BPS states.

BPS states have played special roles in the study of supersymmetric gauge theories and string theories. This is because BPS states are protected from quantum corrections. In other words, quantum corrections may renormalize the value of $Z$, but the BPS inequality holds for all values of the coupling constant. This is basically due to the fact that BPS states are annihilated by certain combination of supercharges and thus their representation have smaller dimensions (i.e. they belong to short representations)
as compared with non-BPS states (which belong to long representations).
Therefore BPS states provide an indispensable tool to study the strong coupling behavior of supersymmetric gauge theories. They are also important for understanding black hole microstates, and also are related to mathematical invariants, as discussed in the introduction.

Let us try to be more specific. In this paper,
we are going to study the counting problem of BPS states arising from type IIA string theory compactified on a Calabi-Yau manifold $X$. 
Depending on whether $X$ is compact or non-compact, we have a 4d $\mathcal{N}=2$ supergravity or 
4d $\mathcal{N}=2$ supersymmetric gauge theory. 

We want to study BPS states which preserve half of the $\scN=2$ supersymmetry in the theory. These states span the BPS Hilbert space \footnote{This space is equipped with a product structure \cite{HarveyMoore}.}
$$
\scH_{\rm BPS},
$$
and we want to study the structure of this Hilbert space.

The BPS states have charges $\gamma$, which takes value in the ($\bZ$-valued) cohomology of $X$%
\beq
\gamma \in \Gamma=H^6(X;\bZ)\oplus H^4(X;\bZ)\oplus H^2(X;\bZ) \oplus H^0(X;\bZ),
\eeq
and this means that the BPS Hilbert states can be decomposed into sectors with specific charge $\gamma$: \footnote{More precisely, the RR-charges of D-branes are classified by K-theory (Refs.~\refcite{MM,Witten98}). For our purposes, however, cohomology is sufficient.}
\beq
\scH_{\rm BPS}=\bigoplus_{\gamma} \scH_{\textrm{BPS}, \gamma}.
\eeq

We define the BPS index $\Omega(\gamma)$ as an important quantity to understand the structure of $\scH_{\textrm{BPS}}$. This index is defined in such a way that the index does not jump when two short multiplets combine into a long multiplet. In four-dimensional $\scN=2$ supersymmetry the answer is unique, and is given by \footnote{The natural candidate for index is the $n$-th helicity supertrace defined by
\beq
\Omega_n(\gamma)=(-1)^{n/2} \frac{1}{n ! }\textrm{Tr}_{\scH_{\textrm{BPS} \gamma}}\,(-1)^{2J_3} (2J_3)^{n}. 
\eeq
However, there are several restrictions one has to worry about. First,
$n$ should be an even integer because we have CPT invariance and CPT
send $J_3$ to $-J_3$. Second, for $n=0$ the helicity supertrace
vanishes, and if $n$ is large then not only the short multiplets, but
also the long multiplets, begin to contribute. In general, if we have
four-dimensional $\scN$ supersymmetry, the only possibility is an even
integer $n$ satisfying $\scN \le n < 2\scN$, and for $\scN=2$, $n=2$ is
the only possible option. The quantity $\Omega(\gamma)$ in the main text
is $\Omega_2(\gamma)$ in this notation.
The situation is different, for example, for $\scN=4$. In this case,
 index is not unique and is given either by $\Omega_4(\gamma)$ or $\Omega_6(\gamma)$, or their linear combination. See appendix E of Ref.~\refcite{Kiritsis} for a nice summary.
}
\beq
\Omega(\gamma)
=-\frac{1}{2}\textrm{Tr}_{\scH_{\textrm{BPS},\gamma}} (-1)^{2 J_3} (2J_3)^2,
\label{indexdef}
\eeq
%
where $J_3$ is the 3rd component of the $SO(3)$ spatial rotation symmetry.
Also, the factor $-\frac{1}{2}$ is just a convention, which will turn out to be useful later.

We can calculate the index explicitly for an $\scN=2$ SUSY multiplet.
For our purposes we only need a massive multiplet, and 
a massive multiplet of $\scN=2$ theory is either\\
(1) a massive long representation,
\beq
L_j: [j]\otimes ([1] \oplus 4[1/2]\oplus 5[0]),\label{L_j}
\eeq
or (2) a short BPS massive multiplet
\beq
S_j: [j]\otimes ([1/2]\oplus 2[0]),
\label{Sj}
\eeq
where $S_0$ is a hypermultiplet and $S_{1/2}$ is a vectormultiplet.
Their contributions to the BPS index are
\beq
\Omega(S_j)=(-1)^{2j} (2j+1),\quad  \Omega(L_j)=0.
\eeq
This explicitly verifies that the long multiplets do not contribute, and that hypermultiplets and vectormultiplets contribute $1$ and $-2$ to the index.

Every BPS particle in an $\scN=2$ theory has a universal half-hypermultiplet obtained from quantizing the fermionic degrees of freedom associated to its center of mass (see \eqref{Sj}). We can therefore write $\scH_{\textrm{BPS},\gamma}=([1/2]\oplus 2[0])\otimes \scH^{'}_{\textrm{BPS},\gamma}$. Writing $J_3'$ for $\scH'_{\textrm{BPS},\gamma}$ we have
\beq
\Omega(\gamma)
&=-2 \ \textrm{Tr}_{\scH^{'}_{\textrm{BPS},\gamma}} (-1)^{2J_3} \left(
2(J_3)^2-(J_3-\frac{1}{2})^2-(J_3+\frac{1}{2})^2
\right)  \nonumber \\
&=\ \textrm{Tr}_{\scH^{'}_{\textrm{BPS},\gamma}} (-1)^{2J_3}.
\eeq

The final result is simply that we have the Witten index $\textrm{Tr} (-1)^F$, the number of boson minus the number of fermions, where bosons/fermions refer to the spin $j$ in \eqref{L_j}. This quantity is important since in many examples 
it is this index which gives the correct microstate degeneracies of black holes \footnote{
There are some discussions in the literature as to whether it is the index or the total degeneracy which gives the correct black hole entropy. 
See Ref.~\refcite{SenBH} for a recent discussion.}.

Let us define the generating function of the BPS index. As we will see later in examples, it is often more useful to study the generating function rather than to study each of the BPS index separately. Decompose the charge lattice $\Gamma$ into electric and magnetic charges lattices:
$$
\Gamma=\Gamma^e\oplus \Gamma^m.
$$
In type IIA compactification of a Calabi-Yau manifold, we have
\beq
\Gamma^e= H^6(X;\bZ)\oplus H^4(X;\bZ), \quad \Gamma^m=H^2(X;\bZ)\oplus H^0(X;\bZ),
\eeq
and when we write 
\beq
\gamma=(q_0,q_I,p^I,p^0)\in H^6(X;\bZ)\oplus H^4(X;\bZ)\oplus H^2(X;\bZ)\oplus H^0(X;\bZ),
\eeq
the electric charges are $q_{\Lambda}=(q_0,q_I)$ and the magnetic ones $p^{\Lambda}=(p^I,p^0)$. Here $I$ and $\Lambda$ run over $I=1,\ldots, \textrm{dim}\, H^4(X;\bZ)$ \footnote{For compact examples, we necessarily have $\textrm{dim}\, H^4(X;\bZ)=\textrm{dim}\, H^2(X;\bZ)$ because of Poincar$\acute{\rm e}$ duality. These can be different in non-compact examples we discuss later.} and $\Lambda=0,\ldots, \textrm{dim}\ H^4(X;\bZ)$, respectively.

We define the generating function $Z_{\rm BPS}$ by a 
\beq
Z_{\rm BPS}(\phi^{\Lambda},p^{\Lambda})=\sum_{q_{\Lambda}} \Omega\left(\gamma=(q_{\Lambda},p^{\Lambda})\right)\ e^{-\phi^{\Lambda} q_{\Lambda}}.
\eeq
This means that we are considering microcanonical
ensemble of magnetic charges $p^{\Lambda}$ and a canonical ensemble for electric charges $q_{\Lambda}$ with chemical potentials $\phi^{\Lambda}$.
Such a mixed microcanonical/canonical ensemble appears in the Ooguri-Strominger-Vafa conjecture \cite{OSV}. In particular, it was shown there that 
\beq
\scF(\phi^{\Lambda},p^{\Lambda}):=\log\left(Z_{\rm BPS}(\phi^{\Lambda},p^{\Lambda}) \right)
\eeq
is the Legendre transform of the entropy $S_{\rm BH}(q_{\Lambda},p^{\Lambda})$ of the black hole
\beq
S_{\rm BH}(q_{\Lambda},p^{\Lambda})=\scF (\phi^{\Lambda},p^{\Lambda})-\phi^{\Lambda} \frac{\partial}{\partial \phi^{\Lambda}}\scF(\phi^{\Lambda},p^{\Lambda}).
\eeq
For this reason, $Z_{\rm BPS}$ is sometimes denoted $Z_{\rm BH}$ in the literature.

\subsection{Wall Crossing Phenomena}\label{sec.WC}
In the previous section we defined the BPS index $\Omega(\gamma)$ and its generating function $Z_{\rm BPS}(\phi^{\Lambda},p^{\Lambda})$. From the viewpoint of counting BPS states all we need to do is to determine these numbers.
However, there is an interesting twist, the wall crossing phenomena. The wall crossing phenomena states that the BPS index $\Omega(\gamma)$ depends on the value of moduli of the Calabi-Yau manifold at infinity \footnote{We added ``at infinity'' here in order to emphasize that the value of the moduli is not the horizon value determined by the attractor mechanism \cite{attractor1,attractor2}.}. More concretely, consider the K\"ahler moduli of the Calabi-Yau manifold. Then wall crossing phenomena states that the BPS index jumps along the real codimension 1 subspace of the K\"ahler moduli space. The codimension 1 subspace is often called the walls of marginal stability (or sometimes simply walls), and the walls of marginal stability divides the K\"ahler moduli space into several regions, called chambers.
The BPS index is piecewise constant inside each of the chambers, but jumps as we cross the walls of marginal stability. We sometimes use the notation $\Omega(\gamma;t)$ to explicitly show the dependence on K\"ahler moduli.

What is the physics behind such a jump? Naively, the jump seems to be inconsistent with the common wisdom that index stays invariant under continuous deformations. The resolution to this puzzle is given by the fact that $\scH_{\textrm{BPS},\gamma}$ used in the definition of the BPS index \eqref{indexdef} is in fact a {\it one-particle} Hilbert space.

\begin{figure}[htbp]
\centering{\includegraphics[scale=0.6]{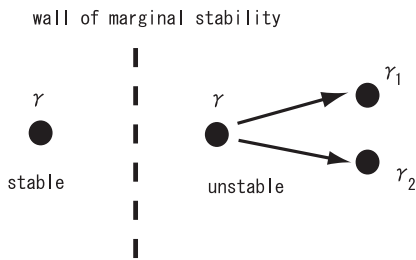}}
\caption[Decay and formation of BPS bound states under the change of moduli.]{Decay and formation of BPS bound states under the change of moduli. The values of moduli are different for left and right.}
\label{fig.decay}
\end{figure}

Consider the following situation: there is a stable particle with charge $\gamma$ when the K\"ahler moduli is $t=t_1$. When we change the K\"ahler moduli to $t=t_2$, the particle becomes unstable and decays into two particles with charges $\gamma_1$ and $\gamma_2$ (see Fig.\,\ref{fig.decay}. Of course, the charge conservation dictates that $\gamma=\gamma_1+\gamma_2$.). This means that the one-particle Hilbert space at $t=t_1$ is larger than that at $t=t_2$. This explains the jump of the BPS index: although the index is deformation invariant, the Hilbert space over which we are taking the trace jumps as we move around the K\"ahler moduli space \footnote{There is a neat supergravity picture for this, which will be explained in \ref{app.multi}.}. Of course, the reverse process is also possible: the stable particles can combine to make a new bound state. 

Now the next question is how to determine the positions of walls of marginal stability. In fact, this is easy. Consider the decay $\gamma\to \gamma_1+\gamma_2$. Then the particle with charge $\gamma$ is stable (unstable) if its mass $M(\gamma)$ is larger (smaller) than the mass or particles $\gamma_1$ and $\gamma_2$. In other words, on the walls of marginal stability we have
$$
M(\gamma)=M(\gamma_1)+M(\gamma_2).
$$
At the same time, since the particles are BPS the mass is given by the central charge \eqref{eq.BPS}
$$
M(\gamma)=|Z(\gamma) |.
$$
Since the central charge is a linear function with respect to charges, we have
$$
Z(\gamma)=Z(\gamma_1)+Z(\gamma_2).
$$
The only way to make the above three equations consistent is to have
\beq
\textrm{Arg}\,\left(Z(\gamma_1)\right)=\textrm{Arg}\,\left(Z(\gamma_2)\right).
\label{wallposition}
\eeq
This is the equation determining the position of walls of marginal stability.
From \eqref{wallposition} it follows that the position of walls of marginal stability for the decay process $\gamma\to \gamma_1+\gamma_2$ is the same as that for the process $\gamma\to N_1\gamma_1+N_2\gamma_2$. 

We can also consider more complicated decay patterns, for example $\gamma\to N_1\gamma_1+N_2\gamma_2+N_3\gamma_3$. However, such a decay occurs in a higher codimension subspace of the K\"ahler moduli, which in this case is determined by
$$
\textrm{Arg}\,\left(Z(\gamma_1)\right)=\textrm{Arg}\,\left(Z(\gamma_2)\right)=\textrm{Arg}\,\left(Z(\gamma_3)\right).
$$
Since these decay patterns are non-generic, we are not going to consider these complicated decay patterns in the rest of this paper.

\bigskip

In this section we have formulated the problem: to determine the BPS index or its generating function in each chamber of the moduli space. In the following sections we will see that this problem is beautifully solved using the crystal melting model.


\section{Crystal Melting}\label{chap.crystal}

In this section we discuss the crystal melting model proposed in Ref.~\refcite{OY1}. The work \cite{OY1} is motivated by and is based on earlier results in mathematics \cite{Szendroi,MR}.

\subsection{Overview: Crystal Melting and Topological Strings}\label{sec.overviewcrystal}

\subsubsection{Formulation of the Problem}

We already discussed BPS state counting in Section\,\ref{sec.index}, but here we would like to be more specific and would like to prepare some notations used throughout the rest of this paper.

There are different versions of BPS counting problems depending on which string theory and what kind of branes we consider. As in the previous section, in this paper we concentrate on the BPS counting problem of Type IIA string theory compactified on a Calabi-Yau manifold $X$ \footnote{By string duality this is sometimes related to the BPS state counting in different string theories. For example, type IIA string theory on $K3\times T^2$ is dual to Heterotic string theory on $T^6$.}. 
We emphasize here that the problem in this paper is the counting in the {\it physical} string theory, not in the topological string theory. Topological string theory comes later into the story, somewhat surprisingly (see Section\,\ref{chap.M-theory} for clarification on this point).

In this section and the following, we assume in addition that $X$ is a toric Calabi-Yau manifold. The advantage of this choice is that thanks to the torus action on $X$ we can often perform explicit computations, therefore making quantitative statements possible. Since $X$ is non-compact, the four-dimensional theory is an $\mathcal{N}=2$ supersymmetric gauge theory.

We consider 1/2 BPS states of this $\mathcal{N}=2$ theory. In string theory, BPS states are realized as supersymmetric BPS bound states of D-branes wrapping holomorphic cycles. We concentrate on the configuration of D-branes which are particles in 4d: in other words, we consider Dp-branes wrapping p-cycles. D-branes which are vortex strings and domain walls in 4d are also of potential interest, and we will see such examples in the discussion of open BPS invariants in Section\,\ref{chap.open}.

In type IIA string theory, we have Dp-branes with $p$ even.
We therefore consider D0/D2-branes wrapping holomorphic compact 0/2-cycles of the Calabi-Yau manifold. Due to the slightly technical reasons explained later, we do not include D4-branes wrapping 4-cycles. We also consider a single D6-brane filling the entire Calabi-Yau manifold. The D6-brane is different from other D-branes in that it wraps the whole $X$, which is non-compact (recall toric Calabi-Yau manifold is non-compact). We will see later that the D6-brane plays crucial roles in the discussion of the wall crossing phenomena.

Let us denote the D-brane charges of the particles by \footnote{We are changing the notation from the previous section. The notation here will be used throughout the rest of this paper.}
\beq
\gamma=(n,\beta,0,1) \in H^6(X;\bZ)\oplus H^4(X;\bZ)\oplus H^2(X;\bZ)\oplus H^0(X;\bZ),
\eeq
where $\beta=\{ \beta_I \}$ collectively denotes an element of $H^4(X;\bZ)$.
As in the introduction, the question is how to compute the BPS index 
$$
\Omega(\gamma),
$$
or its generating function
\beq
Z_{\rm BPS}(q,Q)=\sum_{n,\beta} \Omega\left(\gamma=(n,\beta,0,1)\right)q^n Q^{\beta},
\eeq
where $Q=\{Q_I \}$ and $Q^{\beta}:=\prod_I Q_I^{\beta_I}$.

\subsubsection{The Connection with Topological String Theory} 
Before discussing the crystal melting model, let us summarize the known connection between the BPS partition function and the topological string theory.
In the literature, the partition function $Z_{\rm BPS}$ we define here is related to topological string theory partition function $Z_{\rm top}$ in two different contexts. 

First, as explained in Section\,\ref{sec.index} the quantity $Z_{\rm BPS}$ is the same the black hole partition function $Z_{\rm BH}$ which appears in the context of Ooguri-Strominger-Vafa conjecture \cite{OSV}. The conjecture states that 
 when the D-brane charges are chosen such that bound 
states become large black holes with smooth event 
horizons, the generating 
function $Z_{{\rm BH}}$ of a suitable index for black hole 
microstates is equal to the absolute value squared of the 
topological string partition function $Z_{{\rm top}}$,
\beq Z_{{\rm BH}}=\left| Z_{{\rm top}}\right|^2, \eeq
%
to all orders in the string coupling expansion. 

Second, when there is a single D6 brane 
with D0 and D2 branes bound on it, it has been proposed \cite{INOV}
that the bound states are counted by the Donaldson-Thomas 
invariants \cite{DT,Thomas} of the moduli space of ideal sheaves \footnote{
An ideal sheaf $\scI$ is mathematically defined to be a torsion-free sheaf of rank 1 with trivial determinant. Given such an $\scI$, we can determine the subscheme $Y\subset X$ (roughly speaking, the subspace spanned by the D-brane) by 
$$
0\to \scI \to \scO_X\to \scO_Y.
$$
Donaldson-Thomas invariants $N_{n,\beta}$ counts the ``number'' of such ideal sheaves with
$$\chi(\scO_Y)=n, \quad [Y]=\beta\in H_2(X,\bZ),$$
where $n$ and $\beta$ are D0 and D2-brane charges as defined previously.
In this paper we do not use this explicit definition of (commutative) Donaldson-Thomas invariants, and assume no prior familiarity with them on the side of the reader.
}
on the D6 brane. For a non-compact toric Calabi-Yau manifold, the 
Donaldson-Thomas invariants are related to the topological string partition 
function \cite{ORV,INOV,MNOP1} using
the topological vertex construction \cite{AKMV}. Recently the 
connection between the topological string theory and the Donaldson-Thomas
theory for toric Calabi-Yau manifolds was proven mathematically 
in Ref.~\refcite{MOOP}. Given the conjectural relation between
the counting of D-brane bound states and the Donaldson-Thomas 
theory, it is natural to expect the relation, 
\beq 
Z_{{\rm BH}}=Z_{{\rm top}}. \label{BH=top} 
\eeq
Note that these two conjectures are supposed to 
hold in different regimes of validity, as we will discuss in 
the final section (Section\,\ref{chap.discussion}).

From the viewpoint of counting BPS states in type IIA string theory, the parameters $q$ and $Q$ are just formal variables for the generating function. We will see later that in the context of topological string theory they are related to the topological string coupling constant and the K\"ahler moduli (see \eqref{topvar}).

\subsubsection{What is Discussed in This Section}\label{subsec.Whatis}

The purpose of this section is to understand the case \eqref{BH=top} better.
We start with the 
left-hand side of the relation, namely the counting of BPS states. 
Recently, the non-commutative 
version of the Donaldson-Thomas theory is formulated by 
Szendr\"oi \cite{Szendroi} for the conifold 
and by Mozgovoy and Reineke \cite{MR} for general toric 
Calabi-Yau manifolds \footnote{See Refs.~\cite{Young1,Young2,NN,Nagao1,Nagao2,Nagao3,OY1,OY2,NY,Sulkowski,JM,CJ,CP,Zhou,Gholampour,Krefl} 
for further developments.}.
In this section, we will establish a direct connection between
the non-commutative Donaldson-Thomas theory and 
the counting of BPS bound states of D0 and D2 branes on 
a single D6 brane. Using this correspondence, we will
find a statistical model of crystal melting which counts
the BPS states. The formula we prove will be written as \footnote{As will be discussed in detail in Section\,\ref{chap.M-theory}, this is a specialization of the formula \eqref{Z=Z2}.}
\beq 
Z_{{\rm BH}}=Z_{{\rm crystal}}. \label{BH=crystal} 
\eeq

Some readers will be more familiar with the
crystal melting description of the topological string theory 
on the right-hand side of \eqref{BH=top}. 
It was shown in Refs.~\cite{ORV,INOV} that 
the topological string partition function on 
$\bC^3$, the simplest toric Calabi-Yau manifold,
and the topological vertex can be expressed 
as sums of three-dimensional Young diagrams,
which can be regarded as complements of molten
crystals with the cubic lattice structure \footnote{See Refs.~\refcite{SV,foamtwo,DVVafa,Sulkowskicrystal,Jafferis,HeckmanV,DOR} 
for further developments.}. Since 
the topological vertex can be used to compute
the topological string partition function for
a general non-compact toric Calabi-Yau manifold,
it is natural to expect that a crystal melting 
description exists for any such manifold. To our 
knowledge, however, this idea has not been made
explicit. The crystal melting model defined in 
this section is different from the one 
suggested by the topological vertex construction, and 
the precise relation will be explained in the discussion of wall crossing phenomena in part II.

\bigskip

We should next explain how the crystal arises. The short answer is that it arises from quiver quantum mechanics \footnote{One should keep in mind that the appearance of quantum mechanics (and the computation of the BPS index by it) is more universal than in our toric setup, and should apply to any Calabi-Yau manifold.
The advantage of our setup is that the precise form of the quantum mechanics is known, thanks to the brane tiling techniques. The torus action also helps to compute the index of the quantum mechanics, since we compute the Euler character of the moduli space by localization with respect to the torus action originating from the toric condition. 
}.

Instead of studying the full dynamics of four-dimensional gauge theory, we can use the low energy effective theory on the D-brane worldvolume to compute the index.
The low energy effective theory of D0 and D2 branes
bound on a single D6 brane is a one-dimensional
supersymmetric gauge theory, which is a dimensional reduction
of an ${\cal N}=1$ gauge theory in four dimensions.
Since the index is deformation invariant and thus invariant under the process of dimensional reduction, the BPS index reduces to the Witten index of this 
one-dimensional theory, i.e. quantum mechanics. The Witten index of quantum mechanics is in turn given by the Euler character of the vacuum moduli space \footnote{\label{sign1} More precisely, the correct statement is 
$$
\Omega(\gamma)=(-1)^{\textrm{dim}\, T\scM_{\rm vac}}\chi (\scM_{\rm vac}),
$$
where $\scM_{\rm vac}$ is the vacuum moduli space and $T\scM_{\rm vac}$ is the tangent space of the moduli space \cite{Szendroi}. The sign has the effect of shifting the sign of chemical potentials of the crystal melting model, as will be explained in more detail in \eqref{eq.sign2}  in Section\,\ref{sec.ZBPS}.}. 
This statement will be made more precise in Section\,\ref{sec.quiver}.

The field content of the quantum mechanics is encoded in a
quiver diagram and the superpotential can be found by the so-called
brane tiling techniques \cite{BT1,BT2,BT3,BT4} \footnote{See 
Refs.~\refcite{Kennaway,Yamazaki} for 
reviews of the quiver gauge theory and the brane tiling method.}. 
From these gauge theory data, we define a crystalline structure 
in three dimensions.
The crystal is composed of atoms of different colors, each
of which corresponds to a node of the quiver diagram and carries
a particular combination of D0 and D2 charges. The chemical bond 
is dictated by the arrows of the quiver diagram. There is a special
crystal configuration, whose exterior shape lines up with the toric 
diagram of the Calabi-Yau manifold. Such a crystal corresponds to 
a single D6 brane with no D0 and D2 charges. 
We define a rule to remove atoms from 
the crystal, which basically says that the crystal melts from its
peak. By using the non-commutative Donaldson-Thomas theory \cite{Szendroi,MR}, 
we show that there is a one-to-one correspondence between
molten crystal configurations and BPS bound states carrying 
non-zero D0 and D2 charges. The statistical model of crystal melting 
computes the index of D-brane bound states. 

The number of BPS states depends on the choice of the stability condition,
and the BPS countings for different 
stability conditions are related to each other by
the wall crossing formulas. In this section, we find that,
under a certain stability condition,
BPS bound states of D-branes are counted by 
the non-commutative Donaldson-Thomas theory.
As we will see in Section\,\ref{chap.M-theory}, we can relate non-commutative Donaldson-Thomas theory to the commutative Donaldson-Thomas theory by a series of wall crossings.
Since the topological string theory is
equivalent to the commutative Donaldson-Thomas theory 
for a general toric Calabi-Yau manifold \cite{MOOP},  
the relation \eqref{BH=top} is indeed true for
some choice of the stability condition, as 
expected in Ref.~\refcite{INOV}. 
In general, the topological string partition function and the partition 
function of the  crystal melting model are not identical,
but their relation involves the wall crossing, 
\beq
   Z_{\rm crystal~melting} \sim Z_{\rm top}~~~({\rm modulo~wall~crossings}).
\eeq
This does not contradict with the result in Refs.~\refcite{ORV,INOV} 
since there is no wall crossing phenomenon involved between commutative and non-commutative Donaldson-Thomas invariants for $\bC^3$. This shows that wall crossing phenomena (to be explained in part II) is crucial for understanding the precise relation between crystal melting and topological string theory.

\begin{figure}[htbp]
\centering{\includegraphics[scale=0.4]{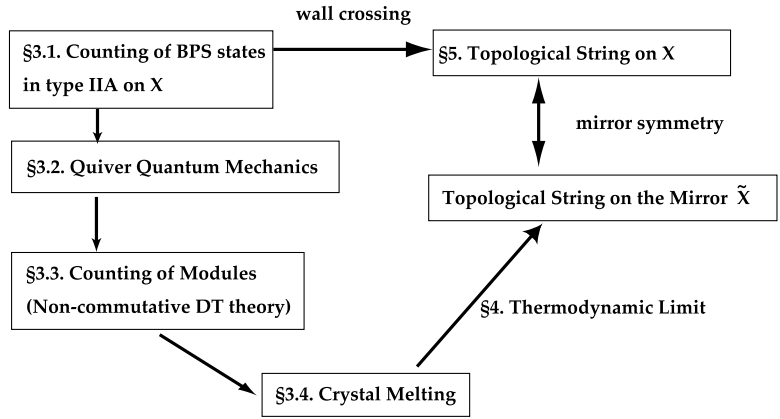}}
\caption[The organization of Section\,\ref{chap.crystal} and the relation to Section\,\ref{chap.thermodynamic} and Section\,\ref{chap.M-theory}.]{The organization of this section and the relation to Section\,\ref{chap.thermodynamic} and Section\,\ref{chap.M-theory}.}
\label{partI}
\end{figure}

The rest of this section is organized as follows (see Fig.\,\ref{partI}).
In Section\,\ref{sec.quiver}, we will summarize the computation of 
D-brane bound states from the gauge theory perspective. 
In Section\,\ref{sec.NCDT}, we will discuss how this is related to
the recent mathematical results on the non-commutative
Donaldson-Thomas invariants. In Section\,\ref{sec.crystal}, we will 
formulate the statistical model of crystal melting for a general
toric Calabi-Yau manifold, whose partition function is computed in Section\,\ref{sec.ZBPS}. 
The final section is devoted to the summary of our result 
and discussion on the wall crossing phenomena.
The equivalence of a configuration 
of molten crystal with a perfect matching of the bipartite graph
is explained in 
appendix \ref{app.PM}.

\subsection{Quiver Quantum Mechanics} \label{sec.quiver}

Let us begin with the description of quiver quantum mechanics. The first problem is to give a precise identification of the quantum mechanics on the worldvolume of D-branes.

In the classic paper by Douglas and Moore \cite{DouglasM}, 
it was shown that the low energy effective theories
of D-branes on a class of orbifolds are described by gauge theories 
associated to quiver diagrams. Subsequently, this result 
has been generalized to an arbitrary non-compact toric 
Calabi-Yau threefold. A toric Calabi-Yau threefold $X_{\Delta}$ 
is a $T^2 \times \bR$-fibration over $\bR^3$, where the
fibers are special Lagrangian submanifolds. The toric diagram 
$\Delta$ tells us where and how the fiber degenerates. For a given 
$X_\Delta$ and a set of D0 and D2 branes on $X_\Delta$, the following 
procedure determines the field content and superpotential
of the gauge theory on the branes. We will add a single D6 brane
to the system later in this section. 

\subsubsection{Quiver Diagram and Field Content}\label{subsec.quiver}

The low energy gauge theory is a one-dimensional theory given 
by dimensional reduction 
of an ${\cal N}=1$ supersymmetric gauge theory in four 
dimensions. The field content of the theory is encoded in 
a quiver diagram, which is determined from the toric data 
and the set of D-branes, as described in the following. 
A quiver diagram $Q=(Q_0,Q_1)$ consists of 
a set $Q_0$ of nodes, 
with a rank $N_i > 0$ associated to each node $i \in Q_0$, 
and a set $Q_1$ of arrows connecting the nodes. 
The corresponding 
gauge theory has a vector multiplet of gauge group $U(N_i)$ 
at each node $i$. There is also a chiral multiplet in the 
bifundamental representation associated to each arrow connecting
a pair of nodes. 

\begin{figure}[htbp]
\centering{\includegraphics[scale=0.3]{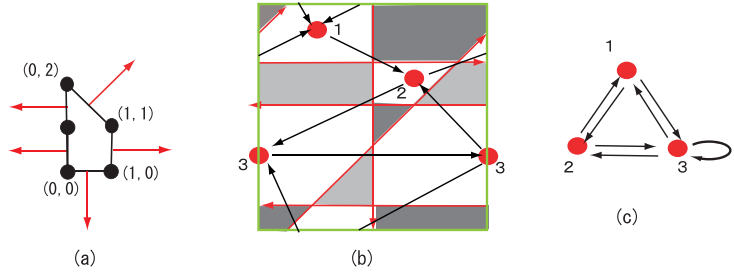}}
\caption[The brane configuration for the Suspended Pinched Point.]{(a) The toric diagram for the Suspended Pinched Point
singularity.
(b) The configuration of D2 and NS5 branes after the T-duality on
$\bT^2$. The green exterior lines are periodically identified. 
The red lines representing NS5 branes 
separate the fundamental domain into several domains. 
The T-dual of D0 branes wrap the entire fundamental domain,
and fractional D2 branes are suspended between the red lines. 
The white domains contain D2 branes only. In each shaded
domain, there is an additional NS5 brane. There are two types 
of shades depending of the NS5 brane orientation. 
The white domains are connected
by arrows through the vertices, and the directions of the arrows are
determined by the orientation of the NS5 branes. (c) The quiver diagram
obtained by replacing the white domains of (b) by the nodes.}
\label{fig.SPPtiling}
\end{figure}

In the following, we will explain how to identify the 
quiver diagram. Readers are encouraged to consult Fig.\,\ref{fig.SPPtiling},
which describes the procedure for the Suspended Pinched Point (SPP) singularity, 
which is a Calabi-Yau manifold defined by the toric diagram
in Fig.\,\ref{fig.SPPtiling}-(a) or equivalently by the equation,
\beq
   xy = z w^2,
\eeq
in $\bC^4$.

To identify the quiver diagram, we take a T-dual of 
the toric Calabi-Yau manifold along the $\bT^2$ fibers \cite{BT2,FHKV}. 
The fibers degenerate at loci specified by the toric 
diagram $\Delta$, and the T-duality replaces the 
singular fibers by NS5 branes \cite{OVBH}. Some of these NS5 branes
divide $\bT^2$ into domains as shown in the red lines
in Fig.\,\ref{fig.SPPtiling}-(b)
\cite{Imamura1,Imamura2,IIKY,Yamazaki}. 
The D0 branes become
D2 branes wrapping the whole $\bT^2$. The original
D2 are still D2 branes after the T-duality, but 
each of them is in a particular domain of $\bT^2$ 
suspended between NS5 branes. In addition, there are
some domains that contain NS5 branes stretched 
two-dimensionally in parallel with D2 branes \footnote{
The NS5 branes are also filling the four-dimensional
spacetime $\bR^{1,3}$ while the D2 branes are localized along a timelike
path in four dimensions.}.
 Let us denote the domains
without NS5 branes by $i \in Q_0$ and the domains with 
NS5 branes by $a \in I$. In Fig.\,\ref{fig.SPPtiling}-(b), the $Q_0$-type
domains are shown in white, and the $I$-type domains are
shown with shade. There are two types of shades, corresponding
to two different orientations of NS5 branes. This distinction
will become relevant when we discuss the superpotential. 

The $Q_0$-type domains
are identified with nodes of the quiver diagram since 
open strings ending on them can contain massless excitations.
The rank $N_i$ of the node $i \in Q_0$ is the number of D2 branes
in the corresponding domain. On the other hand, $I$-type domains
give rise to the superpotential constraints as we shall see below.  
Though two domains $i, j \in Q_0$ never share an edge, they 
can touch each other at a vertex. In that case, open strings 
going between $i$ and $j$
contain massless modes. We draw an arrow from $i\rightarrow j$ 
or $i\leftarrow j$ depending on the orientation of the massless 
open string modes, which is determined by the orientation of 
NS5 branes.  Note that the quiver gauge theory 
we consider in this section are in general chiral. 
This completes the specification of the quiver diagram. 

As another example,
the quiver diagram for the conifold geometry
has two nodes connected by two sets
of arrows in both directions. The ranks of the gauge groups
are $n_0$ and $n_0+ n_2$, where $n_0$ and $n_2$ are the numbers
of D0 and D2 branes \footnote{In the context of 4d $\scN=1$ quiver gauge theories, these branes are called D3 and D5-branes, the latter often called fractional branes.}. The gauge theory is a dimensional reduction
of the Klebanov-Witten theory \cite{KW} when $n_2=0$ and 
the Klebanov-Strassler theory \cite{KlebanovS} when $n_2 >0 $.

\subsubsection{Superpotential and Brane Tiling}\label{tiling.sec}

Each domain $a \in I$ containing an NS5 brane 
is surrounded by domains $i_1, i_2, ..., i_n \in Q_0$ without NS5 branes, 
as in Fig.\,\ref{fig.SPPtiling}-(b). By studying the geometry T-dual to $X_\Delta$ 
in more detail, one finds that the domain is contractible. Since 
open strings can end on the domains $i_1, i_2, ..., i_n$, 
the domain $a$ can give rise to worldsheet instanton
corrections to the superpotential. 

This fact, combined with the requirement that 
the moduli space of the gauge theory agrees with the geometric 
expectation for D-branes on $X_\Delta$, determines the superpotential.
Depending on the NS5 brane orientation, 
the $I$-type domains are further classified into two types, $I_+$ and $I_-$,
and thus the regions of  torus is divided into three types $Q_0$, $I_+$ and $I_-$.
Such a brane configuration, or a classification of regions of $\bT^2$, is called the brane tiling \footnote{In the literature the word brane tiling refers to the bipartite graph explained below.
Here the word brane tiling refers to a brane configuration  
as shown  Fig.\,\ref{fig.SPPtiling}-(b). Such a graph is called the fivebrane diagram in Ref.~\refcite{IKY}.}.
In Fig.\,\ref{fig.SPPtiling}-(b), the brane tiling is shown by the two different shades. 
The superpotential $W$ is then given by
\beq
 W =  \sum_{a\in I_+} {\rm Tr}\left( \prod_{e\in a,\, \textrm{clockwise}}
A_{e}\right)
   -   \sum_{a\in I_-} {\rm Tr}\left( \prod_{e\in a, \,\textrm{counterclockwise}}
A_{e}\right), \label{eq.W}
\eeq
where the relation $e\in a$ means that an edge $e$ is one of the boundary edges of the face $a \in I_\pm$, and the product is over all such edges in a (counter)clockwise manner. 
This formula is tested in many examples. In particular, 
it has been shown that the formula reproduces the toric 
Calabi-Yau manifold $X_{\Delta}$ as the moduli space of the
quiver gauge theory \cite{FV,IU}.

\begin{figure}[htbp]
\centering{\includegraphics[scale=0.3]{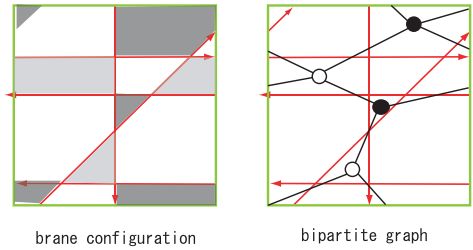}}
\caption[The brane configuration gives the bipartite graph.]
{The correspondence between the brane configuration on 
$\bT^2$  and the bipartite graph. The white (black) vertex of the
bipartite graph corresponds to the region $I_+$ ($I_-$) in light (dark) 
shade. The edge of the bipartite graph corresponds to an
intersection of $I_-$ and in $I_+$. From this construction, it
automatically follows that the graph so obtained is bipartite.}
\label{fig.SPPbipartite}
\end{figure}

\begin{figure}[htbp]
\centering{\includegraphics[scale=0.22]{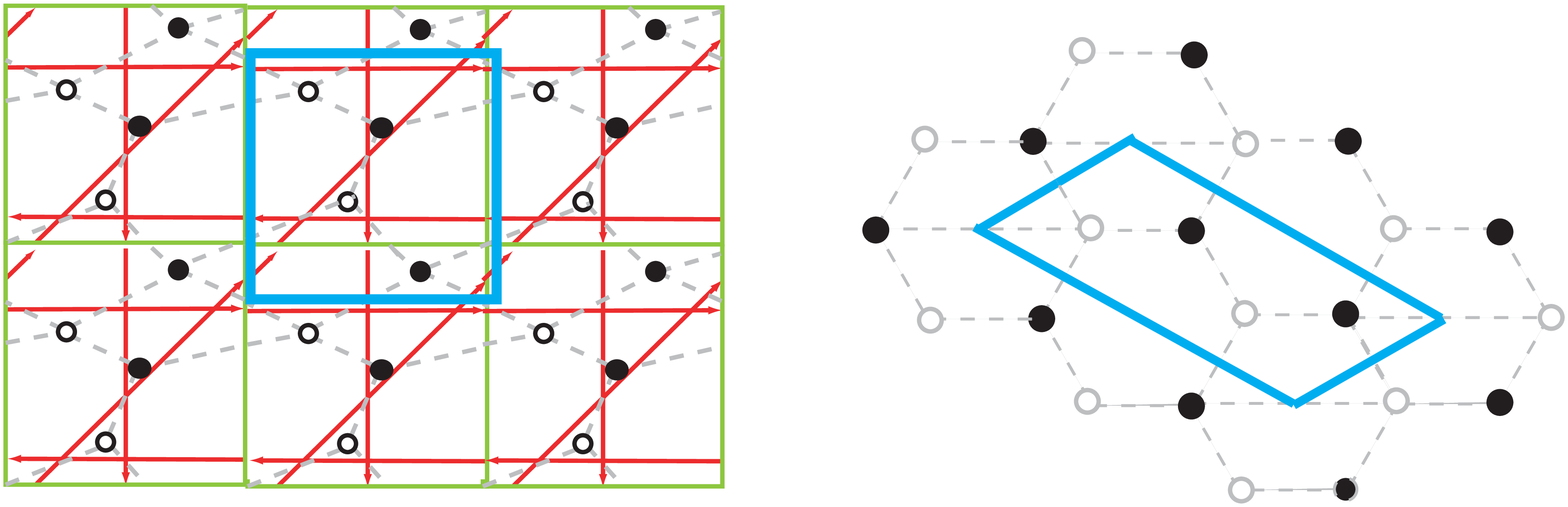}}
\caption[The relation with a bipartite graph in the literature.]{By choosing a different fundamental region
of $\bT^2$, we find a bipartite graph which is more commonly found in the literature.}
\label{fig.SPPregion}
\end{figure}

In the literature of brane tiling, bipartite graphs are often used
in place of brane configurations as in Fig.\,\ref{fig.SPPtiling}-(b). 
A bipartite graph is a graph consisting of vertices colored 
either black or white and edges connecting black and white
vertices. Since bipartite graphs will also play roles in the following
sections, it would be useful to explain how it is related to our story
so far. For a given brane configuration, we can draw a bipartite graph 
on $\bT^2$ as follows. In each domain in $I_{+}$ ($I_-$), place a 
white (black) vertex. Draw a line connecting a white vertex 
in a domain $i\in I_{+}$ and a black vertex in a neighboring 
domain $j\in I_-$.  The resulting graph $\Gamma$ is bipartite.
See Fig.\,\ref{fig.SPPbipartite} for the comparison of the brane configuration and
the bipartite graph in the case of the Suspended Pinched Point
singularity. We can turn this into a form that is more 
commonly found in the literature,
for example in Ref.~\refcite{BT2}, by choosing a different fundamental
region as in Fig.\,\ref{fig.SPPregion}.

\subsubsection{D-term Constraints and the Moduli Space}

The F-term constraints are given by derivatives of the superpotential, 
which can be determined as in the above. 
The moduli space of solutions to the D-term constraints is
then described by a set of gauge invariant observables divided out by
the complexified gauge group $G_{\bC}$ \cite{LT}.  The theorem 
by King \cite{King} states that an orbit 
of $G_{\bC}$ contains a solution to the D-term conditions
if and only if we start with a point that satisfies the 
$\theta$-stability, a condition defined in the next section. 
Thus, we can think of the moduli space as 
a set of solutions to the F-term constraints obeying the
$\theta$-stability condition, modulo the action of $G_{\bC}$. 

\subsubsection{Adding a Single D6 Brane}\label{subsec.addingD6}
To make contact with the Donaldson-Thomas theory, 
we need to include one D6 brane.
Since the D6 brane fills the entire Calabi-Yau manifold, which is
non-compact, it behaves as a flavor brane. In the low energy
limit, the open string between the D6 brane and another D-brane
gives rise to one chiral multiplet in the fundamental 
representation for the D-brane on the other end.
The D6 brane then enlarges the quiver diagram by 
one node and one arrow from the new node. 
To understand why we only get one arrow
from the D6 brane, let us take T-duality along the $\bT^2$
fiber again. The D6 brane is mapped into a D4 brane which is a point 
in some region in $\bT^2$. This means that we only have one new 
arrow from the new node corresponding to the D6 brane to the
node corresponding to the D2 branes in the region \footnote{This intuitive picture holds only in the special chamber where the BPS indices coincide  with the non-commutative Donaldson-Thomas invariants. This is because the basis of fractional D2-branes corresponding to the nodes of the quiver diagram are different in different chambers of the moduli space. See Section\,\ref{subsec.D6}.}. 
See Refs.~\refcite{Jafferis,Szabo} for related discussion in the literature.


\subsection{Non-commutative Donaldson-Thomas Theory} \label{sec.NCDT}

In the previous section, we discussed how to construct
the moduli space of solutions to the F-term and D-term
constraints in the quiver gauge theory corresponding to
a toric Calabi-Yau manifold $X_\Delta$ with a set of D0/D2 branes 
and a single D6 brane. In this section, we will review
and interpret the mathematical formulation of the non-commutative
Donaldson-Thomas invariant in Refs.~\refcite{Szendroi,MR} for $X_\Delta$.
We find that it is identical to the Euler number 
of the gauge theory moduli space.

\subsubsection{Path Algebra and its Module}

For the purpose of this section, modules are the same as representations. 
Consider a set of all open paths on the quiver diagram $Q=(Q_0, Q_1)$. By 
introducing a product as an operation to join a head of a path to a tail 
of another (the product is supposed to vanish if the head and the tail 
do not match on the same node) and by allowing formal sums of paths over $\bC^3$, the set of open oriented paths can be made into an algebra $\bC Q$ called the 
path algebra. Note that we are including both gauge variant as well as gauge invariant operators. In physics we usually do not consider such a huge algebra, but it has turned out to be a useful language for our purposes.

We would like to
point out that there is a one-to-one correspondence between 
a representation of the path algebra and a classical 
configuration of bifundamental fields of the quiver gauge theory. 
Suppose there is a representation $M$ of the path algebra. For each 
node  $i \in Q_0$, there is a trivial path $e_i$ of zero length that
begins and ends at $i$. It is a projection, $(e_i)^2 = e_i$, and satisfies 
 $e_i a_i=a_i$ for any arrow $a_i$ starting from vertex $i$. 
Since every path starts at some node $i$
and ends at some node $j$, 
the sum $\sum_i e_i$ acts as the identity on the path algebra. 
Therefore, 
$M = \oplus_{i\in Q_0} M_i$, where $M_i = e_i M$. 
Let us write $N_i = \dim M_i$. For each
path from $i$ to $j$, one can assign a map from $M_i$ to $M_j$.
In particular, there is an $N_i \times N_j$ matrix for each 
arrow $i \rightarrow j \in Q_1$ of the quiver diagram. By identifying this
matrix as the bifundamental field associated to the arrow
$i \to j$, we obtain a classical configuration 
of bifundamental fields with the gauge group $U(N_i)$ at 
the node $i$. By reversing the process, we can construct
a representation of the path algebra for each configuration of
the bifundamental fields. 

\subsubsection[F-term Constraints and Factor Algebra]{F-term Constraints and Factor Algebra $A$} \label{F-term.subsec}

Let us turn to the F-term constraints. Since the bifundamental
fields of the quiver gauge theory is a representation of the
path algebra, the F-term equations give
relations among generators of the path algebra. It is natural to
consider the ideal ${\cal F}$ generated by the F-term equations
and define the factor algebra $A = \bC Q/{\cal F}$.
The bifundamental fields obeying the F-term constraints then
generate a representation of this factor algebra. Namely,
classical configurations of the quiver gauge theory 
obeying the F-term constraints are in one-to-one correspondence
with $A$-modules. 

As an example, the algebra $A$ for the conifold geometry contains an idempotent ring $\bC[e_1,e_2]$ generated by two elements and is given by the following four generators and relations: \footnote{The center $Z(A)$ of this algebra $A$ is generated by $x_{ij}=a_i b_j+b_j a_i (i,j=1,2)$, and is given by 
\beq
Z(A)=\bC[x_{11},x_{12},x_{21},x_{22}]/(x_{11}x_{22}-x_{12}x_{21}),
\eeq
which is the ring of functions of the conifold singularity.
}
\beq
A=\bC [e_1,e_2]\langle a_1,a_2,b_1,b_2 \rangle / 
\left(a_1 b_i a_2=a_2 b_i a_1, b_1 a_i b_2=b_2 a_i b_1  \right)_{i=1,2},
\eeq
Each $A$-module for this algebra corresponds 
to a choice of ranks of the gauge groups and a
configuration of the bifundamental fields $a_i, b_i$
satisfying the F-term constraints. 

\begin{figure}[htbp]
\centering{\includegraphics[scale=0.3]{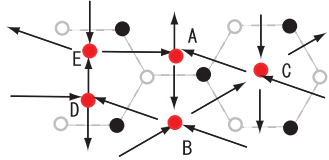}}
\caption[F-term constraints on the quiver diagram. ]{Representation of F-term constraints on the quiver diagram on $\bT^2$. In this example, if we write by $X_{AB}$ the bifundamental corresponding to an arrow starting from vertex $A$ and ending at $B$ etc., 
then the superpotential \eqref{eq.W} contains a term
$
W=-\textrm{tr}(X_{AB}X_{BC}X_{CA})+\textrm{tr}(X_{AB}X_{BD}X_{DE}X_{EA}),
$
and the F-term condition for $X_{AB}$ (multiplied by $X_{AB}$) says that the product of bifundamentals fields along the triangle $ABC$ and that along the square $ABDE$ is the same.
}
\label{fig.SPPFterm}
\end{figure}

F-term constraints have a nice geometric interpretation on 
the quiver diagram, which we will find useful in the next section. 
We observe that each bifundamental field appears exactly twice 
in the superpotential with different signs of coefficients
in the superpotential shown in \eqref{eq.W}. 
By taking a derivative of the superpotential with respect
to a bifundamental field corresponding to a given arrow, 
the resulting F-term constraint states that 
the product of bifundamental fields around a face of 
the quiver on $\bT^2$ on one side of the arrow is equal to that 
around the face on the other side. See Fig.\,\ref{fig.SPPFterm} 
for an example. Therefore, when we have a product of bifundamental
fields along a path, any loop on the path can be moved along
the path and the resulting product is F-term equivalent to the
original one.  In Ref.~\refcite{MR}, it is shown that for any point 
$i, j \in Q_0$, we can find a shortest path $v_{i,j}$ from $i$  to $j$ 
such that any other path $a$ from $i$ to $j$ is F-term equivalent to 
$v_{i,j} \omega^n$ with non-negative integer $n$, where $\omega$ 
is a loop around one face of the quiver diagram.
It does not matter where the loop $\omega$ is inserted along the 
path $v_{i,j}$ since different insertions are all F-term equivalent.
This means that any path is characterized by the integer $n$ and
the shortest path $v_{i,j}$. 

In the next subsection, we will impose the D-term constraints on
the space of finitely generated left $A$-modules, $\module A$. 
Before doing this, however, it is instructive to discuss 
topological aspects of $\module A$ by considering its bounded 
derived category \footnote{See Ref.~\refcite{Aspinwall} 
for an introductory explanation of derived categories 
in the context of string theory.} $D^b (\module A)$.
In mathematics, the algebra $A$ gives 
the so-called ``non-commutative crepant resolution'' \footnote{The origin of the name ``non-commutative'' stems from the fact that the algebra $A$ is non-commutative.} \cite{VdB}.
For singular Calabi-Yau manifolds such as $X_\Delta$, 
the crepant resolution means a resolution that preserves
the Calabi-Yau condition \footnote{Mathematically, we 
mean a resolution $f: Y_{\Delta}\to X_{\Delta}$ such 
that $\omega_Y=f^*\omega_X$, where $\omega_X$ and $\omega_Y$ 
are canonical bundles of $X$ and $Y$, respectively. 
For the class of toric Calabi-Yau threefolds, the existence 
of crepant resolution is known and different crepant resolutions 
 related by flops are equivalent in derived categories \cite{BridgelandFlop}.}.  For a crepant resolution $Y_\Delta$ of $X_\Delta$, we have the following
equivalence of categories \footnote{This is well-known in the case of 
the conifold (cf. Ref.~\refcite{VdB2}). For general toric Calabi-Yau threefolds, 
this is proven in Ref.~\refcite{IU2}. See also Refs.~\refcite{UY2,UY3}.}:
\beq
D^b\left(\coh(Y_{\Delta})\right)\cong D^b (\module A),\label{eq.coh=mod}
\eeq
where $D^b(\coh (Y_{\Delta}))$ is a bounded derived category of coherent 
sheaves of crepant resolution $Y_{\Delta}$, 
and $D^b(\module A)$ is the bounded derived categories of 
finitely generated left $A$-modules.
The equation \eqref{eq.coh=mod} is also interesting from 
the physics viewpoint.
Since $D^b(\coh (Y_{\Delta}))$ gives a topological classification 
of A branes \footnote{Some readers may wonder whether this should be B branes, since it is often stated in the literature that coherent sheaves classify B branes. However, in those references we consider D-branes filling the whole $\bR^{3,1}$, whereas in our context D-branes spread only in the time direction in $\bR^{3,1}$. This means coherent sheaves also describe particle-like D-branes in type IIA string theory.
} on the resolved space $Y_\Delta$, the equivalence
means that $D^b (\module A)$ also classifies D-branes, which 
is consistent with our interpretation above that 
$A$-modules are in one-to-one
correspondence with a configuration of bifundamental fields
obeying the F-term constraints.

We should note that the paper \cite{MR}, which computes the non-commutative
Donaldson-Thomas invariants for general toric Calabi-Yau manifolds,
requires a set of conditions on brane tilings, namely on 
the superpotential. It is easy to show that the conditions specified in lemma 3.5 and conditions 4.12 of Ref.~\refcite{MR} are automatically satisfied for any quiver gauge theories for D-branes on general toric Calabi-Yau manifold. 
To prove the condition 5.3 is more difficult (see Ref.~\refcite{Larjo} for a proof \footnote{See Refs.~\refcite{IU2,Broomhead,Davison} for more on consistency conditions in the mathematics literature.}).

\subsubsection[D-term Constraints and Theta-Stability]{D-term Constraints and $\theta$-Stability}

We saw that the derived category $D^b (\module A)$ of $A$-modules
gives the topological classification of D-branes in the toric Calabi-Yau
manifold $X_\Delta$. To understand the moduli space of D-branes, however,
we also need to understand implications of the D-term constraints.
This is where the $\theta$-stability comes in \footnote{The
$\theta$-stability is a special limit of $\Pi$-stability as 
discussed in Refs.~\refcite{DFR,Bridgeland}.}. 
Let $\theta\in \bN^{Q_0}$ be a vector whose components are real numbers. 
Consider an $A$-module $M$, and recall that this $M$ is decomposed 
as $M=\oplus_{i\in Q_0} M_i$ with $M_i = e_i M$. 
The module $M$ is called $\theta$-stable if
\beq
\sum_{i\in Q_0} \theta_i (\dim e_i M')> 0. \label{eq.thetastability}
\eeq
for every submodule $M'$ of $M$ \footnote{In some literature, 
an additional condition
$\sum_{i\in Q_0} \theta_i (\dim M_i)=0$ is imposed for a choice
of $\theta$. This is trivially satisfied for the choice $\theta=(0,0,\dots,0)$ we choose below.}. 
When $>$ is replaced by $\ge$, 
the module $M$ is called $\theta$-semistable. 

In the language of gauge theory, the stability 
condition \eqref{eq.thetastability} is required by the D-term conditions.
Some readers might wonder why the D-term conditions, which are equality
relations, can be replaced by an inequality as \eqref{eq.thetastability}. 
In fact, the similar story goes for the Hermitian Yang-Mills equations. 
There instead of solving the Donaldson-Uhlenbeck-Yau equations, 
we can consider holomorphic vector bundles with a suitable 
stability condition, the so-called $\mu$-stability or Mumford-Takemoto 
stability \cite{Donaldson,UhlenbeckY}. As we mentioned at the end of section 2, it is known that
a configuration of bifundamental fields is mapped to a solution 
to the D-term equations
by a complexified gauge transformation $G_{\bC}$ if and only if
the configuration is $\theta$-stable. 
Since each $A$-module $M$ gives a representation
of $G_{\bC} = \prod_{i \in Q_0} GL(N_i, \bC)$, where $GL(N_i, \bC)$
is represented by $M_i = e_i M$ at each node, each $A$-module
specifies a particular $G_\bC$ orbit. Thus, 
finding a $\theta$-stable module
is the same as solving the D-term conditions. 

Up to this point we have not specified the value of $\theta$. 
Physically, $\theta$'s correspond to the Fayet-Iliopoulos (FI) parameters, which 
are needed to write down D-term equations \cite{DFR}. 
Although the Euler number of the space of $\theta$-stable $A$-modules
does not change under infinitesimal deformation of $\theta$, 
it does change along the walls of marginal stability \cite{NN,Nagao1}. 
The noncommutative Donaldson-Thomas invariant defined by 
Ref.~\refcite{Szendroi} is in a particular chamber in the 
space of $\theta$'s. Following Ref.~\refcite{MR}, we hereafter take 
$\theta=(0,0,\dots ,0)$ \footnote{This sounds like a rather degenerate choice, making all module to be $\theta$-semistable but not $\theta$-stable according to \eqref{eq.thetastability}. However, as we will see in the next subsection we add an extra node corresponding to the D6-brane and use the stability parameter $\hat{\theta}=(\theta,1)$.}. We will comment more on this issue 
in the final section.

\subsubsection{D6 Brane and Compactification of the Moduli Space}\label{subsec.D6}

We have found that solutions to the F-term and D-term conditions
in the quiver gauge theory are identified with 
$\theta$-stable $A$-modules.
We want to understand the moduli space of such modules
and compute its Euler number. 

Since D-brane charges correspond to the ranks of the gauge groups, 
we consider moduli space of $\theta$-stable modules with  
dimension $\dim M_i = N_i$ ($i \in Q_0$), which we denote by 
$\scM^{N}(A)$. To compute its Euler number, we need to address
the fact that the moduli space of stable $A$-modules is not always 
compact. In mathematics literature, the necessary compactification
is performed by enlarging the quiver diagram by adding
one more node in the following way. 

Let us fix an arbitrary vertex $i_0$, and define a new quiver 
$\hat{Q}=(\hat{Q}_0,\hat{Q}_1)$ by
\beq
\hat{Q}_0=Q_0 \cup \{*\}, \quad \hat{Q}_1=Q_1 \cup \{a_*:*\to i_0\}.
\eeq
Namely, we have added one new vertex $*$ and one arrow $* \to i_0$
to obtain the extended quiver diagram $\hat{Q}$. As in the previous
case for $Q$, we can define the path 
algebra $\bC \hat{Q}$, the ideal $\hat{\cal F}$ generated in $\bC Q$ 
by ${\cal F}$, and the factor algebra $\hat{A}=\bC \hat{Q}/\hat{{\cal F}}$. 
Define 
$\hat{\theta}\in \bN^{Q_0+1}$ by $\hat{\theta}=(\theta,1)$ 
and define $\hat{\theta}$-stable and semistable $\hat{A}$-modules 
using stability parameter $\hat{\theta}$. It is shown in lemma 2.3 of Ref.~\refcite{MR} 
that $\hat{\theta}$-semistable $\hat{A}$-modules are always 
 $\hat{\theta}$-stable \footnote{It is easy to see this fact from the definition \eqref{eq.thetastability}. Suppose that a module $\hat{M}$ is $\hat{\theta}$-semistable but not $\hat{\theta}$-stable. This means that $\hat{M}$ has a submodule $\hat{M}'$ such that
\beq
\sum_{i\in \hat{Q}_0} \hat{\theta}_i (\textrm{dim}\, e_i \hat{M}')=0.
\eeq
This is impossible, however, since the left hand side is $\textrm{dim}\, e_{*} \hat{M}'=1$.
}, and the moduli space $\hat{\scM}_{i_0}^{N}(A)$ of
 $\hat{\theta}$-stable modules with specified dimension 
vector $\hat{N}\in \bN^{Q_0+1}$ is compact (the boundary of moduli space of $\hat{\theta}$-stable modules are given by $\hat{\theta}$-semistable modules which are not $\hat{\theta}$-stable).

Adding the extra node allows us to compactify the moduli space.
In the language of D-branes, this corresponds to adding a
single D6 brane filling the entire Calabi-Yau manifold, which is 
necessary to interpret the whole system as a six-dimensional $U(1)$
gauge theory related to the Donaldson-Thomas theory. 
As we mentioned in Section\,\ref{subsec.addingD6}, the D6 brane serves
as a flavor brane and adds an extra node exactly in the way 
described in the above paragraph. 
Note that, in the above paragraph, the ideal $\hat{{\cal F}}$ is generated
by the original ideal ${\cal F}$. 
In the quiver gauge theory, this corresponds to
the fact that the flavor brane does not introduce a new gauge invariant
operator to modify the superpotential.
In this way, we arrive at the definition of non-commutative 
Donaldson-Thomas invariant as the Euler characteristic $\chi
(\hat{\scM}_{i_0}^{N}(A))$
of cohomologies \footnote{More precisely, this invariant 
should be defined from
the Euler class of the obstruction bundle over the moduli space. The
existence of obstruction theory guarantees the equivalence of these two
quantities up to a possible sign \cite{Szendroi,BehrendFantechi}.}.
With our identification of $\hat{\scM}_{i_0}^{N}(A)$ with the moduli 
space of solutions to the F-term and D-term conditions, 
$\chi(\hat{\scM}_{i_0}^{N}(A))$ computes the Witten index of
bound states of D0 and D2 branes bound on a single D6 brane
ignoring the trivial degrees of freedom corresponding the center of mass 
of D-branes in $\bR^{1,3}$ (see Section\,\ref{sec.index}).

\bigskip

We have chosen a specific vertex $i_0$ to define the non-commutative 
Donaldson-Thomas invariant. The $i_0$ dependence drops out in simple 
cases such as $\mathbb{C}^3$ and conifold, but in general
$\chi(\hat{\scM}_{i_0}^N(A))$ depends on the choice of the $i_0$. We note that 
the quiver gauge theory discussed in Section\,\ref{sec.quiver} also has an 
apparent dependence on $i_0$.

The most systematic method for computing the number and direction of arrows in the quiver diagram is to use fractional branes, or more mathematically exceptional collections \footnote{Exceptional collection appear in the work of Ref.~\refcite{HoriIV} and in Ref.~\refcite{CFIKV}. See Refs.~\refcite{Herzog,Rudakov}.}. The resolved conifold $X$ is a normal bundle $\scO_C(-1)\oplus \scO_C(-1)$ over a curve $C\simeq \bP^1$, and the fractional branes are given by $\scO_C$ and $\scO_C(-1)[1]$ \cite{AK}, where $[1]$ means the shift of the complex by one \footnote{Recall that fractional branes are in the derived categories of coherent sheaves, which consists of complexes of coherent sheaves (localized with respect to the quasi-isomorphism).}. These sheaves correspond to the nodes of the quiver diagram, and we have 4 (=2+2) arrows since
\beq
\textrm{Ext}^1(\scO_C,\scO_C(-1)[1])\simeq \textrm{Ext}^1(\scO_C(-1)[1],\scO_C)\simeq \bC^2.
\eeq
In this language, the D6-brane is given by the structure sheaf $\scO_X[1]$ of the whole $X$, with a suitable shift (in this case, 1). By computing the Ext groups, we can determine the quiver diagram including the D6-brane node and the arrows starting from it \footnote{Computation of the superpotential by this method is in general difficult, although possible in some examples. See Ref.~\refcite{AK}.}.

\begin{figure}[htbp]
\centering{\includegraphics[scale=0.4]{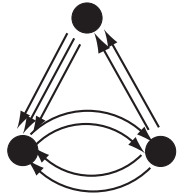}}
\caption[The quiver with a different choice of fractional branes.]{The quiver with a different choice of fractional branes, for $n=2$. In general, we have $n+1$ ($n$) arrows starting from (ending at) the D6-brane node,  corresponding to the chamber $C_n$ discussed later in Section\,\ref{sec.conifold}.}
\label{fig.mutatedquiver}
\end{figure}

As a digression, this understanding gives a bonus. As is well-known, the choice of the fractional branes is far from unique. For example, instead of $\scO_X[1], \scO_C$ and $\scO_C(-1)[1]$, we can choose $\scO_X[1], \scO_C(-n)$ and $\scO_C(-n-1)[1]$, where $n$ is a positive integer. Computing the Ext group of the quiver diagram, we have a different quiver, shown in Fig.\,\ref{fig.mutatedquiver}.

Physically, the quivers are related by Seiberg dualities (called mutations in mathematics \footnote{There is a conjecture that Seiberg duality is the same as the mutation of exceptional collections \cite{Herzog}.}). Note that in the construction given below, different quivers (and different superpotential) give a different path algebra, and thus a different crystal. This means that the BPS index changes, which is reminiscent of the wall crossing phenomena. Indeed, this is not an analogy; as clarified in Refs.~\refcite{CJ,NN}, in the quiver language wall crossing phenomena can be understood as Seiberg dualities on the quiver. We will return to the issue of wall crossing in part II. But before that, let us describe the structure of the crystal.

\subsection{Crystal Melting}\label{sec.crystal}

In this section, we define a statistical mechanical model
of crystal melting and show that the model reproduces the 
counting of BPS bound state of D-branes. Using
the quiver diagram and the superpotential of the gauge theory,
we define a natural crystalline structure 
in three dimensions.
We specify a rule to remove atoms from the crystal
and show that each molten crystal corresponds to a particular
BPS bound state of D-branes. We use the result of 
Ref.~\refcite{MR} to show that all the relevant BPS states are 
counted in this way. 

\subsubsection{Crystalline Structure}\label{crystal.subsec}

Mathematically, the three-dimensional crystal we define here is
equivalent to a set of basis for $A e_{i_0}$, where $A$ is the factor algebra
$A = \bC Q/{\cal F}$ of the path algebra $\bC Q$ divided by the 
ideal ${\cal F}$ generated by the F-term constraints and
$e_{i_0}$ is the path of zero length at the reference node $i_0$,
which is also the projection operator to the space of paths
starting at $i_0$. Colloquially, the crystal is a set of paths
starting at $i_0$ modulo the F-term relations. 
As we shall later, it corresponds to a single D6 brane with no D0 and D2 charges.
We interpret $A e_{i_0}$ in terms of a three-dimensional crystal as follows. 

The crystal is composed of atoms piled up 
on nodes in the universal covering $\tilde{Q}$ on $\bR^2$. By using 
the projection, $\pi: \tilde{Q} \to Q$, each atom is assigned with 
a color corresponding to the node in the original quiver diagram $Q$. 
The arrows of the quiver diagram determines the chemical bond between 
atoms. We start by putting one atom on the top of the reference node 
$i_0$. Next attach an atom at an adjacent node $j\in \tilde{Q}_0$ that is
connected to $i_0$ by an arrow going from $i_0$ to $j$. The atoms at such nodes 
are placed lower than the atom at $i_0$. In the next step, start with 
the atoms we just placed and follow arrows emanating from them
to attach more atoms at the heads of the arrows. 

\begin{figure}[htbp]
\centering{\includegraphics[scale=0.23]{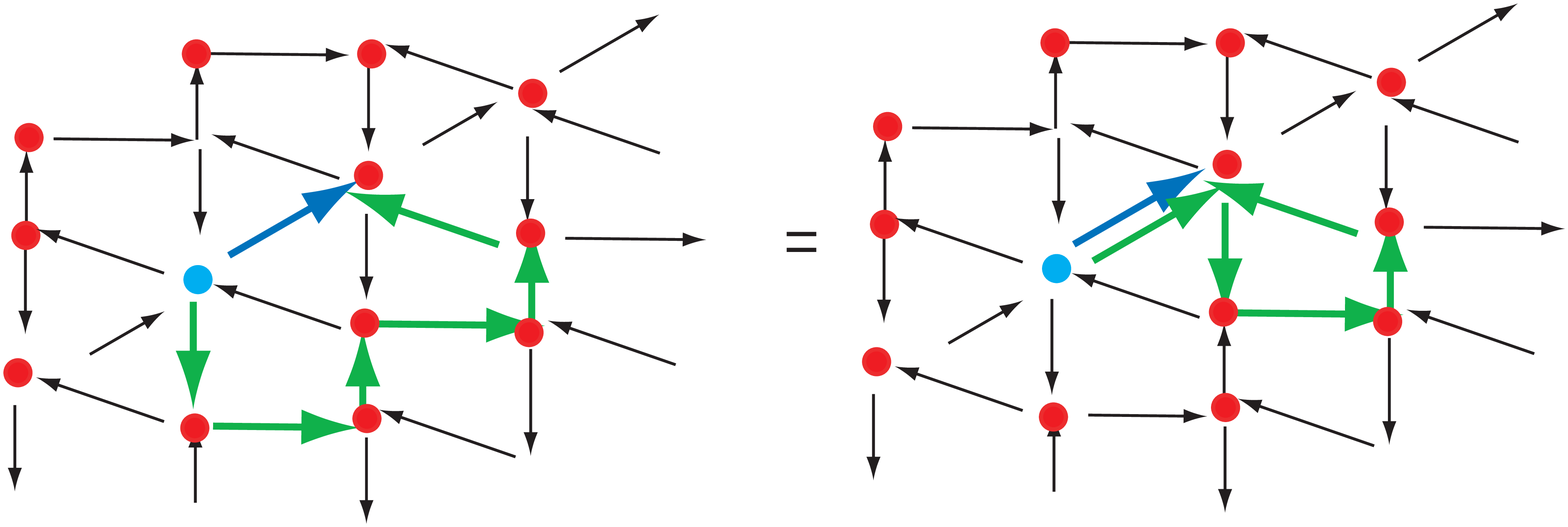}}
\caption[Every path is F-term equivalent to a shortest path times a loop]{This example shows that the green path is F-term equivalent to the shortest path (blue path) plus a loop.}
\label{SPPpathEx}
\end{figure}

\begin{figure}[htbp]
\centering{\includegraphics[scale=0.15]{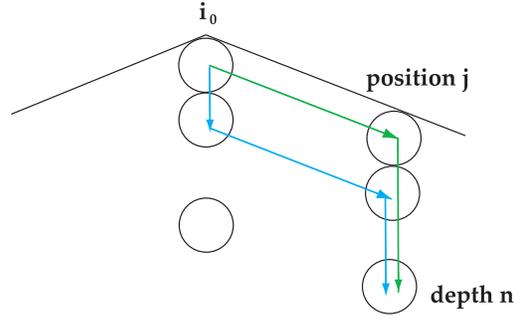}}
\caption[Structure of the crystal, representing an atom $a$ at position $j$ and depth $n$.]{Structure of the crystal, representing an atom $a$ at position $j$ and depth $n$. Different paths from the top atom to the atom $a$ (for example, the green path and the green path) are all F-term equivalent.}
\label{RuleEx}
\end{figure}

As we repeat this procedure, we may return back to a node where 
an atom is already placed. In such a case, 
we use the following rule. As we explained in
Section\,\ref{F-term.subsec}, modulo F-term constraints,
any oriented path $a$ starting at $i_0$ and ending at $j$ can 
be expressed as $v_{i_0,j}\omega^n $, where $\omega$ is 
the loop around a face in the quiver
diagram
 and $v_{i_0,j}$ is one of the shortest
paths from $i_0$ to $j$ (Fig.\,\ref{SPPpathEx} for an example). This defines an integer $h(a)=n$
for each path $a$. The rule of placing atoms is that, 
if a path $a$ takes $i_0$ to $j$ and if $h(a)=n$, 
we place an atom at the $n$-th place under the first atom 
on the node $j$ (Fig.\,\ref{RuleEx}). If there is already an atom at the 
$n$-th place, we do not place a new atom.

By repeating this procedure, we continue to attach atoms
and construct a pyramid consisting of infinitely many atoms.
Since atoms are placed following paths from $i_0$ modulo 
the F-term relations, it is clear that atoms in the crystal 
are in one-to-one correspondence with basis elements of $A e_{i_0}$. 
Note that by construction the crystal has a single peak at the 
reference node $i_0$.

\begin{figure}[htbp]
\centering{\includegraphics[scale=0.18]{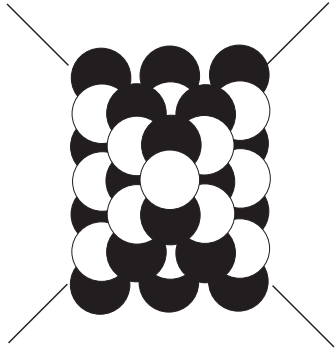}}
\caption[The crystal for the resolved conifold.]{The crystal for the resolved conifold. The pyramid consists of infinite layers of atoms, each layer colored alternatingly by white and black. The four ridges of the pyramid are shown by four black lines extending to infinity.}
\label{fig.conifoldatom}
\end{figure}

\begin{figure}[htbp]
\centering{\includegraphics[scale=0.15]{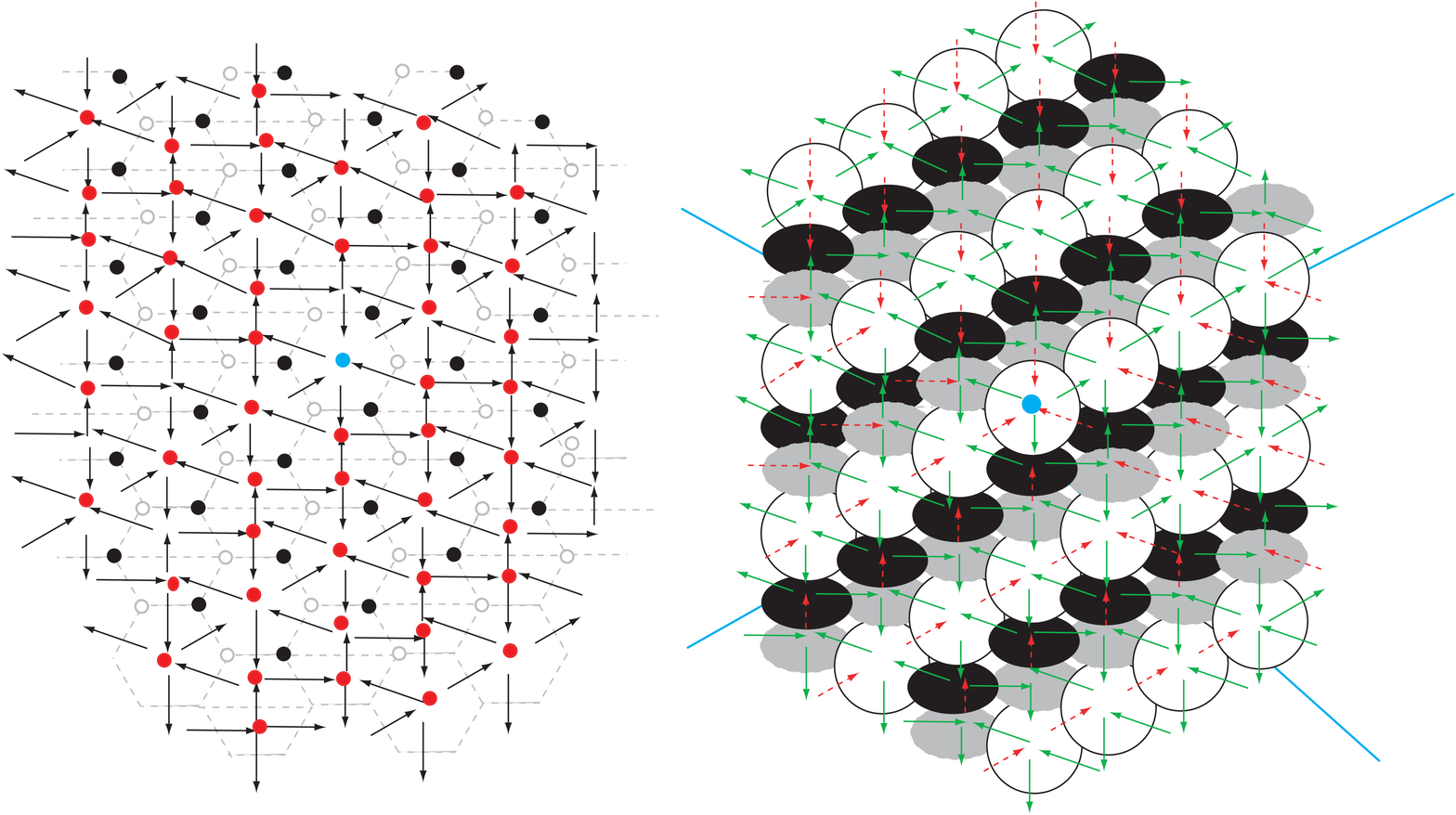}}
\caption[The crystal for the Suspended Pinched Point.]{Starting from the universal cover $\tilde{Q}$ of quiver $Q$ shown on the left, we can construct a crystal on the right. 
Each atom carries a color
corresponding to a node in $Q$, and they are connected by
arrows in $\tilde{Q}_1$. The green arrows represent arrows on the surfaces of the crystal, whereas the red ones are not.
In the case of the Suspended Pinched Point singularity, 
the atoms come with 3 colors (white, black and gray),
corresponding to the 3 nodes of the original quiver diagram $Q$ on $\bT^2$ shown in Fig.\,\ref{fig.SPPtiling}.}
\label{fig.SPPatom}
\end{figure}

This defines a crystalline structure
for an arbitrary toric Calabi-Yau manifold. 
In particular, it reproduces the crystal for $\bC^3$
discussed in Refs.~\refcite{ORV,INOV}, and the one for conifold 
in Ref.~\refcite{Szendroi}.  See Fig.\,\ref{fig.conifoldatom} (Fig.\,\ref{fig.SPPatom})
for the crystalline structure corresponding to
the resolved conifold (Suspended Pinched Point singularity). In these two examples, the ridge of the crystal (shown as blue lines in 
Fig.\,\ref{fig.SPPatom}) coincides with the $(p,q)$-web of 
the toric geometry. As we will discuss later, this is a 
general property of our crystal.

\subsubsection{BPS State and Molten Crystal}\label{molten.subsec}

In the forthcoming discussions, the crystal defined above will be 
identified with a single D6 brane with no D0 and D2 charges. Bound
states with non-zero D0 and D2 charges are obtained by removing
atoms following the rule specified below. 

In Ref.~\refcite{Szendroi,MR}, the 
Donaldson-Thomas invariants $\chi(\hat{\scM}_{i_0}^{N}(A))$
are computed by using the $U(1)^{2}$ symmetry of the 
moduli space  $\hat{\scM}_{i_0}^N$ corresponding to the translational
invariance of $\bT^2$. By the standard localization techniques, 
the Euler number can be evaluated at the fixed point
set of the moduli space under the symmetry. Correspondingly, 
in the gauge theory side, BPS states counted by the index are those
that are invariant under the global $U(1)^{2}$ symmetry acting on 
bifundamental fields preserving the F-term constraints since those
do not have extra zero modes and do not contribute to the index. 
We are interested in counting such BPS states. 

\begin{figure}[htbp]
\centering{\includegraphics[scale=0.18]{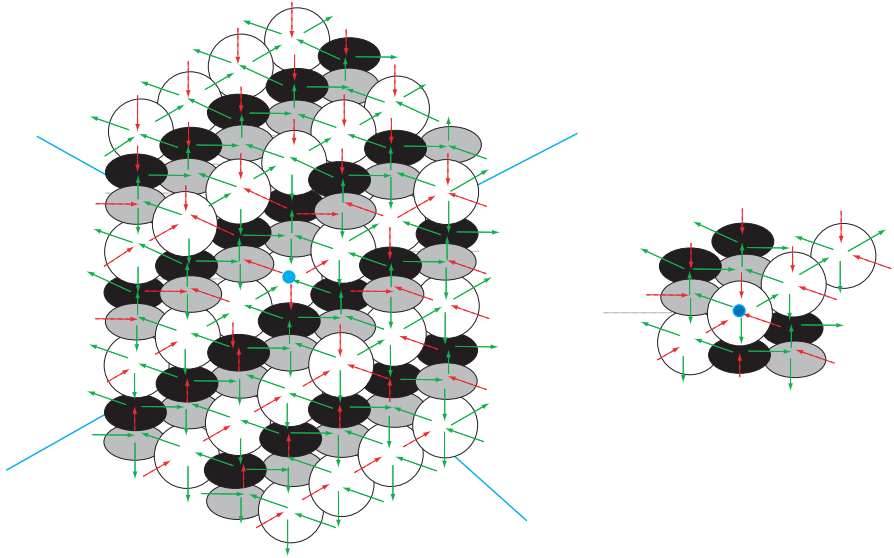}}
\caption[Example of a molten crystal and its complement $\Omega$.]{Example of a molten crystal and its complement $\Omega$. In this example $\Omega$ contains 12 atoms, one hidden behind an atom on the reference point represented by a blue point.
It is easy to check that $\Omega$ satisfies the melting rule mentioned in the text.}
\label{fig.SPPmolten}
\end{figure}

In order for a molten crystal to correspond to $U(1)^{2}$ 
invariant $\hat{\theta}$-stable $A$-modules, we need to impose 
the following rule to remove atoms from the crystal. Let $\Omega$
be a finite set of atoms to be removed from the crystal. 
\medskip\\
{\bf The Melting Rule}: 
\beq
\textrm{If}\quad a\alpha \in \Omega\quad \textrm{for some} \quad a \in A, \quad 
\textrm{then} \quad \alpha\in \Omega. 
\label{meltingruleold}
\eeq
\noindent
Since atoms of the crystal correspond to elements of $A e_{i_0}$, 
we used the natural action of $A$ on $A e_{i_0}$ to define $a\alpha$ in
the above. This means that crystal melting starts at the peak at $i_0$
and takes place following paths in $A e_{i_0}$. 
An example of a molten crystal satisfying 
this condition is shown in Fig.\,\ref{fig.SPPmolten}.

The melting rule means that a complement ${\cal I}$ 
of the vector space spanned by $\Omega$ in $A e_{i_0}$ gives
an ideal of $A$. To see this, we just need to take the contraposition of 
the melting rule. It states:
For any $\beta \in {\cal I}$ and
for any $a \in A$, $a \beta$ is also in ${\cal I}$. 

\medskip
Generally speaking, an ideal of an algebra defines a module.
To see this, consider a vector $|{\cal I} \rangle$ which is 
annihilated by all elements of the ideal ${\cal I}$. 
From $|{\cal I} \rangle$,
we can generate a finite dimensional
representation of the algebra $A$ by acting with elements 
of $A$ on it. However, the converse is not always true. 
Fortunately, when modules are $\hat{\theta}$-stable
and invariant under the 
$U(1)^{2}$ symmetry, it was shown in Refs.~\refcite{Szendroi,MR} that 
there is a one-to-one correspondence between ideals and modules.
It follows that our molten crystal configurations are
also in one-to-one correspondence with $A$-modules and therefore with
relevant BPS bound states of D-branes. This proves that the statistical 
model of crystal melting computes the index of D-brane bound states.

It would be instructive to understand explicitly 
how each molten crystal configuration corresponds to a BPS bound state.  
Starting from a molten crystal specified by $\Omega$,
prepare a one-dimensional vector space $V_{\alpha}$  
with basis vector $e_{\alpha}$ for each atom $\alpha\in \Omega$. 
For each arrow $a$ of $\tilde{Q}$, define the action of $a$ on $V_{\alpha}$ by
$a(e_{\alpha})=e_{\beta}$ when the arrow $a$ begins from $\alpha$ and ends an another atom $\beta \in \Omega$. Otherwise $a(e_{\alpha})$ is defined to be zero. 
Since an arbitrary path is generated by concatenation of arrows, we have defined an action of $a\in A$ on each $V_{\alpha}$. By linearly extending the action of $a$ onto the total space $M=\oplus_{\alpha\in \Omega} V_{\alpha}$, we obtain
a $A$-module $M$. 

There are several special properties concerning this module $M$. 
First, the F-term relations are automatically
satisfied. This is because when there exist two different paths $a,b\in A$ 
starting at $\alpha$ and ending at $\beta$, 
$a(e_{\alpha})$ and $b(e_{\alpha})$ are both defined to be $e_{\beta}$.
Second, by construction $M$ is generated by the action of 
the algebra $A$ on 
a single element $e_{i_0}\in V_{i_0}$. 
In such a case $M$ is called a cyclic $A$-module, and by lemma 2.3 of Ref.~\refcite{MR} is also $\hat{\theta}$-stable. 
Third, by the cyclicity of the module it follows that $M$ is $U(1)^{2}$ invariant up to gauge transformations.
Therefore, $M$ is a $U(1)^{2}$ invariant 
$\hat{\theta}$-stable module. It follows from the 
result of Ref.~\refcite{MR} discussed at the beginning of this section 
that $M$ indeed corresponds a bound state of D-branes 
contributing to the Witten index.

At the beginning of this subsection, we stated without 
explanation that the original crystal corresponds to 
a single D6 brane with no D0 and D2 charges, and removing
atoms correspond to adding the D-brane charges. To understand
this statement, let us recall that, in Section\,\ref{sec.quiver}, 
we started with a configuration of D0 and D2 branes on 
the toric Calabi-Yau manifold and took a T-duality along the fiber
to arrive at the brane configuration. Thus, the number of D2 branes
at each node $j$ of the quiver diagram $Q$ is a combination of D0 and D2
charges before T-duality. It is this number that 
is equal to the rank of the gauge group at $j$.

By using the projection $\pi: \tilde{Q}_o \rightarrow Q_0$, 
the $A$-module $M$ is decomposed as 
$M=\oplus_{j \in Q_0} M_j$ as we saw in Section\,\ref{sec.NCDT}, where
\beq
M_j=\bigoplus_{\alpha \in \Omega, \pi(\alpha)=j } 
V_{\alpha}\ . \label{backtomodule}
\eeq
In particular, the formula \eqref{backtomodule} means that 
the rank of the gauge group $N_j = {\rm dim}\, M_j$ at the node $j$
is equal to the number of atoms with the color corresponding
to the node $j$ that have been removed from the crystal. 
Thus, removing an atom at the node $j$ is equivalent to 
adding D0 and D2 charges carried by the node $j$. 
It is interesting to note that each atom in the crystal does not correspond
to a single D0 brane or a single D2 brane, but each of them carries
a specific combination of D0 and D2 charges. In the crystal melting picture,
fundamental constituents are not D0 and D2 branes but the atoms. 
This reminds us of
the quark model of Gell-Mann and Zweig \cite{GellMann,Zweig}, where the fundamental
constituents carry combinations of quantum numbers of hadrons, as
opposed to the Sakata model \cite{Sakata}, where existing elementary particles
such as the proton, neutron and $\Lambda$ particle are
chosen as fundamental constituents.

\subsubsection{Observations on the Crystal Melting Model}\label{subsec.observations}

We would like to make a few observations on the statistical model 
of crystal melting that counts the number of BPS bound states of D-branes. 

We have studied several examples of toric Calabi-Yau manifolds and
found that the crystal structure in each case matches with 
the toric diagram. In 
particular, the ridges of the crystal, when projected onto the 
$\bR^2$ plane, line up with the $(p,q)$ web of the maximally degenerate
toric diagram. This phenomenon is discussed in \ref{app.PM}. There, we
also explain the correspondence between molten crystal configurations
and perfect matching of the bipartite graph introduced in Section\,\ref{tiling.sec}.

In the last subsection, we found it useful to describe BPS bound
states using ideals of the algebra $A$. In the case when the toric
Calabi-Yau manifold is $\bC^3$, 
ideals are closely related to  the quantization of the
toric structure as discussed in Ref.~\refcite{INOV}. The
gauge theory for $\bC^3$ 
is the dimensional reduction of the ${\cal N}=4$ 
supersymmetric Yang-Mills theory
in four dimensions down to one dimension, and the
bifundamental fields are three adjoint fields. The F-term and D-term conditions
require that they all commute with each other. Thus chiral ring is generated 
by three elements $x, y, z$ which commute with each other without any further 
relation. In this case, any ideal ${\cal I}_\pi$ is characterized by 
the three-dimensional Young diagram $\pi$. 
Locate each box in the 3d Young diagram 
$\pi$ by the Cartesian coordinates $(i,j,k)$ ($i,j,k = 1, 2, 3, ...$)
of the corner of the box most distant
from the origin, and define $\Omega_\pi$ to be a set of 
the 3d Cartesian coordinates $(i,j,k)$ for boxes in $\pi$. We can then
define the ideal $\mathcal{I}_\pi$ of the chiral ring by,
\beq
{\cal I}_\pi = \{ x^{i-1} y^{j-1} w^{k-1} | (i,j,k) 
\notin \Omega_\pi
\}. \label{ideal}
\eeq
In Ref.~\refcite{INOV}, this description was obtained
by quantizing the toric geometry by using its canonical K\"ahler form
and by identifying $x^{i-1} y^{j-1} w^{k-1}$ as states in the Hilbert
space. 

This can be generalized to an arbitrary toric Calabi-Yau manifold
$X_\Delta$ as follows. One starts with the quiver diagram corresponding to
$X_\Delta$ and use the brane tiling to identify the F-term equations. 
This gives the chiral ring generated by bifundamental fields obeying
the F-term and D-term relations. As we saw in Section\,\ref{subsec.observations}, each BPS
bound state is related to an ideal of the chiral algebra. 
We expect that such ideals arise from quantization of the toric structure.
BPS bound states of D-branes emerging from the quantization of 
background geometry is reminiscent of the bubbling AdS space of 
Ref.~\refcite{LLM} and Mathur's conjecture on black hole microstates
\cite{Mathur}.

\subsection{The BPS Partition Function}\label{sec.ZBPS}

We have seen that a molten crystal is in one-to-one correspondence with a BPS state contributing to the index. It automatically follows that the BPS partition function, which is defined in Section\,\ref{sec.index} by summing over BPS states contributing to the index, is the same as the statistical partition function of the crystal melting model.

From the explanation around \eqref{backtomodule}, the partition function expressed in terms of the crystal is given by
\beq
Z_{\rm BPS}(q_0,\ldots, q_N)=\sum_{\Omega} \pm  q_0^{w_0(\Omega)} q_1^{w_1(\Omega)}\ldots q_N^{w_N(\Omega)},
\label{eq.ZBPS}
\eeq
where $w_i(\Omega)$, which is the same as $\textrm{dim}\, M_j$, is the number of atoms in $\Omega$ with the $i$-th color.
The parameters $q_i$, counting the rank of the dimension of the vector space at vertex $i$, is a combination of D0/D2-brane chemical potentials. 
For example, in the conifold example dimension vectors are related to the D-brane charges $n_0, n_2$ by (see the discussion at the end of Section\,\ref{subsec.quiver})
\beq
\textrm{dim}\, M_1= n_0, \quad \textrm{dim}\, M_2= n_0+n_2.
\eeq
This means that the relation between parameters are given by
\beq
q= q_0 q_1, \quad Q=q_1.\label{paramchg}
\eeq

In the discussion so far we have not discussed the sign in \eqref{eq.ZBPS}, which should be properly taken into account when we discuss the BPS index. Namely the correct statement is that the BPS index is the same as the number of configuration of molten crystals times a sign, which mathematically comes from the 
dimension of the tangent space of the vacuum moduli of the quantum mechanics, as mentioned in footnote \ref{sign1} in Section\,\ref{subsec.Whatis}.
Recall that we are actually computing the BPS index, which is defined as the number of boson and that of fermions, and therefore can take negative as well as positive values. In general, for a configuration of molten crystal $\Omega$, Theorem 7.1 of Ref.~\refcite{MR} says that the sign is given by
\beq
(-1)^{w_{i_0}(\Omega)+\langle w(\Omega), w(\Omega) \rangle},
\label{eq.sign2}
\eeq
where $\langle \alpha, \beta \rangle$ for $\alpha=(\alpha_i)_{i\in Q_0}, \beta=(\beta_i)_{i\in Q_0}$ is the Ringel form of the quiver, given by
\beq
\langle \alpha, \beta \rangle =\sum_{i\in Q_0} \alpha_i \beta_i-\sum_{a\in Q_1:i\to j}\alpha_i \beta_j.
\eeq

In general we have to insert this sign to each term of the crystal
melting partition function separately. When the toric Calabi-Yau manifold $X$ does not have a compact
4-cycle, however, these signs are simply taken care of by the change of the
signs of the chemical potentials in the partition function. To show
this, we begin with the fact that in these examples 
the quiver diagram takes the following form \footnote{This statement is well-known in the brane tiling literature. See for example Fig.\,6 of Ref.~\refcite{IIKY} for the basic building blocks of the bipartite graph (fivebrane diagram), from which we can construct the bipartite graph. The dual graph gives a quiver diagram.}. 

Suppose that the quiver has $N$ nodes, together with one node
 corresponding to the D6-brane. Then we can number the $N$ nodes by $i=1\ldots, N$ such that (a) there is always exactly one arrow from node $i$ to node $i+1$, (b) there is always exactly one arrow from node $i$ to node $i-1$, and 
 (c) there may or may not be an arrow from node $i$ to node $i$.
In this situation contributions from arrows of type (a), (b) cancel out in the second term of \eqref{eq.sign2}, and the only remaining contribution is from (c):\begin{align*}  
w_{i_0}(\Omega)+\langle w(\Omega), w(\Omega) \rangle &\equiv 
w_{i_0}(\Omega)+\sum_{i\in Q_0} w_i(\Omega)^2- \sum_{i:\exists a\in Q_1,
 a:i\to i} w_i(\Omega)^2 \\
& \equiv 
w_{i_0}(\Omega)+\sum_i \Omega_i-\sum_{i:\exists a\in Q_1, a:i\to i} \Omega_i
 \quad ~({\rm mod }\, 2).
\end{align*}
Thus the total sign 
can also be written as 
\beq
  (-1)^{\sum_{i\in S} \Omega_i}
\eeq
for a set $S=\{i_0 \}\cup \{i |\, \exists\, a\in Q_1, a:i\to i\}$. This means that the sign is conveniently taken care of by the replacement \footnote{As emphasized previously, this is not the case in general, e.g. for the canonical bundle of $\bP^1\times \bP^1$.} (we use $p_i$ for the parameters in the generating functions of the crystal without signs)
\beq
q_i\to p_i=-q_i ~~({\rm if }\,\, i \in S), \quad q_i\to p_i= q_i ~~({\rm if }\,\, i \notin S).
\label{eq.pqsign}
\eeq
In other words, the precise relation between the BPS partition function \eqref{eq.ZBPS} and the crystal partition function (defined by \eqref{eq.ZBPS} without signs) is given by
\beq
Z_{\rm BPS}(q_0,\ldots, q_N)=Z_{\rm crystal}(p_0,\ldots,p_N)
\eeq
under the identification in \eqref{eq.pqsign}.

As an example, for the resolved conifold there is no arrow of type (c), and we have $\{ S\}=\{1 \}$ and
\beq
p_0= q_0, \quad p_1= -q_1.
\label{eq.conifoldpq} 
\eeq
This concludes the discussion of signs in the BPS partition function \eqref{eq.ZBPS}.

\begin{figure}[htbp]
\centering{\includegraphics[scale=0.23]{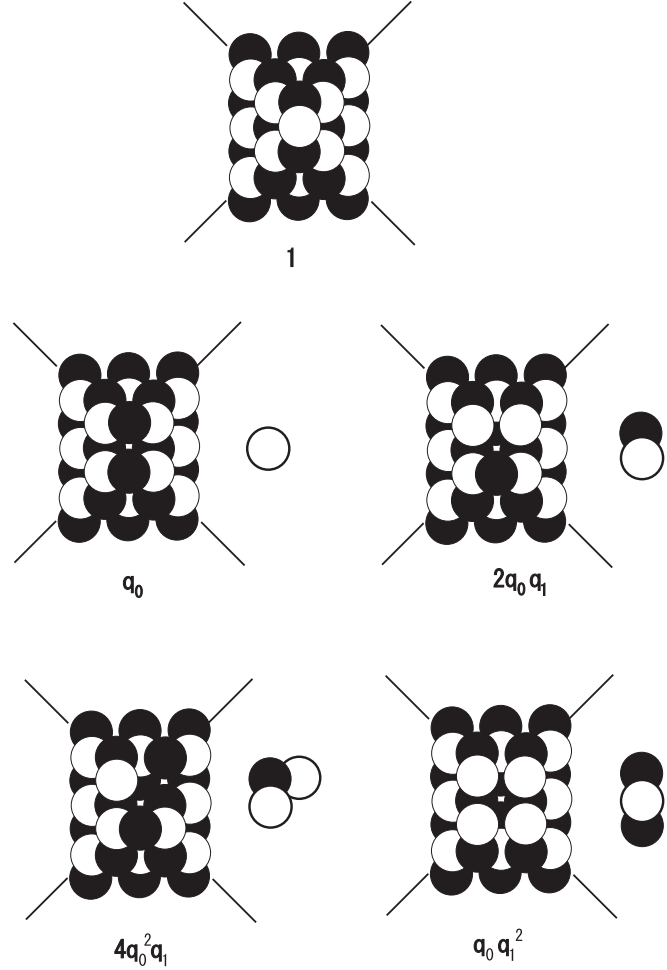}}
\caption{The examples of the states of the conifold crystal, starting with the ground state.}
\label{fig.conifoldcomp}
\end{figure}

\bigskip
Since we already know the explicit form of the crystal and the sign, we can compute $Z_{\rm BPS}$ from lowest order terms. For example, for the conifold example we have (see Fig.\,\ref{fig.conifoldcomp})
\beq
Z_{\rm BPS}(q_0,q_1)=1+q_0-2 q_0 q_1-4 q_0^2 q_1+ q_0 q_1^2+\ldots,
\eeq
which becomes, under the parameter identification \eqref{paramchg},
\beq
Z_{\rm BPS}(q,Q)=1+q Q^{-1}-2 q- 4 q^2 Q^{-1}+ q Q+\ldots,
\eeq
We can perform similar computations for any crystal, however complicated it is.

In some examples we can write down a closed expression for the BPS partition function, instead of computing each term separately. For example, for the resolved conifold Young \cite{Young1} has proven combinatorially that
\beq
Z_{\rm crystal}(p_0,p_1)=M(p_0 p_1)^2 \prod_n (1+ (p_0 p_1)^n p_1)^n \prod_n (1+(p_0 p_1)^n p_1^{-1})^n.
\eeq
This means, using \eqref{eq.conifoldpq} and \eqref{paramchg},
\beq
Z_{\rm BPS}(q,Q)=M(-q)^2 \prod_n (1-(-q)^n Q)^n \prod_n (1-(-q)^n Q^{-1})^n,
\eeq
where $M(q)$ is the MacMahon function 
\beq
M(q)=\prod_{n>0} (1-q^n)^{-n}.
\label{MacMahondef}
\eeq
This is very similar to the Gopakumar-Vafa \cite{GV1,GV2} expansion of the topological string partition function
\beq
Z_{\rm top}(q,Q)=M(-q) \prod_n (1-(-q)^n Q)^n,
\label{conifoldZGV}
\eeq
where the parameters $q$ and $Q$ in the above expression are related to the topological string coupling constant $g_s$ and the K\"ahler moduli $t$ by \footnote{For notational simplicity, in the following sections we will be sloppy and often absorb the minus sign in \eqref{topvar} into the definition of $q$ to write $q=e^{-g_s}$.}
\beq
q=-e^{-g_s}, \quad Q=e^{-t}.
\label{topvar}
\eeq
In fact, we can write 
\beq
Z_{\rm BPS}(q,Q)=Z_{\rm top}(q,Q) Z_{\rm top}(q,Q^{-1}).
\label{question}
\eeq
Similar expressions hold in all the toric Calabi-Yau 3-folds without compact 4-cycles. To explain the appearance of the topological string theory, as well as the relation \eqref{question}, will be the topic of Section\,\ref{chap.M-theory}.




\subsection{Summary}\label{sec.crystalsummary}

In this section, we established the connection between the counting
of BPS bound states of D0 and D2 branes on a single D6 brane to
the non-commutative Donaldson-Thomas theory. We studied the moduli space 
of solutions to the F-term and D-term constraints of the quiver
gauge theory which arises as the low energy limit of the brane
configuration. We found the direct correspondence between
the gauge theory moduli space and the space of modules of the factor 
algebra of the path algebra for the quiver diagram quotiented
by its ideal related to the F-term constraints, subject
to a stability condition to enforce the D-term constraints. 
Using this correspondence, we found
a new description of BPS bound states
of the D-branes in terms of the statistical model of crystal
melting. The crystalline structure is determined by
the quiver diagram and the brane tiling which characterize 
the low energy effective theory of D-branes. The crystal 
is composed of atoms of different colors, each of which 
corresponds to a node of the quiver diagram, and the chemical 
bond is dictated by the arrows of the quiver diagram. BPS states 
are constructed by removing atoms from the crystal. 

We naturally encounter the wall crossing phenomena, the theme of the second half of this paper, by studying the relation between the commutative and non-commutative Donaldson-Thomas invariants. 
The degeneracy of D-brane bound states changes when the value of 
$\theta$, used to define the stability condition, which is the wall crossing phenomena discussed in Section\,\ref{sec.WC} \footnote{The FI parameters $\theta$ is a function of the K\"ahler moduli.}. 
The jump in the degeneracy can be computed by the wall crossing formula 
\cite{DenefM,KontsevichS}, and if we start from a particular chamber and 
apply the wall crossing formula, we can obtain the value of 
$\chi(\hat{\scM}_i^{N}(A))$ in any chamber we want.
We will see this explicitly in the conifold example in Section\,\ref{chap.WCF}. 
As will be discussed there, wall crossing relates non-commutative Donaldson-Thomas invariants to commutative Donaldson-Thomas invariants and to new invariants defined by Pandharipande and Thomas \cite{PT}. 
This story is further generalized by Ref.~\refcite{Nagao1} when $\Delta$ has 
no internal lattice point. 

It has been proven recently that the topological 
string theory is  equivalent to the commutative Donaldson-Thomas theory 
for a general toric Calabi-Yau manifold \cite{INOV,MOOP}.
Since the commutative Donaldson-Thomas theory
count BPS states for some choice of stability
condition, 
\beq
Z_{\rm BH} = Z_{\rm top},
\eeq
is indeed true in some chamber. 
On the other
hand, our result shows that the relation,
\beq
    Z'_{\rm BH} = Z_{\rm crystal~melting},
\eeq
holds in another chamber, where $Z'_{\rm BH}$ is the BPS state counting
for another choice of the stability condition. 
Combining these two results, we find that 
the topological string theory and the 
statistical model of crystal melting 
are related by the wall crossing, and we have
\beq
   Z_{\rm crystal~melting} \sim Z_{\rm top}~~~({\rm modulo~wall~crossings}).
\eeq
Since there is no wall crossing phenomenon between  the  commutative and non-commutative Donaldson-Thomas
theory on $\bC^3$, this result does not contradict with Ref.~\refcite{INOV},
where a direct identification of the topological string theory
and the crystal melting is made for $\bC^3$.
In general, we expect that a proper understanding of the relation 
between the topological string theory and the crystal melting
requires that we take the wall crossing phenomena into account. 
This motivates the study of wall crossing phenomena in part II.


\section{Thermodynamic Limit}\label{chap.thermodynamic}

In this section we study the thermodynamic limit of the crystal melting model defined in the previous section. This section is based on the work \cite{OY2}.

\subsection{Overview: Why Thermodynamic Limit?}\label{sec.thermintro}

In the previous section we have seen that the BPS counting problem is beautifully solved by the crystal melting model, making an explicit computation possible. However, the question still remains whether the crystal expansion is just a technical tool to obtain the answer, or whether it has much deeper significance.

There is one notable feature of the crystal melting model: it is a strong coupling expansion with respect to the (topological) string coupling constant. In the usual string theory, we consider a perturbative, weak coupling ($g_s\ll 1$) expansion
\beq
Z_{\rm top}=\exp\left(\sum_{g=0}^{\infty} g_s^{2g-2} \scF_g \right),
\eeq
whereas the crystal expansion gives
\beq
Z_{\rm crystal} = \sum_{n,\beta} \Omega(n, \beta) e^{- g_s n-t \beta},
\label{partitionfunction}
\eeq
where we used the parameter identification \eqref{topvar}, which holds for any toric Calabi-Yau manifold.
Due to the exponential factor, this expansion is good when $g_s\gg 1$.

This in particular means that crystal melting provides a tool to study the physics at the Planck scale \footnote{The following explanation is a bit imprecise in that $g_s$ here is the topological string coupling, and is different from the physical string coupling constant which really determines the relation between the string scale and the Planck scale. However, the topological string theory is relevant for counting of microstates of black holes in the superstring theory \cite{OSV}, and we expect that our result sheds some light on quantum nature of spacetime in the superstring theory also.}. When $g_s$ is small, as is usually the case, the string scale $l_s$ is larger than the Planck scale $l_p$. This means that we see stringy corrections first, and quantum corrections are obscured. When $g_s$ is large, however, $l_s$ is smaller than the Planck scale $l_p$
kicks in first. This means that we can explore the `quantum' aspects of geometry.

The natural question is to understand the relation between the two descriptions. In this section, following the result of Ref.~\refcite{OY2} we will show how the smooth Calabi-Yau geometry emerges from the discrete structure of
the crystal melting model in the limit  $g_s \rightarrow 0$. 
Since $g_s$ is the chemical potential for the total number of atoms removed, $g_s\to 0$ means that we are in the thermodynamic limit (high temperature limit), where an infinite number of atoms are removed.
It is reasonable to expect that a classical geometric picture emerges
in the limit of large D0 and D2 charges since it can represent
a large black hole in the superstring theory. The corresponding
thermodynamic limit in the dimer model was studied in Ref.~\refcite{KOS}. 

\bigskip

\begin{figure}[htbp]
\centering{\includegraphics[scale=0.7]{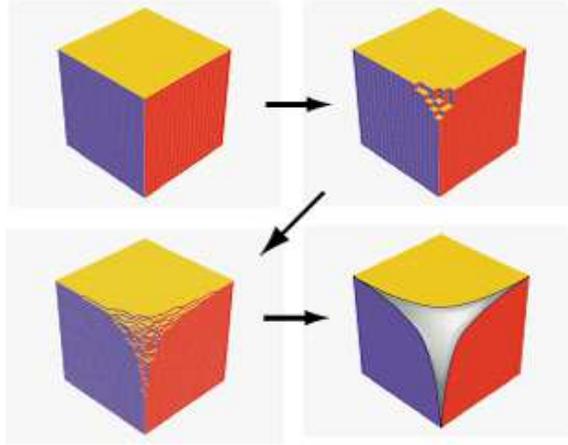}}
\caption[Limit shape of the crystal for $\bC^3$.]{Limit shape of the crystal, in the $\bC^3$ example. In the thermodynamic limit, the boundary of the molten region (where the atoms are taken away) becomes a smooth curve, which is called the limit shape. The claim is that this curve is the same as the amoeba of the mirror curve, which is shown in Fig.\,\ref{fig:Ronkin-eg} (b).}
\label{fig.limitshape}
\end{figure}

There are two main results proved in this section. First, the limit shape
(see Fig.\,\ref{fig.limitshape}) of the crystal in the thermodynamic limit coincides with a projection (amoeba) of the mirror Calabi-Yau manifold. 

This is proven as follows. A toric Calabi-Yau 3-fold $X$ is a K\"ahler quotient of ${\bC}^{F+3}$ by $U(1)^{F}$, and its mirror manifold $\widetilde{X}$
is defined by the polynomial equation \cite{HoriV,HoriIV},
\beq
u v + P(z,w) =0, ~~~ u,v \in {\bC},~~~ z,w \in {\bC}^{\times}.
\label{mirror}
\eeq
Here $P(z,w)$ is a Newton polynomial of the form,
\beq
 P(z,w) = \sum_{i=1}^{F+3} c_i(t) z^{n_i} w^{m_i},
\label{mirrorP}
\eeq
and  $c_i(t)$'s are functions of the K\"ahler
moduli $t$ of the original toric 3-fold $X$. 
The exponents $(n_i, m_i) \in {\bZ}^2$ correspond to
lattice points of the toric diagram. For example,
for the mirror of $\bC^3$, $P(z,w)$ is given by
\beq
P(z,w) = 1 + z +w,
\eeq
and for the mirror of the resolved conifold (${\cal O}(-1)\oplus {\cal O}(-1)$ bundle over $\bP^1$) by
\beq
P(z,w) = 1 + z +w + e^{-t} zw.
\eeq

In this section, we will show that the Newton polynomial $P(z,w)$
for the mirror of a Calabi-Yau manifold is identical to the characteristic 
polynomial of the corresponding dimer model, which is the partition 
function of the model on a torus. The relation between $P(z,w)$ and 
the characteristic polynomial had been discussed earlier in Ref.~\refcite{FHKV}. 
Here, we will prove their precise equality including the dependence on the 
moduli $t$. As discussed in \ref{app.PM}, the dimer model is equivalent to the crystal melting model discussed in the previous section.
Since we can show from the result of Ref.~\refcite{KOS} that the amoeba of the characteristic polynomial gives the limit shape of the crystal, this proves our result.

As the second important result in this section, we will show that the partition function of the
dimer model evaluated in the thermodynamic limit is equal to 
the genus-$0$ limit of the partition function of the topological
string theory on $X$. As discussed in Section\,\ref{sec.crystalsummary}, the dimer model has been formulated to 
describe the non-commutative Donaldson-Thomas theory
\cite{Szendroi,MR}, while the topological string theory for
a general toric Calabi-Yau manifold is
equivalent to the commutative Donaldson-Thomas theory \cite{DT}, 
as shown by Ref.~\refcite{MOOP}. Despite the non-trivial wall crossing factors between the two theories discussed in Section\,\ref{sec.crystalsummary}, the thermodynamic limit of the crystal
partition function gives the topological string partition function.

Our results suggests that the atoms in the crystal are a discretization of the Calabi-Yau geometry at the Planck scale, and the thermodynamic limit is a process where a smooth geometry emerges from a collection of such discretized piece of the geometry. The emergence of Calabi-Yau geometry from the thermodynamic
limit has been observed in Ref.~\refcite{ORV} in the case of $\bC^3$, and is interpreted in Ref.~\refcite{INOV} from the viewpoint of K\"ahler gravity Ref.~\refcite{BershadskyS}. 
In this section,
we make the connection sharper and more explicit 
by showing the direct connection between the partition functions 
of the crystal melting model and the topological theory
for a general toric Calabi-Yau 3-fold. 

\subsection{Thermodynamic Limit of the Crystal Melting Model}\label{sec.thermlimit}
The main object of study in this section is the partition function,
\beq
Z = \sum_{n, \beta_I} \Omega(n, \beta_I) e^{- g_s n-t^I \beta_I},
\label{partitionfunction2}
\eeq
where $\Omega(n, \beta)$ is the Witten index for
bound states of 
$n$ D0 branes and $\beta=(\beta_I)$ D2 branes on the $I$-th 2-cycle
($I = 1,...,\textrm{dim}\,H_2(M)$) with a single D6 brane on a toric Calabi-Yau
manifold. According to the dictionary in the previous section, 
the Witten index is equal (up to a sign) to the number of molten
crystal configurations where $n$ is the total number of
atoms removed,
whereas the relative numbers of different types of atoms 
removed from the crystal are specified by $\beta_I$'s. 
The variables $g_s$ and $t^I$ are chemical potentials for D0 and D2 charges, and as we have seen in \eqref{paramchg} \footnote{For simplicity we here neglect the sign in \eqref{paramchg}.} we will identify $g_s$ as the topological string coupling constant and $t^I$ as the K\"ahler moduli of the toric Calabi-Yau manifold $X$. 

The behavior of $Z$ as $g_s\rightarrow 0$ can be evaluated by 
the result of Ref.~\refcite{KOS}. Consider a finite covering the original quiver 
diagram, $N$ times in one direction and $N$ times in another direction
on $\bT^2$. $N$ is introduced as an infrared regulator, and 
we will take $N \rightarrow \infty$ at the end of the computation
so that we have the dimer model on the plane $\bR^2$.
The surface of the crystal 
is determined by the height function $h$ over the plane.
To define
$h$, we start with the canonical perfect matching $m_0$ 
of the dimer model corresponding to the initial crystal configuration
with no D0 or D2 charges (in this section, we heavily use the language of dimer models, whose equivalence with crystal melting is explained in \ref{app.PM}.). For any other 
perfect matching $m$, the superposition of $m_0$ and $m$ gives
a set of closed loops on the dimer graph. If $m$ corresponds to
a bound state with finite D0 and D2 charges, $m$ and $m_0$ differ only
in a finite region on the graph. The value of the height function $h$ at $i$-th  node of the $N\times N$ cover is defined so that it is $0$ 
far away from the region where $m$ and
$m_0$ differ, and it increases by $1$ every time we cross 
a closed loop as we move inside of the region (see Fig.\,\ref{fig.SPPheight} for an example). 
The corresponding molten crystal configuration is obtained
by removing $h(i)$ atoms of the initial crystal over the node $i$. 
In particular,
\beq
 \sum_i h(i) = n,
\eeq
where $n$ is (as before) the D0-brane charge, or equivalently the total number of atoms removed.

To take the thermodynamic limit,  
it is useful to introduce the Cartesian coordinates $(x,y)$,
$0 \leq x, y \leq 1$, on the $N\times N$ covering of $\bT^2$.
In the limit where $g_s \ll 1$ and $1 \ll N$, 
the height $h(i)$ becomes a smooth function
$h(x,y)$. We rescale the height function by the factor of 
$1/N$ so that 
\beq
  N^2 \int_0^1 dx dy ~h(x,y) = \frac{n}{N},
\label{rescaledh}
\eeq
to take into account the large $N$ scaling of the 
partition function discussed in Ref.~\refcite{Sheffield} and quoted as
 Theorem 2.1 in Ref.~\refcite{KOS} \footnote{This statement is consistent with our intuition that the height diverges in the same order as the size of the enlarged torus.}. 
The statistical weight in the thermodynamic limit
is given by an integral of a surface tension 
$\sigma(\partial h)$, which is a function of the gradient of $h$,
as \footnote{Here we only show the $g_s$ dependence explicitly 
  and the dependence on the K\"ahler moduli $t^I$ is in $\sigma$.}
\beq
  Z \sim \exp\left[ N^2\, {\rm max}_h\! \int_0^1 dx dy  \left[-
\sigma(\partial h)
- g_sN\, h(x,y)\right]\right]. 
\label{thermlimit}
\eeq
The integral of $g_s N\, h(x,y)$ in the exponent comes from 
the weight factor $e^{-g_s n}$ 
in (\ref{partitionfunction}), and we used (\ref{rescaledh}). 
Note that interestingly the combination $g_s N$ appears, which is the 't Hooft parameter.
In the thermodynamic limit, we 
look for a height function $h(x,y)$ which maximize the
exponent. 

\bigskip

In order to derive the macroscopic surface tension from the microscopic crystal melting model, we need a few definitions.
We first define 
a characteristic polynomial $\widetilde{P}(z,w)$ as the partition function per fundamental domain, i.e. the sum of 
perfect matchings with weights assigned to edges of the dimer model 
on $\bT^2$ \cite{KOS}. The parameters $z$ and $w$ specify the height difference in the language of dimer models.

\begin{figure}[htbp]
\centering{\includegraphics[scale=0.18]{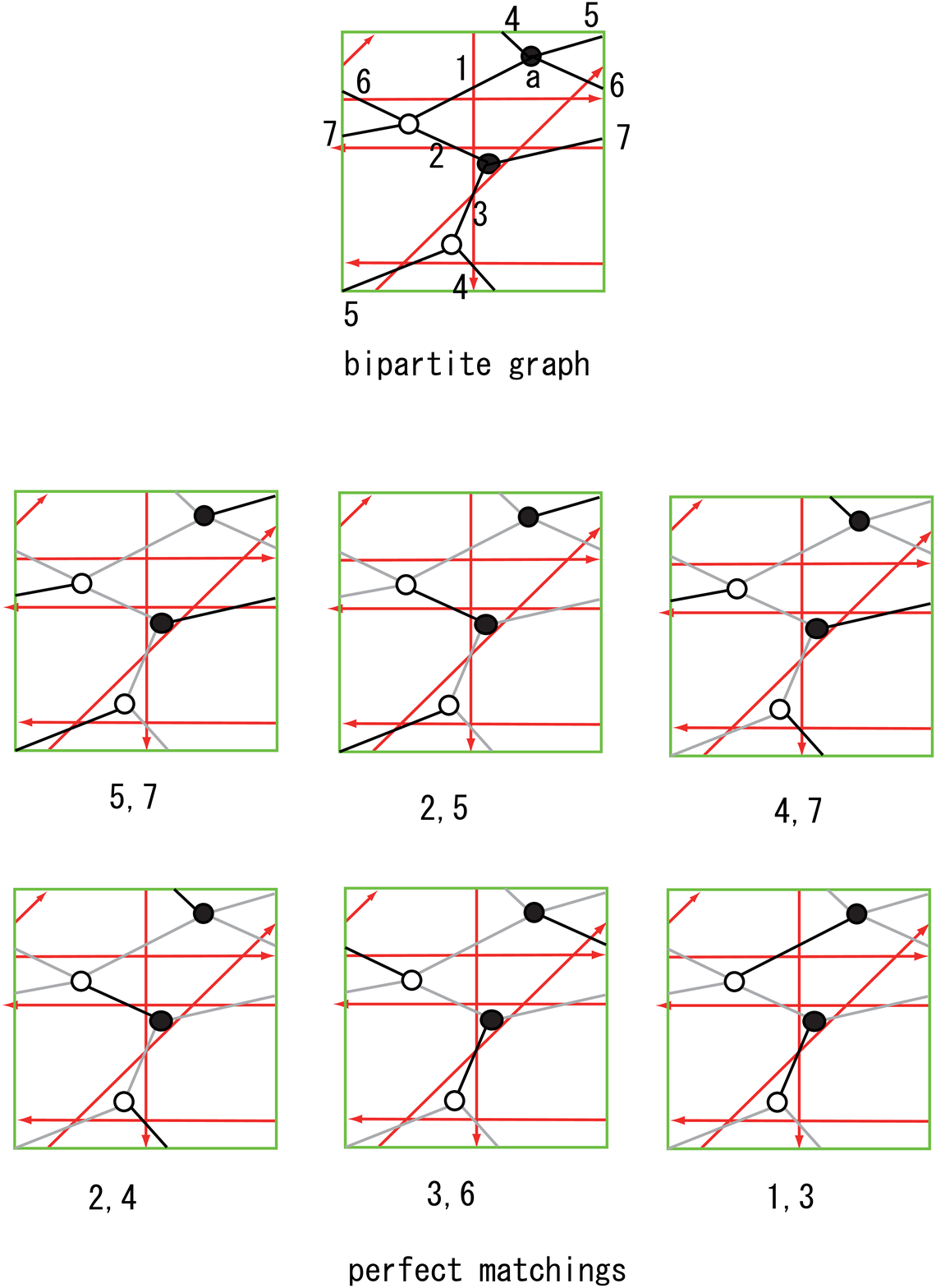}}
\medskip
\caption[The perfect matchings for Suspended Pinched Point bipartite graph.]{The 6 perfect matchings for Suspended Pinched Point bipartite graph (Fig.\,\ref{fig.SPPbipartite}). The numbers assigned to each perfect matching shows which edge is included in the matching.}
\label{SPPPM}
\end{figure}

Let us explain the meaning of parameters $z$ and $w$ in more detail,
using the Suspended Pinched Point shown in Fig.\,\ref{SPPPM} as an example.
In this example, the bipartite graph has 7 edges and 6 perfect matchings.
When we denote the weight for the $i$-th edge by $e^{w_i}$, the characteristic polynomial is given by
\beq
e^{w_5+w_7}+ e^{w_2+ w_5}+e^{w_4+ w_7}+e^{w_2+ w_4}+e^{w_3+ w_6}+e^{w_1+ w_3}.
\eeq
This is a function of seven variables, but there is a redundancy in the description. We call this redundancy ``gauge symmetry''. For example, we can simultaneously multiply by a constant $c$ all the weights for edges incident to the vertex $a$ (i.e. $w_1, w_4, w_5, w_6$), and the partition function stays the same up to the overall multiplication; this is because every perfect matching has exactly one edge incident to vertex $a$. This means that we can use this gauge degrees of freedom to set $w_5=0$. If we repeat the similar procedure for the other vertices, we can also set
$w_3=w_7=0$, and we have fixed all gauge degrees of freedom. The characteristic polynomial now simplifies to
\beq
1+ e^{w_2} +e^{w_4}+e^{w_2+w_4}+e^{w_6}+e^{w_1}.
\label{tempeq}
\eeq

Instead to introducing redundant variables with gauge symmetry as above, we can be more economical and express this result in terms of gauge invariant variables. The gauge invariant observables are given by the ``magnetic flux'' \cite{KOS}. To define this, let us fix an oriented loop
$\gamma=\{w_1,b_1,\ldots, w_{k-1}, b_{k-1}, w_k \},~~ w_k=w_0$, where $w_i$ ($b_i$) are white (black) vertices.
 We then define the magnetic flux along $\gamma$ to be
\beq
\int_{\gamma} e^w\equiv e^{\sum_{i=1}^{k-1} (w_{(w_i,b_i)} -w_{(w_{i+1},b_i)})},\label{magneticflux}
\eeq
where $w_{(w_i,b_i)}$ is the weight assigned to an edge $(w_i,b_i)$ connecting a white vertex $w_i$ and a black vertex $b_i$.
This is clearly gauge invariant. In general, when we denote the number of faces of the bipartite graph by $F$ (this is the same as the number of nodes of the quiver diagram), we have $F+1$ independent gauge invariants: $2$ of them comes from $\alpha$ and $\beta$-cycles of the torus, and $F-1$ comes from the flux around each of the $F$ faces, with one condition that the sum of the fluxes of all faces sum up to zero. 

In the Suspended Pinched Point example, we have $F+1=4$ independent parameters, and correspondingly we define
\begin{eqnarray*}
& &z=e^{w_2-w_7},\quad  w=e^{w_1-w_2+w_3-w_4},\\  
& &e^{-t_1}=e^{-w_3+w_4-w_5+w_3-w_2+w_7} \quad e^{-t_2}=e^{-w_1+w_2-w_7+w_6},
\end{eqnarray*}
where the parameters $z$ and $w$ correspond to the fluxes along the $\alpha$ and $\beta$-cycles, and $t_1$ and $t_2$ to those along cycles around a face.
In these variables the previous equation \eqref{tempeq} becomes (after setting $w_3=w_5=w_7=0$ as before)
\beq
\tilde{P}(z,w)=1+ z +e^{-t_1} z+e^{-t_1} z^2 + e^{-t_1-t_2}  z w+e^{-t_1} z^2 w. \label{SPPK1}
\eeq

Given $\tilde{P}(z,w)$, we next define its Ronkin function \cite{Ronkin} $R(x,y)$  by
\beq
  R(x,y) = \int_0^{2\pi} \ln \widetilde{P}(e^{x+i\theta}, e^{y+i\phi}) 
\frac{d\theta
d\phi}{(2\pi)^2}.
\label{ronkin}
\eeq
The integrand in this definition diverges when $\widetilde{P}(e^{x+i\theta}, e^{y+i\phi})=0$, however the function itself is well-defined. To see this, let us compute the its derivative:
\begin{eqnarray}
  \partial_x R(x,y) &=& \int_0^{2\pi} \frac{\partial_z \widetilde{P}(e^{x+i\theta}, e^{y+i\phi}) } {\widetilde{P}(e^{x+i\theta}, e^{y+i\phi}) } e^{x+i\theta}
\frac{d\theta
d\phi}{(2\pi)^2} \nonumber \\
&=&
\int_{|z|=e^x, |w|=e^y} \frac{\partial_z \widetilde{P}(z,w) } {\widetilde{P}(z,w)} z \frac{1}{(2\pi)^2}
\frac{dz}{z}\frac{dw}{w}.
\label{Ronkinx}
\end{eqnarray}
This can be evaluated as a residue integral. For example, suppose that $(x,y)$ is outside amoeba 
\footnote{See Ref.~\refcite{Mikhalkin} for a survey of aspects of amoebae, and Refs.~\refcite{MN,FNOSY} 
for their uses in other aspects of supersymmetric gauge theories.},
which is a subset of $\bR^2$ defined by 
\footnote{There is a similar notion, called alga or coamoeba, which is defined by 
\begin{equation}
{\rm Alga} = \{ (\theta,\phi)\in \bT^2 :~\widetilde{P}(e^{x +i\theta},
e^{y+i\phi})=0 {\rm ~for~some~}(x,y)\}.
\label{alga}
\end{equation}
It is known that coamoebas are equivalent to fivebrane diagrams, and therefore yield the brane configurations and the bipartite graph (see Ref.~\refcite{Yamazaki} for detailed discussions). They also have interesting connections with mirror symmetry \cite{FHKV,UY2,UY3,FutakiU}.
}
\begin{equation}
{\rm Amoeba} = \{ (x, y)\in \bR^2 :~\widetilde{P}(e^{x +i\theta},
e^{y+i\phi})=0 {\rm ~for~some~}(\theta,\phi)\}.
\label{amoeba}
\end{equation}
Then the only pole contributing to \eqref{Ronkinx} is located at infinity. When the leading contribution of $\tilde{P}(z,w)$ at infinity is given by $z^m w^n$, we can easily obtain $(\partial_x R(x,y), \partial_y R(x,y))=(m,n)$.
In other words, Ronkin function is linear in the complement of amoeba \cite{Ronkin,PR}.

\begin{figure}[htbp]
\begin{center}
\begin{tabular}{cc}
\includegraphics[width=30mm]{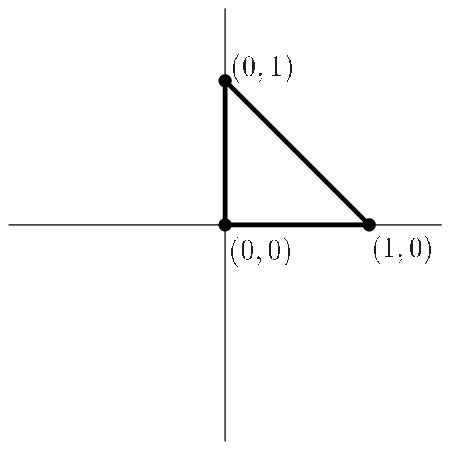} &
\includegraphics[width=30mm]{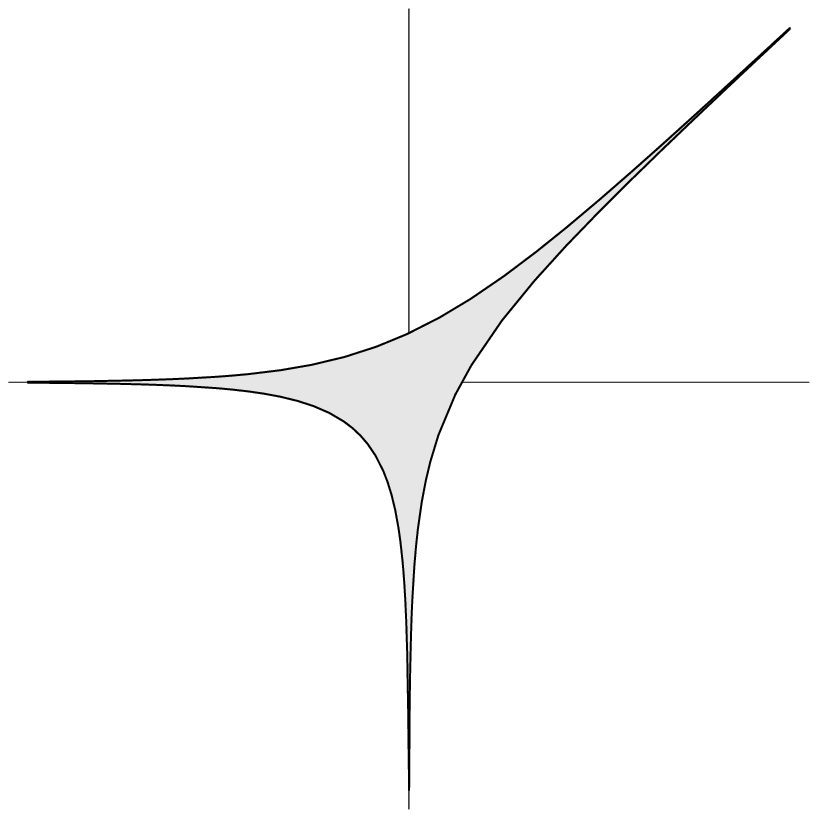} \\
(a) toric diagram & (b) amoeba 
\end{tabular}
\end{center}
\caption[The toric diagram
(a), and amoeba (b) for $\bC^3$]{The toric diagram
(a), and amoeba (b) for $\bC^3$, adapted from Ref.~\refcite{FNOSY}.}
\label{fig:Ronkin-eg}
\end{figure}

\begin{figure}[htbp]
\begin{center}
\begin{tabular}{ccc}
\includegraphics[width=40mm]{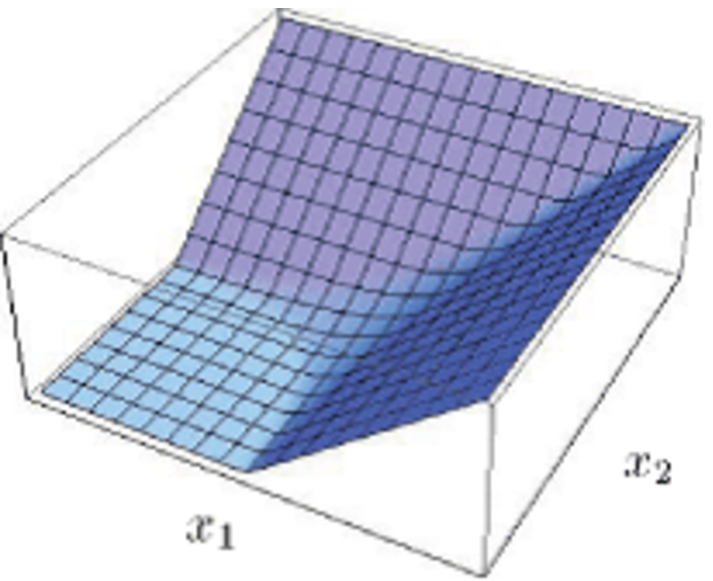} & \hs{10} &
\includegraphics[width=40mm]{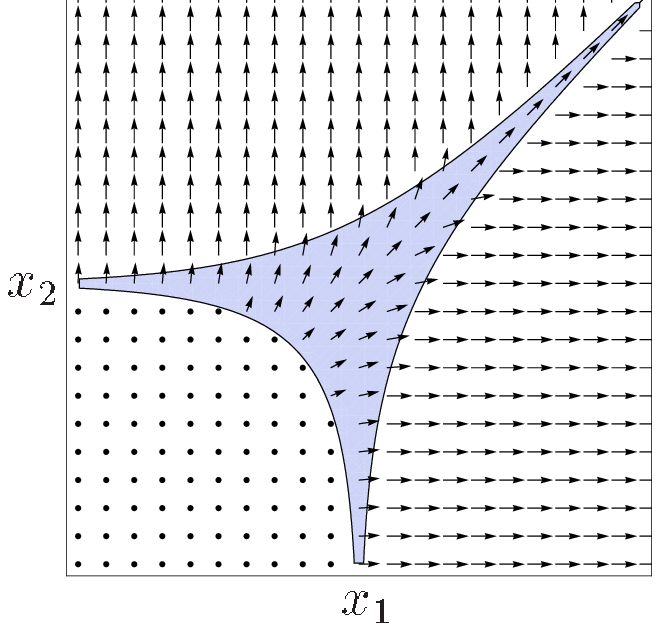} \\
(a) Ronkin function & & (b) gradient of Ronkin function 
\end{tabular}
\end{center}
\caption[(a) Ronkin function and (b) the gradient flow of the Ronkin function.]{(a) Ronkin function and (b) the gradient flow of the Ronkin function, from Ref.~\refcite{FNOSY}.}
\label{fig:Ronkin-plot2}
\end{figure}

Let us here give the simplest example: $\bC^3$, whose toric diagram is given in Fig.\,\ref{fig:Ronkin-eg} (a). 
The characteristic polynomial is given by
\beq
P(z,w) = z+w+ 1,
\label{eq:example-P}
\eeq
and its amoeba is shown in Fig.\,\ref{fig:Ronkin-eg} (b).
The derivatives of the corresponding Ronkin function are computed to be 
\beq
\partial_x R(x,y) = 
\left\{
\begin{array}{ccl}
0 & \text{for} & x < \displaystyle \log \left| e^{y} - 1 \right| \\
\vs{2} \displaystyle 1 - \frac{1}{\pi}\cos^{-1} \left( \frac{e^{2 x} - e^{2 y} - 1}{2 e^{y}} \right) & \text{for} & \displaystyle \log \left| e^{y} - 1 \right| \leq x \leq \displaystyle \log \left| e^{y} + 1 \right| \\
\displaystyle 1 & \text{for} & x > \displaystyle \log \left| e^{y} + 1 \right|
\end{array}
\right.,
\nonumber
\eeq
\beq
\partial_y R(x,y) = 
\left\{
\begin{array}{ccl}
0 & \text{for} & y < \displaystyle \log \left| e^{x} - 1 \right| \\
\vs{2} \displaystyle 1 - \frac{1}{\pi}\cos^{-1} \left( \frac{e^{2 y} - e^{2 x} - 1}{2 e^{x}} \right) &  \text{for} & \displaystyle \log \left| e^{x} - 1 \right| \leq y \leq \displaystyle \log \left| e^{x} + 1 \right| \\
\displaystyle 1 &  \text{for} & y > \displaystyle \log \left| e^{x} + 1 \right|
\end{array}
\right. ,
\nonumber
\eeq
and their plots are given in Fig. \ref{fig:Ronkin-plot2}. 
Note that the derivatives of the Ronkin function takes  
a constant value at each  
complement of 
amoeba, 
as expected.

\medskip

We now come back to the thermodynamic limit of the partition function \eqref{thermlimit}. The reason we introduced the Ronkin function of the characteristic polynomial is that the surface tension 
$\sigma(\partial_xh, \partial_y h)$ is the Legendre transform
of the Ronkin function with respect 
to $(s,t)=(\partial_x h, \partial_y h)$ as (Theorem 3.6 of Ref.~\refcite{KOS})
\footnote{
Let us given a rough outline of why the Ronkin function arises here.
\eqref{thermlimit} shows that in the thermodynamic limit the surface tension is related to
\beq
\lim_{N\to \infty} -\frac{1}{N^2} \log Z_N,
\label{limitN}
\eeq
where $Z_N$ is defined for dimer model with fundamental domain enlarged by $N\times N$. Now since the characteristic polynomial can be written as a determinant of the Kasteleyn matrix \cite{Kasteleyn} it follows that \cite{KOS}
\beq
Z_N(z,w)=\prod_{z_0^N=z}\prod_{w_0^N=w} \tilde{P}(z_0,w_0).
\label{Kasteleyn}
\eeq
Plugging \eqref{Kasteleyn} into \eqref{limitN}, the product is turned into a Riemann sum, which is nothing but the integral in the definition of the Ronkin function in \eqref{ronkin}.
}
\beq
R(x,y)=  {\rm max}_{s,t} \left[ -\sigma(s,t)
+xs + yt\right].
\label{legendre}
\eeq

The first step in relating the dimer model to the topological
string theory is to show that
the characteristic polynomial $\widetilde{P}(z,w)$ of the dimer model 
is equal to the Newton polynomial $P(z,w)$ \eqref{mirrorP} for the mirror Calabi-Yau 
manifold (\ref{mirror}),
\beq
   \widetilde{P}(z,w) = P(z,w).
\label{agree}
\eeq

According to Ref.~\refcite{FV}, 
there is a one-to-one correspondence between 
perfect matchings of the dimer model on $\bT^2$ and bi-fundamental fields 
of gauged linear sigma model appearing in the K\"ahler quotient 
construction of the toric Calabi-Yau manifold $X$. 
They are then related, by change of variables described in Ref.~\refcite{FV},
to lattice points of the toric diagram and to terms $z^{n_i}w^{m_i}$ in the
Newton polynomial (\ref{mirrorP}). This shows that
there is a one-to-one correspondence between terms in $P(z,w)$ and $\tilde{P}(z,w)$,
as pointed out by Ref.~\refcite{FHKV}.
Furthermore we can show that their coefficients agree.
The K\"ahler moduli of $X$ are
the Fayet-Iliopoulos (FI) parameters
of the quiver quantum mechanics. 
According to the dictionary 
between quiver gauge theories and dimer models (Section\,\ref{sec.quiver}),
FI parameters are associated with nodes of the quiver diagram, or equivalently
to faces of the dimer model.
Thus, we can identify the FI parameters with the magnetic fluxes through
faces of the dimer model \eqref{magneticflux}, which parametrize the energy of each perfect 
matching of the dimer model (recall the example of the Suspended Pinched Point discussed previously; there the magnetic fluxes along faces are denoted by $t_1$ and $t_2$). 
Each perfect matching appears in 
$\widetilde{P}(z,w)$ with the weight given by an exponential of the
fluxes. On the other hand, the Newton polynomial is a sum of
$z^{n_i}w^{m_i}$, each of which corresponds to a lattice point of the toric diagram
and is weighted by an exponential of the K\"ahler moduli $t$ \cite{HoriV}.
We have verified that the weight for perfect matchings and
the weight for lattice points of the toric diagram agree, and this proves 
the identity (\ref{agree}). 

Combining (\ref{thermlimit}) and (\ref{legendre}) and discarding a term
in total derivative in $(x,y)$, which is justified by the subtraction
of the linear piece in $R(x,y)$ discussed in the next paragraph, 
we find
$$
  Z \sim \exp\left[ N^2 \int dx dy R\left(
\frac{g_sN}{2} x, \frac{g_sN}{2} y\right) \right]. 
$$
By rescaling $(x,y)$ by the factor of $g_sN/2$, this becomes 
\beq
  Z \sim \exp\left[ \frac{4}{g_s^2} \int dx dy R\left(x, y\right) \right]. 
\label{largeN}
\eeq
Note that the $N$ dependence has disappeared except that the range of the $(x,y)$ integral has been rescaled by the factor of $g_sN/2$. For 
$N \rightarrow \infty$ with small but fixed $g_s$, we have an integral over
the whole $(x,y)$ plane. 

The integral (\ref{largeN}) in the large $N$ limit is divergent.
To identify and subtract the divergent part, we use the amoeba of the characteristic polynomial defined in \eqref{amoeba}. 
In the thermodynamic limit, the amoeba corresponds to the liquid
phase of the crystal \cite{KOS} \footnote{The paper Ref.~\refcite{KOS} gives three different phases of the dimer model: solid, liquid and gas. In the solid phase (frozen phase), the height difference between points arbitrary far away are deterministic. Non-solid phases are classified into either a gas phase (rough phase) or a liquid phase (smooth phase) depending on whether or not the height difference between the two point is bounded independently of the choice of the points.
It was shown in Ref.~\refcite{KOS} that an unbounded exterior region, an interior region and a bounded exterior region of the amoeba correspond to a solid, liquid and gas region, respectively.
}. If there are no interior points in 
the toric diagram, the complement of the amoeba is the solid phase,
where the crystal retains its original shape. There, the Ronkin function
$R(x,y)$ is linear. If there are interior lattice
points in the toric diagram,
the amoeba acquires holes, inside of which are in the gas phase,
where the Ronkin function is again linear but the slope of the crystal
surface is different from the original one. 
The integral (\ref{largeN}) becomes finite if
we subtract the linear piece of the Ronkin function
in the solid phase so that the partition function is normalized
to be $1$ for the initial crystal configuration.

\subsection[Topological String at Genus 0]{Topological String at Genus $0$}\label{sec.genus0}

Our next task is to compute the genus-$0$ topological string partition 
function $\scF_0$ of the toric Calabi-Yau manifold $X$ and compare it with the 
thermodynamic limit of the partition function of the crystal melting model (\ref{largeN}). 
For this purpose, it is convenient to use the mirror Calabi-Yau manifold
$\widetilde{X}$  
defined by the equation (\ref{mirror}) since
$\scF_0$ can be
evaluated by the classical period integral as,
\beq
\scF_0=\int_{\scC_0} \Omega,
\label{period}
\eeq
where $\Omega$ is the holomorphic 3-form on the mirror, 
and $\scC_0$ is the
Lagrangian 3-cycle which is the mirror of the 6-cycle filling 
the entire toric Calabi-Yau manifold.

According to the microscopic derivation of the mirror symmetry by 
Hori and Vafa \cite{HoriV}, the sigma-model on the toric Calabi-Yau manifold
is equivalent to the Landau-Ginzburg model with the superpotential
\beq
 W(u, x, y) = e^u P(e^x,e^y).
\eeq
It was shown in Ref.~\refcite{HoriIV} that an integral of $e^W$ in a 3-dimensional
subspace of the $(u,x,y)$ plane can 
be transformed into a period integral of the holomorphic 3-form $\Omega$
on the mirror Calabi-Yau manifold. Thus, we should be able to evaluate
(\ref{period}) as an integral of $e^W$. To do so, we need to identify the
contour of the integral. 

Since $g_s$ and $t^I$'s in (\ref{partitionfunction}) are 
taken to be real in the
dimer model, the Newton polynomial $P(x,y)$ in our case is with real 
coefficients. The mirror manifold (\ref{mirror}) has the 
complex conjugation involution, and thus the fixed point set is a natural
candidate for $\scC_0$. In fact, following the mirror symmetry 
transformation as described in Ref.~\refcite{HoriV}, we find that the 6-cycle 
of the original toric Calabi-Yau manifold $X$ corresponds to the real section in the
$(u, x, y)$ space in the mirror $\widetilde{X}$. Thus, we find
\begin{equation}
\scF_0=\int_0^{\infty} du \int_{-\infty}^\infty
 dx dy\, e^{e^u P(e^x,e^y)} 
= - \int dx dy \ln P(e^x,e^y).
\label{log}
\end{equation}
The divergent part of the integral in
the $(x,y)$-plane can be removed by subtracting
a linear term in $\ln P(e^x, e^y)$ 
for $x,y \rightarrow \infty$ as we did for
(\ref{largeN}). The integral \eqref{log} is almost equal to 
the exponent of the partition function 
(\ref{largeN}) of the crystal melting model, except that 
we do not have the averaging over the phases $(\theta,\phi)$ to
define the Ronkin function as in (\ref{ronkin}). It turns out that
the integral over $(x,y)$ in (\ref{log}) removes the dependence 
on $(\theta,\phi)$, and thus the averaging process is not necessary.

To see this, let us define a generalization of $\scF_0$ for an
integral of $(x,y)$ with arbitrary phases as
\beq
\scF(\theta,\phi) 
= \int dx dy \ln P(e^{x+i\theta},e^{y+i\theta}).
\eeq
Taking derivatives of $\scF$ with
respect to $(\theta,\phi)$, we find the integrand becomes
a total derivative in $(x,y)$ as in 
\begin{equation*}
\left(\alpha {\partial\over \partial \theta}+
 \beta {\partial\over \partial \phi}\right)\scF 
=\int dx dy \,i\left(\alpha {\partial\over \partial x} 
+\beta {\partial\over \partial y} \right) \ln P(e^{x+i \theta},e^{y+i\phi}).
\end{equation*}
If we choose $(\alpha,\beta)$ so that it is not in the directions
of the tentacles of the amoeba (\ref{amoeba}), the
boundary term is removed by the regularization and we find $(\alpha \partial_\theta + 
\beta \partial_\phi)\scF=0$. 
Since $\alpha, \beta$ is arbitrary except in the directions of tentacles of amoeba, $\scF$ is independent of $(\theta,\phi)$ and
agrees with its average. Namely, 
\beq
\scF_0= - \int dx dy \ln P(x,y) = -\int dx dy\, R(x,y).
\eeq

Thus, we found that the thermodynamic limit of the partition function 
 of the crystal melting model given by (\ref{largeN}) is equal to $\exp(-\frac{4}{g_s^2} 
\scF_0)$, which is the genus-$0$ partition function of the topological
string theory. This is what we wanted to show.

The fact that that crystal melting gives the topological string theory in the thermodynamic limit, despite the non-trivial wall crossing phenomena between the two, poses a puzzle. We will see later in Section\,\ref{chap.M-theory} how this puzzle is solved in the analysis of the wall crossing phenomena.



\section{Wall Crossing and M-theory}\label{chap.M-theory}

In section \ref{chap.crystal} we succeeded in computing the BPS partition function as a statistical partition function of crystal melting. However, the crystal melting gives only an expansion of the BPS partition function, and whether the partition function can be expressed in a closed expression is a different issue. In Section\,\ref{sec.ZBPS}, we have seen that in the example of the resolved conifold the BPS partition function takes a simple expression, which is essentially a square of the topological string partition function. The question we would like to address in this section is to understand this curious fact. Interestingly, this also gives a nice intuitive derivation of the wall crossing phenomena.
This section is based on the results of Ref.~\refcite{AOVY}.

\subsection{Overview: Topological Strings and Wall Crossings}\label{sec.M-theoryintro}

Since one of the goals of this section is to relate the BPS partition function with the topological string partition function, let us begin with an overview of topological string theory.

The topological string theory gives solutions to a variety of 
counting problems in string theory and M-theory (see Refs.~\refcite{MarinoReview,NeitzkeV,OoguriReview} for review).
From the worldsheet perspective, the A-model topological string partition
function  $Z_{\rm top}$ generates the Gromov-Witten invariants, which
count holomorphic curves in a Calabi-Yau 3-fold $X$. On the other hand,
from the target space perspective, $Z_{\rm top}$ computes
the Gopakumar-Vafa (GV) invariants, which count BPS states of
spinning black holes in 5 dimensions
constructed from M2 branes in M-theory on $X$ \cite{GV1}.
Moreover, the absolute-value-squared $|Z_{\rm top}|^2$ has been
related to the partition function
of BPS extremal black holes in 4 dimensions, which are bound states of D-branes
in type II string theory
on $X$ \cite{OSV}.

The topological string partition function $Z_{\rm top}$ also
counts the numbers of D0 and D2
brane bound states on a single D6 brane on $X$, namely the
Donaldson-Thomas (DT) invariants defined
in Refs.~\cite{DT,Thomas}. The relation between the Gopakumar-Vafa invariants
 and DT invariants was suggested and
formulated in Refs.~\refcite{INOV,MNOP1}, and its physical explanation was given in Ref.~\refcite{DVVafa} using the 4D/5D
connection \cite{GSY}. More recently, a mathematical proof of the GV/DT correspondence was given in
Ref.~\refcite{MOOP} when $X$ is a toric Calabi-Yau 3-fold.

However, the number of BPS states has background dependence. 
As we vary moduli of
the background geometry and cross a wall of marginal stability, 
the number can jump (this is the wall crossing phenomena discussed in \ref{sec.WC}).
This means that the K\"ahler moduli is implicitly assumed in most of the above computations, 
and we need to understand at which value of the K\"aher moduli the statement holds and
what happens if we change the value of the moduli.

In this section we will generalize the results of Ref.~\refcite{DVVafa} to include 
the background dependence of
the M-theory computation. 
We show BPS bound states are organized into a free field Fock space, whose 
generators correspond to BPS states of spinning M2 branes in M-theory 
compactified down to 5 dimensions by a Calabi-Yau 3-fold $X$. This
enables us to write the generating function 
$Z_{\rm BPS}$ of BPS bound states of D-branes
as a reduction of the square of the topological string partition function,
\beq
   Z_{\rm BPS}(q,Q) = Z_{\rm top}(q,Q) Z_{\rm top}(q,Q^{-1}) |_{\rm chamber}, \label{theformula}
\eeq
in an appropriate sense described in the following, 
in all chambers of the K\"ahler moduli space. 
Our results apply to the BPS counting
for an arbitrary Calabi-Yau (whether toric or non-toric) without compact 4-cycles. 

For the conifold, the change
of the numbers of BPS states across a wall of marginal stability
has been studied by physicists in Refs.~\refcite{JM,CJ} (see also Ref.~\refcite{DG}) and
mathematicians \cite{Young1,NN}. The case of generalized conifold geometries was
studied in Ref.~\refcite{Nagao1}. 
The formula \eqref{theformula} derived 
from the perspective of M-theory reproduces these results. 
Our results also provide a simple derivation of the ``semi-primitive'' wall
crossing formula of Denef and Moore \cite{DenefM} (discussed in Section\,\ref{chap.WCF}), in the present context.

The rest of this section is organized as follows. In Section\,\ref{sec.idea}, we will explain the basic idea:
to use M-theory to count bound states of a single D6 brane with D0 and D2 branes on a Calabi-Yau
3-fold. In Section\,\ref{sec.derivation}, we will describe the counting procedure in more detail and derive
the generating function for the numbers of BPS bound states using a free field Fock space
in any chamber of the background K\"ahler moduli space. 
In Section\,\ref{sec.examples}, we will compare the Fock space picture with
the known results for the the resolved
conifold and its generalizations. We also discuss in Section\,\ref{sec.WCcrystal} the representation of the wall crossing phenomena in the language of the crystal melting model.

\subsection{The Basic Idea}\label{sec.idea}

In this section we will explain the basic idea.  We will apply this
idea, in the following sections, to find a concrete expression for
BPS state degeneracies in various chambers for Calabi-Yau 3-folds with no
compact 4-cycles. As we will see, this technical assumption about
compact 4-cycles will ensure that M2 brane particles are mutually BPS,
 and thus the partition function takes an infinite product form \footnote{See Ref.~\refcite{AganagicSchaffer} for cases with compact 4-cycles.}.

We are interested in counting the BPS partition function of one D6 brane
bound to an arbitrary number of D2 and D0 branes.  The idea is
the following:  In M-theory, the D6 brane lifts to the Taub-NUT space with
the unit charge.
D2 branes are M2 branes transverse to the $S^1$, and D0 branes are
gravitons with Kaluza-Klein momenta along the $S^1$.
The Taub-NUT space is an $S^1$ fibration over $\bR^3$,
and $S^1$ shrinks at the position of D6. Thus the problem of
finding bound states to the D6 brane becomes simply the problem of finding BPS
states in the Taub-NUT geometry.  Suppose we have BPS states for flat
$\bR^{4,1}$ background.  Then for each such BPS state we can consider the
corresponding possible BPS states in the Taub-NUT geometry.  For each
single particle BPS state we can consider its normalized wave functions
in this geometry.  Such states would constitute BPS states which in the type IIA
reduction correspond to BPS particles bound to the D6 brane.  However,
this would only constitute single particle BPS states bound to the D6 brane.

Now consider multiple such particles in the Taub-NUT background.  This problem
may sound formidable, because now we will have to consider the interaction
of such particles with each other and even their potentially forming
new bound states.  We will now make the following two assumptions:

\medskip
\noindent
{\bf Assumption 1}:  We can choose the background moduli of Calabi-Yau as well
as the chemical potential so that a maximal set of
BPS states have parallel central charge and thus exert
no force on one another.  Therefore, at far away separation, the bound states
correspond to single particle wave function in the Taub-NUT geometry.

\noindent
{\bf Assumption 2}:  The only BPS states in 5D are particles.  In other words
there are no compact 4-cycles in the Calabi-Yau and thus we can ignore BPS {\it string
states} obtained by wrapping 5 branes around 4-cycles \footnote{From the viewpoint of geometric engineering \cite{KatzKV,KatzMV}, this means that the corresponding gauge theory is Abelian.}.
\label{Assumptions}
\medskip

Assumption 1 can be satisfied as follows:  Consider the Euclidean geometry
of M-theory in the form of Taub-NUT times $S^1$, where we have compactified
the Euclidean time on the circle.  The BPS central charge for M2 branes wrapping
2-cycles of Calabi-Yau, but with no excitation along the Taub-NUT, is given by
$$Z(M2)=iA(M2)- C(M2),$$
where $A(M2)$ denotes the area of the M2 brane and $C(M2)$ corresponds to the
coupling of the M2 brane to the 3-form potential turned on along the Calabi-Yau 2-cycles
as well as the $S^1$ of the Taub-NUT. However we need to include excitations
along the Taub-NUT.  As discussed in Ref.~\refcite{DVVafa} these are given by the momenta
along the Taub-NUT circle.  Let us denote the total momentum along the circle
by $n$ (as we will review in Section\,\eqref{sec.derivation}, this can arise both due to internal spin
as well as the orbital spin in the $SU(2)_L\subset SO(4)_{\rm rotation}$). Let us denote the radius of the Taub-NUT
circle by $R$.  In this case the central charges
of the BPS M2 brane becomes \footnote{
Note that $C(M2)$ is periodic with period $1/R$. To see this, note that we can view it as a holonomy of the gauge field obtained by reducing the 3-form on the 2-cycle of Calabi-Yau, around the Taub-NUT circle. The holonomy of a gauge field on a circle of radius $R$ is periodic, with period $1/R$. In terms of the IIA quantities, we have
\beq
C(M2) = B(D2)/R,
\eeq
where $B(D2)$ is the NS-NS B-field through the 2-cycle in IIA on the Calabi-Yau wrapped by the corresponding D2 brane (which has periodicity $B\to B+1$). We are denoting by $1/R$ the central charge of the D0 brane. As such, it does not have to be positive, and in fact it does not have to be real either. A better way to think about this is that the quantity we denote by $1/R$ is proportional, up to a complex constant, to the inverse radius of the M-theory circle. For simplicity of the notation, we will identify the two, but this fact has to be kept in mind.
Note also that relative to Ref.~\refcite{JM} we are keeping the D6 brane charge fixed 
(for example, to Z(D6)=1), and varying the D0
and D2 brane central charges.}
\beq
Z(M2,n)=iA(M2)-C(M2)-n/R.
\eeq

To satisfy assumption 1, we need to make sure that differently wrapped
M2 branes all have the same phase for $Z$.  This in particular means that
we need to choose the K\"ahler classes so that the 2-cycles of Calabi-Yau have
all shrunk to zero size, $i.e.$
$A(M2)=0$ for all the classes.  Even though this may
sound singular and it could lead to many massless states, by turning
on the $C(M2)$ we can avoid generating massless states in the limit.
The condition that different states have the same central charge
is simply that
\beq
C(M2)+n/R>0.
\eeq
Note that, in going to type IIA, this condition is simply the statement
that the $B$ fields are turned on along 2-cycles of Calabi-Yau and the M2 branes
wrapping them will have $B(D2)$ 0 branes induced.  Moreover $n$, being
the momentum along the Taub-NUT translates to D0 brane charge and as long
as the net number of 0 branes is positive, they correspond to BPS states
of the same type, $i.e.$ preserving the same supersymmetry.

Now we are ready to put together
all these mutually BPS states as a gas of particles
in the Taub-NUT geometry.  By the fact that they are mutually BPS, they
will exert no force on one another.  Moreover, as long as they are far away,
we can simply consider the product of the individual wave functions.
One may worry what happens if they come close together.  Indeed
they can form bound states, but that is already accounted for by
including all single particle bound states of M2 brane.  Here
is where the assumption 2 becomes important:  If we in addition had
4-cycles, then wrapped M5 branes along 4-cycles, which also wrap the
$S^1$ of Taub-NUT can now form new bound state with the gas of M2
brane particles on the Taub-NUT \footnote{From gravity viewpoint, we have black strings.}.  But in the absence of 4-cycles
of Calabi-Yau, we can simply take the single particle wave functions
(taking their statistics into account) and write the total degeneracy
of such BPS states, by taking suitable bosonic/fermionic creation
operators, one for each state satisfying $C(M2)+n/R>0$. Finally, while the assumption 1 is satisfied only for special backgrounds where $A(M2)$ vanishes, the degeneracies are guaranteed to be the same everywhere within a given chamber, and are independent of this choice. This is all we need to compute all the degeneracies of BPS states in various chambers
as we will show in the following sections.

\subsection{BPS State Counting and Wall Crossing}\label{sec.derivation}

We will use this section to spell out, in a little more detail,
how to use M-theory to compute the degeneracies of
one D6 brane on $X$ bound to D2 branes wrapping 2-cycles in $X$ and D0 branes.
The D6-D2-D0 partition function is the Witten index \footnote{
Here we are ignoring the fermionic zero modes in the 4 non-compact directions. Otherwise, additional factors need to be inserted to absorb these. See the definition of BPS index in Section\,\ref{sec.index}.}

$$
{\rm Tr}[(-1)^F e^{-\epsilon H}]
$$
of the theory on
$$
X \times \bR^3 \times S_t^1,
$$
where we have compactified the Euclidean time on a circle of radius 
$\epsilon$. The type IIA geometry with one D6 brane lifts to to M-theory on
$$
X \times {\rm Taub} \hbox{-} {\rm NUT} \times S_t^1,
$$
where the asymptotic radius of the Taub-NUT circle $R$ is related to IIA string coupling.
Since the D6 brane is geometrized,
the computation of the BPS bound states of D2 branes and D0 branes with D6 brane
lifts to a question of computing the degeneracies of
$M2$ branes with momentum around the Taub-NUT circle.

\begin{figure}[htbp]
\centering{\includegraphics[scale=0.25]{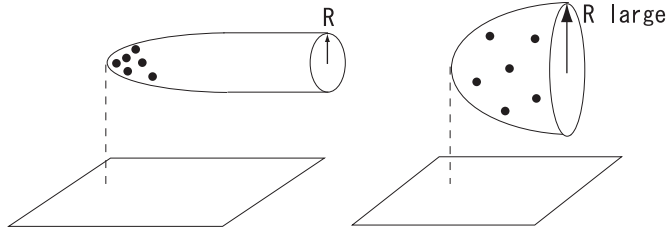}}
\caption[The pictorial representation of 4d/5d correspondence. ]{The pictorial explanation of 4d/5d correspondence. When we send $R\to\infty$, the Taub-NUT space, which is asymptotically $\bR^3\times S^1$ as shown on the left, becomes flat $\bR^4$ as shown on the right. The M2-brane particles, which are originally bound to the tip of the Taub-NUT, moves freely in $\bR^4$ after this process. This figure is the reproduction of Fig.\,1 in Ref.~\refcite{DVVafa}.}
\label{4d5d}
\end{figure}

Suppose we know the degeneracies of M-theory in the
$$
X \times \bR^4 \times S^1.
$$
This corresponds to taking the
$R\to \infty$
limit, where the Taub-NUT just becomes $\bR^4$ (Fig.\,\ref{4d5d}).
As is clear from the previous section, at {\it fixed} $B$ the degeneracies are {\it unchanged} by varying
$R$ since no states decay in the process --- all the central charges simply get re-scaled.
Thus, the knowledge of these allows us to compute the degeneracies on
$X \times$Taub-NUT$\times S^1$ background as well.

The Kaluza-Klein momentum around the Taub-NUT circle gets identified, in terms of the theory in the $R\rightarrow \infty $ limit, with the total spin of the M2 brane. This can be understood by comparing the isometries of the finite and the infinite $R$
theory, as explained in Ref.~\refcite{DVVafa}.
We can view taking $R$ to infinity as zooming in to the origin of the Taub-NUT. The isometry group is the rotation group $SO(4) = SU(2)_L \times SU(2)_R$ about the origin of $\bR^4$. The $SU(2)_R$ is identified with the $SO(3)$ that rotates the sphere at infinity of the $\bR^3$ base of the Taub-NUT. Moreover, the rotations around the $S^1$ of the Taub-NUT, end up identified with the
$$
U(1) \subset SU(2)_L.
$$
Thus, the Kaluza-Klein momentum, is identified with the total $J^{L}_z$ spin of the M2 brane on $\bR^4$.

Now, let
$$N_{\beta} ^{(m_L, m_R)}$$
be the degeneracy of the {\it 5-dimensional BPS} states of M2 branes of charge $\beta$ and spin the intrinsic $(2j^z_L, 2 j^z_R) =(m_L, m_R )$
(where the spin refers to the spin of the highest state of the multiplet).
To get an index, we will be tracing over the $SU(2)_R$
quantum numbers, so we get a net number
\beq
N_{\beta}^{m_L} = \sum_{m_R} (-1)^{m_R} N_{\beta}^{(m_L, m_R)}
\label{Nindex}
\eeq
of 5D BPS states, of the fixed $SU(2)_L$ spin $m_L$. These integer invariants are the Gopakumar-Vafa invariants of Refs.~\refcite{GV1,GV2}.

Each such 5D BPS particle can in addition have excitations on $\bR^4$. Namely, for each 5D particle we get an analytic field
$$\Phi(z_1,z_2)$$
on $\bR^4$ with $z_{1,2}$ as the complex coordinates. In the usual way, the modes of this field
\beq
\Phi(z_1,z_2) = \sum_{\ell_1, \ell_2} \alpha_{\ell_1,\ell_2} z_1^{\ell_1} z_2^{\ell_2}
\eeq
correspond to the ground-state wave functions of the particle with different momenta on $\bR^4$. (We are suppressing a Gaussian factor that ensures the wave functions are normalizable).
Since $U(1)\subset SU(2)_L$ acts on $z_1,z_2$ with charge $1$, the particle corresponding to
$$
{\alpha}_{\ell_1,\ell_2}
$$
carries, in addition to the M2 brane charge $\beta$ and intrinsic momentum $m$, a total angular momentum \footnote{In the present case, we are restricting to Calabi-Yau manifolds with no compact 4-cycles. When the Calabi-Yau is furthermore toric, as in the cases discussed in Section\,\ref{sec.conifold} and \ref{sec.gc}, the genus of the target space curve wrapped by the M2 branes vanishes. This means that the intrinsic spin of all the M2 branes vanishes as well.}.
\beq
n = \ell_1+\ell_2+m.
\label{totaln}
\eeq

\begin{figure}[htbp]
\centering{\includegraphics[scale=0.2]{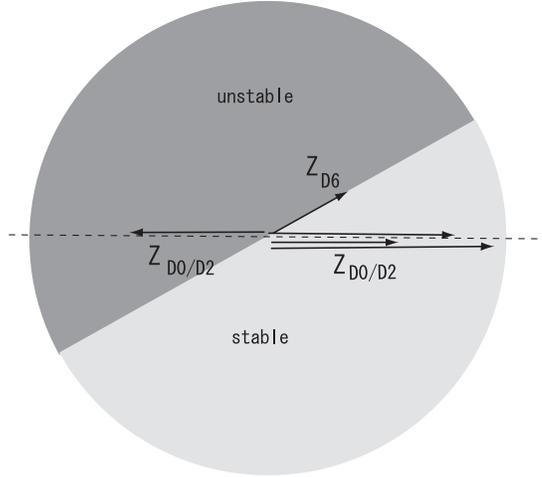}}
\caption[The central charges of the D6-brane and the D0/D2-brane particles.]{The central charges of the D6-brane and the D0/D2-brane particles, expressed in the complex plane. The D0/D2 particles, whose central charge lie on the real axis, is stable (unstable) in the light (dark) gray region.}
\label{Zalignment}
\end{figure}

Which of these 1-particle states are mutually BPS? The answer depends on the background, and a priori, we need to consider particles in four dimensions coming from both the M2 branes and the anti-M2 branes in M-theory.
Along the slice in the moduli space we have been considering, the central charges of the particle with M2 brane charge $\beta$ and total spin $n$ is
\beq
Z(\beta, n) = C(\beta) + n/R =  (B(\beta) + n)/R,
\label{Z}
\eeq
where $C(\beta), B(\beta)$ are the $C$-field ($B$-field) flux through
the 2-cycle $\beta$.
The states with
\beq\label{cc}
Z(\beta,n)>0
\eeq
all preserve the same supersymmetry and bind to the D6 brane (we could have picked the opposite sign, and than the particles would bind to anti-D6 branes).
See Fig.\,\ref{Zalignment} for illustration. Note that although all the D0/D2 particles are mutually BPS, they preserve different supersymmetry from that of the D6-brane, which is why we have a stable bound state.

It is easy to analyze the condition \eqref{cc}. For example, for 
$$
B>0, \qquad R>0
$$
along side M2 branes with $\beta>0$, and for sufficiently large $n$ also the anti-M2 branes with $\beta<0$ have positive $Z>0$ and contribute to the BPS partition function. So in general we need to consider both signs of $\beta$. It is important to note that the degeneracies $N_{\beta}^{m}$ of the 5D particles are independent of the background. The choice of background only affects which half of the supersymmetry the states preserve.

Now, we can put these all together and compute the BPS partition function in a given chamber. Simply, in each chamber, the BPS partition function is the character in the Fock space of single particle states
preserving the same supersymmetry! In fact, a useful way to go about computing the partition function is in the following steps:
\vskip 0.3cm
\noindent
{\bf Step 1.\;}  Start with the unrestricted partition function -- the character
\beq
Z_{\rm Fock} = {\rm Tr}_{\rm Fock} \; q^{Q_0} Q^{Q_2}
\eeq
in the full Fock space.
The oscillators of charge $\beta$ and intrinsic spin $m$ and arbitrary 4d momenta contribute a factor
\beq\label{factor}
\prod_{\ell_1+\ell_2=l} (1- q^{\ell_1+\ell_2 +m } Q^{\beta})^{N_{\beta}^m}
=
(1-q^{l+m} Q^{\beta})^{l N_{\beta}^m}.
\eeq
In addition, both the M2 branes, and the anti-M2 branes contribute, and the total character is
\beq
Z_{\rm Fock} = \prod_{\beta,m} \prod_{l=1}^{\infty}\,
(1-q^{l+m} Q^{\beta})^{l N_\beta^{m}}
\eeq
\vskip 0.3cm

\noindent
{\bf Step 2.\;} The 5d degeneracies $N_{\beta}^m$ of M-theory on $X\times \bR^{4,1}$ are computed by the topological string partition function on $X$ \cite{GV1,GV2} .  This allows us to write
\beq
Z_{\rm Fock} = Z_{\rm top}(q,Q) Z_{\rm top}(q,Q^{-1}).
\eeq
In particular, the knowledge of topological string amplitude allows us to compute the BPS degeneracies in any chamber.

The topological string partition function has an expansion
\beq
Z_{\rm top}(q,Q) = M(q)^{\chi(X)/2} \prod_{\beta>0, m} \prod_{l=1}^{\infty}
(1-q^{m+l} Q^{\beta})^{l N_m^{\beta}}.
\eeq
where $q$ and $Q$ are determined by the string coupling constant $g_s$
and the K\"ahler moduli $t$ by $q=e^{-g_s}$ and
$Q = e^{-t}$ \footnote{As compared with \eqref{topvar}, we are here absorbing the minus sign in \eqref{topvar} into the definition of $q$. We will use this notation throughout the rest of this paper. Of course, the topological strings partition function itself is the same in both notations.}. The MacMahon function $M(q)$ was defined previously in \eqref{MacMahondef}, and 
$\chi(X)$ is the topological Euler characteristic of $X$. Note that topological string involves only the M2 states with positive $\beta>0$. On the other hand, the full Fock space includes also anti-M2 branes. Since M2 branes and anti-M2 branes are CPT conjugates in 5d, this gives another factor of $Z_{\rm top}$ with $Q \rightarrow Q^{-1}$. 

Note that we also have states with $\beta=0$. 
These are the pure KK modes, the particles with no M2 brane charge. 
To count the number of BPS states of this type,
we note that, for each $\mathbb{R}^4$ momentum $(\ell_1,\ell_2)$ we get a classical particle 
whose moduli space is the Calabi-Yau $X$. Quantizing this, we get a particle 
for each element of the cohomology of $X$. On a $(p,q)$ form $SU(2)_R$ acts 
with the Lefshetz action, and $SU(2)_L$ acts trivially. We get that $m_R$ 
eigenvalue of a $(p,q)$ form on $X$ is $m_R = p+q - 3$. Therefore,
the pure KK modes contribute with $N_{\beta=0} = -\chi(X)$. This agrees
with the power of the MacMahon function $M(q)$ we get from 
$Z_{\rm top}(q,Q) Z_{\rm top}(q,Q^{-1})$.

\vskip 0.3cm

\noindent
{\bf Step 3.} We identify the walls of marginal stability as places where,
the central charge vanishes \footnote{This is a degenerate limit of \eqref{wallposition}.} 
for one of the oscillators contributing to $Z_{\rm top}(q,Q)$ or  
$Z_{\rm top}(q,Q^{-1})$.

\vskip 0.3cm
\noindent
{\bf Step 4.\;} In any chamber, the BPS partition function is a restriction 
of $Z_{\rm Fock}$ to the subspace 
of states that satisfy $Z(\beta,n)>0$ in that chamber.
\begin{align}
\begin{split}
Z_{\rm BPS}({\rm chamber}) =& Z_{\rm Fock}|_{\rm chamber}, \\
 = & Z_{\rm top}(q,Q) Z_{\rm top}(q,Q^{-1})|_{\rm chamber} \label{ZChamber}
\end{split}
\end{align}

There is a simple way to keep track of the chamber dependence. For the book-keeping purposes, it is useful to identify the central charge with the chemical potentials. Then, in a given chamber, the BPS states are those for which
\beq
q^n Q^{\beta}<1
\eeq
where $n=l+m=\ell_1+\ell_2+m$ is the total spin \eqref{totaln}. As we vary the background, and cross into a chamber where this is no longer satisfied for some $(n,\beta)$ in $Z_{\rm top}(q,Q)$ or in $Z_{\rm top}(q, Q^{-1})$, we drop the contribution of the corresponding oscillator. 
\vskip 0.3cm

For example, consider some special cases. When
\beq 
R>0, \qquad B \rightarrow \infty, \label{DTChamber}
\eeq
for all K\"ahler classes, $Z(\beta,n)= (\beta B +n)/R>0$ implies that
$$
\beta>0.
$$
In this case, only M2 branes contribute to the partition function.
This is the chamber discussed in Ref.~\refcite{DVVafa}. By taking the limit
\eqref{DTChamber} in \eqref{ZChamber}, we find
\beq
Z_{\rm BPS}(R>0, B \rightarrow \infty)  = 
Z_{\rm DT}(q,Q) = M(q)^{\chi(X)/2} Z_{\rm top}(q,Q).
\eeq
The partition function in this chamber computes DT invariants.
In Ref.~\refcite{MOOP}, it was shown that, for a toric Calabi-Yau,
$Z_{\rm BPS}$ is equal to $Z_{\rm top}$ up to a factor
which depends only on $q$. 
Here we derived the relation between $Z_{\rm BPS}$ and
$Z_{\rm top}$ including the factor of $M(q)^{\chi(X)/2}$. 

On the other hand, when $0 < B \ll 1$, the BPS partition function 
is given by 
\beq
   Z_{\rm BPS}(q,Q) =
Z_{\rm NCDT}(q,Q) =  Z_{\rm top}(q,Q)
Z_{\rm top}(q,Q^{-1}). \label{NCDTgeneral}
\eeq
This gives the non-commutative DT invariants studied in Section\,\ref{chap.crystal}. When
$X$ is toric, the partition function is computed by the crystal
melting picture \cite{MR,OY1}, generalizing the previous result of Refs.~\cite{ORV,INOV} for $\bC^3$. In Section\,\ref{chap.thermodynamic}, we have seen that the
thermodynamic limit of the partition function of the crystal
melting model gives the genus-$0$ topological
string partition function. This result was mysterious since the relation
between $Z_{\rm top}$ and $Z_{\rm BPS}$ was supposed to hold
in the DT chamber discussed in the previous paragraph, not in the NCDT chamber. 
We now understand why there is such a relation in the non-commutative
DT chamber also as in \eqref{NCDTgeneral}.

\subsection{Examples}\label{sec.examples}

In this section we give some examples
of geometries without compact 4-cycles. 
We first study toric cases, namely resolved conifold and generalized conifolds.
We also give a simple example of a non-compact, non-toric Calabi-Yau manifold as well. In each of these cases, we will use our methods to lay out the chamber structure, identifying walls where BPS states jump, and the BPS partition function in each chamber.
In some of the cases we study, the jumps were studied by other means. We will show that they agree with the M-theory results.

\subsubsection{Resolved Conifold}\label{sec.conifold}

The topological string partition function for the resolved conifold
was given previously in \eqref{conifoldZGV}, which
 means that the only non-vanishing Gopakumar-Vafa
invariants are 
\beq
N_{\beta=\pm 1}^0=1,\quad N_{\beta=0}^0=-2,
\label{conifoldGV}
\eeq
and that all BPS states in 5 dimensions has no intrinsic spin 
\cite{GV1,GV2}. Our formula \eqref{ZChamber} then implies that BPS states
are counted by
\begin{align}
\begin{split}
 Z_{\rm BPS} (q, Q) &=Z_{\rm top}(q, Q)Z_{\rm top}(q, Q^{-1})|_{\rm chamber} \\
& =  \prod_{(\beta, n): Z(\beta, n)> 0} (1 - q^n Q^\beta)^{n N_\beta^0}.
\label{conifold}
\end{split}
\end{align}
The product is over $\beta =0, \pm 1$ and $n=1,2,...$ such that
$Z(\beta, n)> 0$.

The chamber structure is easy to identify in this case since the K\"ahler 
moduli space is one-dimensional. When
\beq
   R>0 ~ {\rm ~~and~~}~ m-1 < B < m
\eeq
with some $m \geq 1$, the formula \eqref{conifold} gives
\beq
  Z_{\rm BPS} (q, Q) = M(q)^2
\prod_{n=1}^\infty(1-q^nQ)^n \prod_{n=m}^\infty
(1-q^n Q^{-1})^n.
\label{ZCnresult}
\eeq
In particular, the chamber at $m=\infty$ counts the DT invariants
\cite{DT}, 
\beq
Z_{\rm BPS} (q, Q) = Z_{\rm DT} (q, Q) = M(q)^2
\prod_{n=1}^\infty(1-q^nQ)^n,
\eeq
while the chamber at $m=1$ counts the non-commutative DT invariants
\cite{Szendroi},
\beq
Z_{\rm BPS} (q, Q) =Z_{\rm NCDT} (q, Q) = M(q)^2
\prod_{n=1}^\infty(1-q^nQ)^n \prod_{n=1}^\infty
(1-q^n Q^{-1})^n.
\eeq

On the other hand, when 
\beq
   R<0 ~ {\rm ~~and~~}~ -m-1 < B < -m
\eeq
with $m \geq 1$, we have
\beq
Z_{\rm BPS} (q, Q) = \prod_{n=1}^m(1-q^nQ)^n. 
\eeq
In particular, the chamber at $m=\infty$ counts the 
Pandharipande-Thomas invariants \cite{PT},
\beq
Z_{\rm BPS} (q, Q) =  
Z_{\rm PT} (q, Q) = \prod_{n=1}^\infty(1-q^nQ)^n. 
\eeq

\begin{figure}[htbp]
\centering{\includegraphics[scale=0.5]{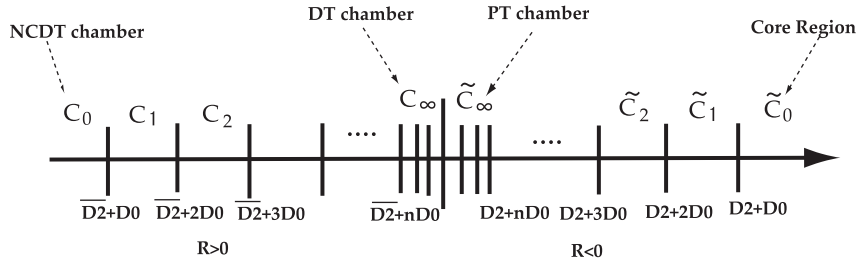}}
\caption[Chamber structure for the resolved conifold.]{Chamber structure for the resolved conifold. The charge $\gamma_2$ for the decay pattern $\gamma\to \gamma_1+\gamma_2$ is shown for each wall.}
\label{conifoldchamber}
\end{figure}

Summarizing, wall crossing of the resolved conifold is essentially a one-dimensional problem (see Fig.\,\ref{conifoldchamber}) \footnote{
Strictly speaking, only half of the chambers are shown in Fig.\,\ref{conifoldchamber}. There are also chambers for the flopped geometry, which correspond to 
$R>0, -n-1<B<-n$ and $R<0, n-1<B<n$. This means that the there are two different ways to go from the core region to the NCDT chamber, and the chambers actually line up on the $S^1$, rather than on $\bR$ as depicted in Fig.\,\ref{conifoldchamber}.
}. 
For later reference, we defined by chamber $\tilde{C}_n$ and $C_n$ as a region where $R<0, -n-1<B<-n$ and $R>0, n-1<B<n$, respectively.
The partition function is trivial in the chamber $\tilde{C}_0$ (the so-called core region \cite{DenefM}), and each time we cross the wall we obtain one factor. On the left most side, we have $Z_{\rm NCDT}(q,Q)=Z_{\rm top}(q,Q)Z_{\rm top}(q,Q^{-1})$.

As we will see in the next section, these results agree with the predictions from the wall crossing formula in all chambers.

\subsubsection{Toric Calabi-Yau without Compact 4-cycles}\label{sec.gc}

We can also test our formula \eqref{ZChamber} for a more general
toric Calabi-Yau without compact 4-cycles. 
A toric Calabi-Yau is characterized by a convex polygon on a square 
lattice, and the absence of compact 4-cycles means that there is no
internal lattice point in the polygon \cite{Fulton,Oda}. 
By $SL(2, \mathbb{Z})$ transformations 
of the lattice, one can move one of the edges of the polygon along the
positive $x$-axis, and one of the vertices to $(x,y)$ with 
$-y < x \leq 0$. If we require that there is no internal lattice point,
there are essentially two possibilities: $(x,y) = (0,1)$ and $(0,2)$.
In the former case, the polygon is a trapezoid of height 1, and the
corresponding Calabi-Yau is
the so-called generalized conifold, which has $L-1$ $\bP^1$'s 
where $L$ is the area of the trapezoid. We will describe the 
resolved geometry in 
more detail below. In the latter case, 
we have a isosceles right triangle with two legs of length 2,
which corresponds to $\mathbb{C}^3/\mathbb{Z}_2\times \mathbb{Z}_2$.

For the generalized conifold, the topological string partition function 
has been computed in Ref.~\refcite{IK} using the topological vertex \cite{AKMV}.
The counting of BPS states has been carried out in all chambers in 
Ref.~\refcite{Nagao1}. Thus, we will use this case to test our formula \eqref{ZChamber}. 
For $\mathbb{C}^3/\mathbb{Z}_2\times \mathbb{Z}_2$, the counting
in the non-commutative DT chamber has been done in 
Ref.~\refcite{Young2}. 

Homology 2-cycles of the generalized conifold
correspond to the simple roots  $\alpha_1, \cdots, \alpha_{L-1}$
of the $A_{L-1}$ algebra. To identify them in the toric diagram, 
we divide the trapezoid into $L$ triangles of area 1 and label
the internal lines dividing the triangles as $i = 1, \cdots, L-1$. 
Each line $i$ corresponds to the blowing up $\bP^1$ at $\alpha_i$. 
We will denote the D2 charge by
\beq
   \beta = \sum_i n_i \alpha_i.
\eeq
In general, there are several ways to 
divide the trapezoid, and they correspond to different crepant resolutions 
of the singularity. If the two triangles across the line $i$ form 
a rhombus, we have a resolution by $\mathcal{O}(-1,-1)$. On the other hand,
if the two triangles form a triangle of area 2, the resolution is by 
$\mathcal{O}(-2,0)$. Both the topological string partition function 
and the BPS counting depend on the choice of the
resolution. 

The topological string partition function for this geometry is given
by 
\beq
 Z_{\rm top} (q, Q)
= M(q)^{L/2} \prod_{n=1}^\infty \prod_{i\leq j} 
(1-q^n Q_i Q_{i+1} \cdots Q_j)^{n N_{ij}}, \label{generaltop}
\eeq
where
\beq
  N_{ij} = (-1)^{1 + \# \{ k \in I| i \leq k \leq j\}}, \label{positiveroot}
\eeq
and
$I$ is the set of internal lines of the toric diagram 
corresponding to the resolution by
$\mathcal{O}(-1,-1)$. Thus, the only non-vanishing
Gopakumar-Vafa invariants are 
\beq
 N_{\beta=\pm (\alpha_i\ldots \alpha_j)}^{m=0} = (-1)^{1 +  \# \{ k \in I| i \leq k \leq j\}}, \label{root}
\eeq
and
\beq
N_{\beta=0}^{m=0} = -L.
\eeq
All BPS states in 5 dimensions carry 
no intrinsic spin. 

The central charge $Z(\beta, n)$ is given by
\beq
   Z(\beta, n) = R^{-1}\left(n + \sum_i n_i B(\alpha=i)\right).
\eeq
The formula \eqref{ZChamber} predicts that BPS
states in the chamber characterized by $B_i$'s are
counted as
\beq
  Z_{\rm BPS}(q, Q) = 
  M(q)^L\prod_{(\beta,n): Z(\beta, n)>0} (1-q^n Q^\beta)^{nN_\beta^0}.
\eeq
Here the product is over all roots $\beta$ of $A_{N-1}$ and $n=1,2,...$
such that $Z(\beta, n) > 0$. This agrees with the result in Ref.~\refcite{Nagao1}.

\subsubsection{A Non-toric Example}\label{subsec.nontoric}
\begin{figure}[htbp]
\centering{\includegraphics[scale=0.45]{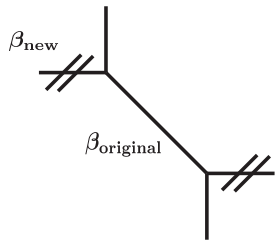}}
\caption[An example of a non-toric CY.]{The non-toric Calabi-Yau which arises by identifying two external legs of the $(p,q)$-web of the resolved conifold.}
\label{nontoric}
\end{figure}

Our discussion in Section\,\ref{sec.idea} and \ref{sec.derivation} are not limited to toric Calabi-Yau's, and applies to any Calabi-Yau without compact 4-cycles. In order to illustrate this point in a concrete setting, let us describe the geometry shown in Fig.\,\ref{nontoric}.
This geometry arises by identifying two of the four external legs of the $(p,q)$-web of the resolved conifold. This is one of the simplest the non-toric (albeit formally toric) geometries studied in Ref.~\refcite{HollowoodIV}, and 
it is straightforward to repeat the following analysis to other non-toric geometries discussed in Ref.~\refcite{HollowoodIV}.

In addition to the $\bP^1$ of the resolved conifold, the geometry of Fig.\,\ref{nontoric} has another compact $\bP^1$ which arises from identification. Let us denote their homology classes by $\beta_{\rm original}$ and $\beta_{\rm new}$, respectively. As a basis of the homology class, we choose $\beta_1=\beta_{\rm original}$ and $\beta_2=\beta_{\rm original}+\beta_{\rm new}$.

The topological string partition function is given by \cite{HollowoodIV}
\beq
Z_{\rm top}(q,Q_1,Q_2)=M(q) \left( \prod_{n=1}^{\infty} (1-Q_1 q^n)^n \right) 
\prod_{k,n=1}^{\infty} \left( \frac{(1-q^n Q_1 Q_2^k ) (1-q^n Q_1^{-1} Q_2^k)}{(1-q^{n-1} Q_2^k) (1-q^{n+1} Q_2^k)}  \right)^n,
\eeq
where $Q_1$ and $Q_2$ are the variables corresponding to $\beta_1$ and $\beta_2$.
The Gopakumar-Vafa invariants are therefore given by
\beq
N^0_{\beta=0}=-2,\quad  N_{\pm \beta_1}^0=1, \quad N^0_{\pm \beta_1+k\beta_2}=1, \quad
N^{\pm 1}_{k\beta_2}=-1 \quad (k\in \bZ \backslash \{ 0 \}). 
\eeq
Notice that genus 1 Gopakumar-Vafa invariants are nonvanishing in this non-toric example. This arises because our geometry has an extra cycle compared with the resolved conifold.

Again, the general formula gives (notice that $m\ne 0$ in this case)
\begin{align}
 Z_{\rm BPS} (q, Q) &=Z_{\rm top}(q, Q)Z_{\rm top}(q, Q^{-1})|_{\rm chamber} \nonumber \\
& =  \prod_{(\beta, l,m): Z(\beta, n=m+l)> 0} (1 - q^{l+m} Q^\beta)^{l N_\beta^m}.
\end{align}
The formula for the central charge is
\beq
Z\left(\beta=\sum_{i=1,2} n_i \beta_i,n\right)=R^{-1}\left(\sum_i n_i B(\beta_i) +n \right),
\eeq
and the equation 
\beq
Z(\beta,n=l+m)=0,
\eeq
with corresponding Gopakumar-Vafa invariants nonvanishing, determines the position of walls of marginal stability. This is a new result which has not been discussed in the literature to the best of our knowledge.

\subsection{Wall Crossing and Crystal Melting}\label{sec.WCcrystal}

We finally explain how the wall crossing phenomena described in this section is recast in the language of crystal melting described in Section\,\ref{chap.crystal}. Let us describe the example of the resolved conifold for simplicity; more general discussions, applicable to any chamber for generalized conifolds, is included in Section\,\ref{chap.open} as a special case of more general open BPS invariants.

\begin{figure}[htbp]
\centering{\includegraphics[scale=0.15]{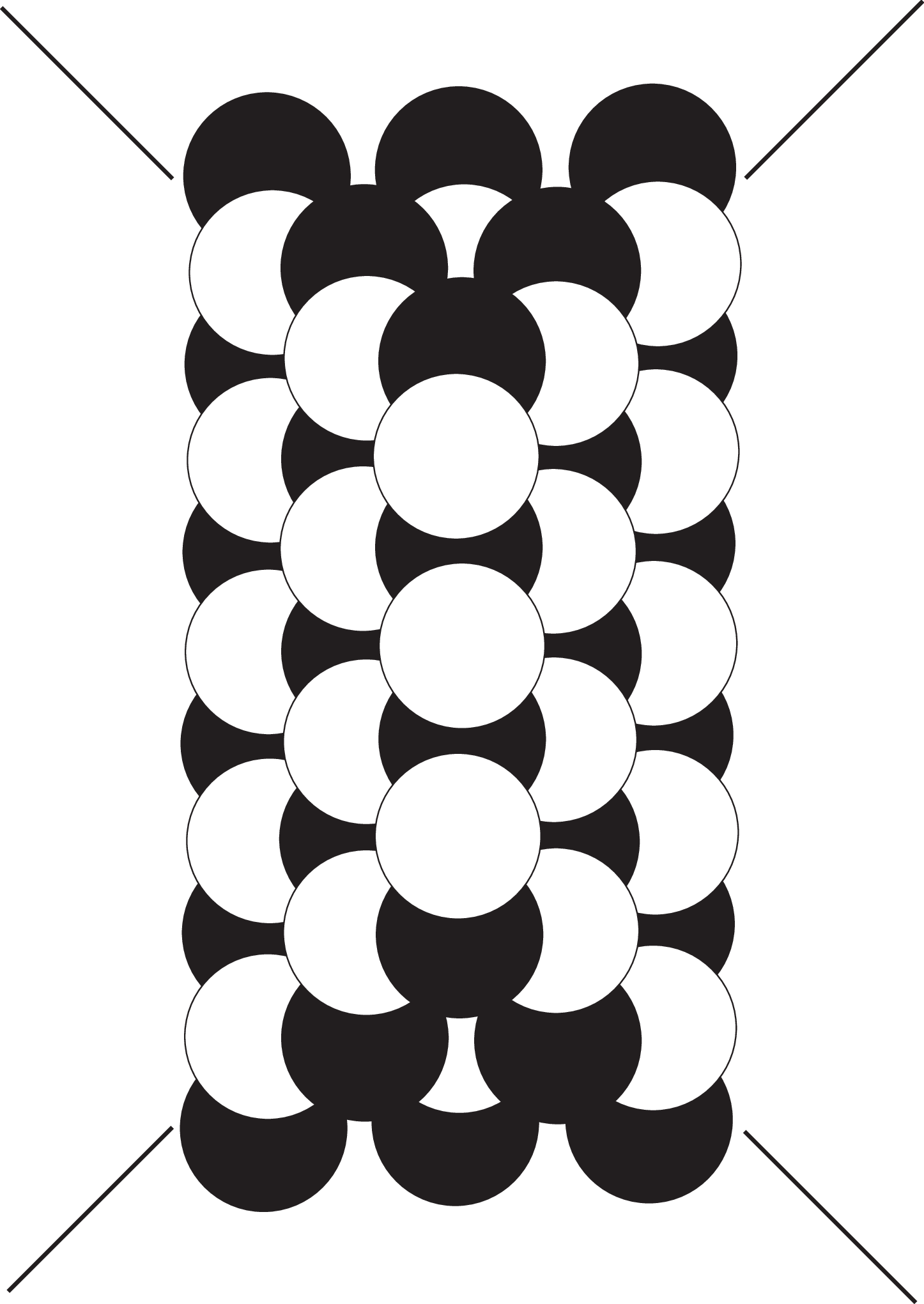}}
\caption[The conifold crystal in the chamber $C_2$.]{The conifold crystal in the chamber $C_2$. We have 3 atoms on the top. Compare with Fig.\,\ref{fig.conifoldatom}.}
\label{newgroundstate}
\end{figure}

The crystal melting model described in Section\,\ref{chap.crystal} is for the chamber $C_0$ (see Fig.\,\ref{fig.conifoldatom}). In the chamber $C_n$, we have to change the ground state of the crystal. The new ground state (see Fig.\,\ref{newgroundstate}) has $n+1$ atoms on top. 

Using the same melting rule, we have the function $Z_{\textrm{crystal},C_n}$. In Ref.~\refcite{Young1} it was shown combinatorially (see Section\,\ref{sec.WCvertex} for derivation from vertex operators) that the partition function takes a simple form 
\beq
\begin{split}
&Z_{\textrm{crystal},C_n}(q_0^{(n)}, q_1^{(n)}) =  M(q_0^{(n)}q_1^{(n)})^2  \\
&\quad \times
 \prod_{k\ge 1}
\left(1 + (q_0^{(n)})^k
(q_1^{(n)})^{k-1} \right)^{k+n}
\prod_{k\ge 1}
\left(1 + (q_0^{(n)})^k
(q_1^{(n)})^{k+1}
\right)^{\textrm{max}(k-n,0)},
\end{split}
\label{ZcrystalCn}
\eeq
where $q_{0,1}^{(n)}$ are weights in the $n$-th chamber $C_n$.
If we take 
\beq
q_0^{(n)}=q_0^{n+1} q_1^n , \quad q_1^{(n)}=q_0^{-n} q_1^{-n+1}, 
\label{wtchg}
\eeq
and substitute back into \eqref{ZcrystalCn}, we have 
\beq
M(q_0 q_1)^2 \prod_{k>n} (1+q_0^k q_1^{k-1})^k \prod_{k>0} (1+q_0^k q_1^{k+1})^k,
\eeq
which coincides with the result of \eqref{ZCnresult}
 under the parameter identification \eqref{paramchg}.
This means that the wall crossing is represented as (1) a change of the ground state and (2) change of the weight in the crystal melting model.
This is consistent with the discussion at the end of Section\,\ref{sec.NCDT}, 
where different choice of the fractional brane leads to Seiberg duality of the quiver. The fact we have 3 atoms on the top of the crystal corresponds to 3 arrows starting from the D6-brane node in Fig.\,\ref{fig.mutatedquiver}.

When we send $n\to\infty$, the ridges of the crystal approaches to the $(p,q)$ web of the resolved conifold with larger and larger $\bP^1$ inside, and 
we begin to see two different corners of the crystal, each corresponding to the one for $\bC^3$. This is how the combinatorial expression \eqref{ZcrystalCn} has been derived, and clarifies the connection between the crystal melting model of Section\,\ref{chap.crystal} and those in the topological vertex model. In other words, the limit $n\to\infty$ of our crystal gives the crystal melting model appearing in topological vertex constructions.

We can also construct crystal melting models for the chambers $\tilde{C}_n$,
where the partition function is a finite product. See Refs.~\refcite{NN,CJ} for discussion.

\subsection{Wall Crossing and Free Fermions}\label{sec.WCvertex}

There is yet another representation of the wall crossing phenomena, from the viewpoint of free fermions \footnote{See Refs.~\refcite{ADKMV,NekrasovO,DHSV,DHS} for connections between topological string theory and free fermions.}. This is well-known for the $\bC^3$ crystal; the topological vertex can be computed in the transfer matrix approach \cite{ORV}. In this section we are going to see that this can be generalized to our more general toric geometries. We moreover see that this gives another perspective on the wall crossing phenomena . This section is based on the papers \cite{Young2,Nagao3,Sulkowski}.
For simplicity we specialize the conifold example in this section. More complete discussion, including the open BPS invariants, will be given in Section\,\ref{sec.closedexpression}. See also Ref.~\refcite{YamazakiKokyuroku} for combinatorial description.

We begin with two-dimensional free fermions \footnote{See Refs.~\refcite{JimboM,Macdonald} for the formalism used here.}, which have the mode expansion
\beq
\psi(z)=\sum_k \psi_{k+1/2} z^{-k-1}, \quad 
\psi^*(z)=\sum_k \psi^*_{k+1/2} z^{-k-1},
\eeq
satisfying the commutation relations 
\beq
\{\psi_{k+1/2},\psi_{-l-1/2} \}=\delta_{k,l}.
\eeq
Starting from the vacuum $| 0 \rangle $ annihilated by all positive modes
\beq
\psi_{k+1/2}|0 \rangle =\psi^*_{k+1/2}|0\rangle =0, \quad k\ge 0,
\eeq
we can construct a state
\beq
|\mu\rangle =\prod_{i} \psi^*_{-a_i-1/2}\psi_{-b_i-1/2} |0\rangle ,
\quad a_i=\mu_i-i, \quad b_i=\mu_i^t-i,
\eeq
corresponding to a two-dimensional partition $\mu=(\mu_1,\mu_2 \ldots)$ \footnote{See \ref{app.core} for basic definitions about partitions.}.
The mode expansion of the bosonized field $\partial \phi(z)=: \psi(z)\psi^*(z):$
\beq
\partial\phi(z)=\sum_n \alpha_n z^{-n-1} 
\eeq
satisfy the Heisenberg commutation relation
\beq
[\alpha_n,\alpha_m]=n\delta_{n+m,0}.
\label{Heisenberg}
\eeq
Define vertex operators \footnote{We follow the notation of Ref.~\refcite{Nagao3}, which is different from those of Ref.~\refcite{Young2,Sulkowski}. The notation here makes it easier to generalize our results here to those in Section\,\ref{subsec.openvertex}.
Compared with Ref.~\refcite{Young2} we have
\begin{eqnarray*}
\Gamma^{\rm here}_{-}(q)=\Gamma^{\rm there}_+(q),\quad
\Gamma^{\rm here}_{+}(q)=\Gamma^{\rm there}_-(q^{-1}), \quad
\Gamma^{' \textrm{here}}_{-}(q)=\Gamma^{' \textrm{there}}_+(q),\quad
\Gamma^{' \textrm{here}}_{+}(q)=\Gamma^{' \textrm{there}}_-(q^{-1}).
\end{eqnarray*}
}
\beq
\Gamma_{\pm}(q)=\exp \sum_n \frac{q^{\mp n}}{n} \alpha_{\mp n},
\eeq
and
\beq
\Gamma^{'}_{\pm}(q)=\exp \sum_n \frac{(-1)^{n-1} q^{\mp n}}{n} \alpha_{\mp n},
\eeq
They have the property \cite{Young2}
\beq
\Gamma_- (q) |\mu \rangle =\sum_{\lambda\prec \mu }q^{|\mu|-|\lambda|} |\lambda \rangle, \quad  \Gamma_+ (q) |\mu \rangle= \sum_{\lambda\succ \mu }q^{|\mu|-|\lambda|} |\lambda\rangle,
\eeq
and
\beq
\Gamma^{'}_- (q) |\mu \rangle =\sum_{\lambda^t \prec \mu^t }q^{|\mu|-|\lambda|} | \lambda\rangle , \quad 
\Gamma^{'}_+ (q) |\mu \rangle =\sum_{\lambda^t \succ \mu^t}q^{|\mu|-|\lambda|} | \lambda\rangle,
\eeq
where $\lambda\succ \mu$ denotes the interlacing condition
\beq
\lambda_1\ge \mu_1 \ge \lambda_2\ge \mu_2\ge \ldots . 
\eeq
From the commutation relation \eqref{Heisenberg}, we have
\beq
\begin{split}
& [\Gamma_-(p),\Gamma^{'}_+(q)]=1+p^{-1}q, \quad
[\Gamma^{'}_-(p),\Gamma_+(q)]=1+p^{-1}q, \\
& [\Gamma_-(p),\Gamma_+(q)]=\frac{1}{1-p^{-1}q}, \quad
[\Gamma^{'}_-(p),\Gamma^{'}_+(q)]=\frac{1}{1-p^{-1}q},
\end{split}
\label{GammaCR1}
\eeq
where the commutator means the {\it multiplicative} commutator $[x,y]=x y x^{-1} y^{-1}$ here.
This can be concisely summarized as 
\beq
[\Gamma^{\epsilon_1}_{\iota_1}(p_1), \Gamma^{\epsilon_2}_{\iota_2}(p_2)]
=(1-\epsilon_1 \epsilon_2 p_1^{\iota_1} p_2^{\iota_2})^{-\iota_1 \epsilon_1 \epsilon_2 \delta_{\iota_1+\iota_2}}
\label{GammaCR2}
\eeq
for $\epsilon_{1,2}=\pm 1, \iota_{1,2}=\pm 1$, if we write $\Gamma^+_{\pm}(p)\equiv \Gamma_{\pm}(p)$ and $\Gamma^-_{\pm}(p)\equiv \Gamma^{'}_{\pm}(p)$.

Now the claim is that the BPS partition function in non-commutative Donaldson-Thomas chamber can be expressed in the following form (as before, we define $q\equiv q_0 q_1$):
\beq
\begin{split}
Z_{\rm BPS}(q_0,q_1)=\langle \emptyset | 
\ldots
\Gamma_-(q^{-2})& \Gamma^{'}_-(q^{-2} q_1) 
\Gamma_-(q^{-1}) \Gamma^{'}_-(q^{-1} q_1) \\
\times 
& \Gamma_+(1) \Gamma^{'}_+(q_1) 
\Gamma_+(q) \Gamma^{'}_+(q q_1) 
\ldots
| \emptyset \rangle ,
\end{split}
\label{Zconifoldvertex}
\eeq
i.e. 
\beq
Z_{\rm BPS}(q_0,q_1)=\langle \emptyset | 
\ldots
A_{-}(q^{-2}) A_{-}(q^{-1}) 
A_{+}(1) A_{+}(q) 
\ldots
| \emptyset \rangle ,
\label{Zconifoldvertex2}
\eeq
when we introduce 
\beq
A_+(x)=\Gamma_+(x)\Gamma^{'}_+(x q_1), \quad A_-(y)=\Gamma_-(y) \Gamma^{'}_-(y q_1)
\eeq
This can also be expressed in a suggestive form 
\beq
Z_{\rm BPS}=\langle \Omega_- | \Omega_+\rangle 
\eeq
when we introduce 
\beq
\langle \Omega_- | = \langle \emptyset |
\ldots
 A_{-}(q^{-3}) A_{-}(q^{-2}) A_{-}(q^{-1}), \quad 
| \Omega_+\rangle  =A_{+}(1) A_{+}(q) A_{+}(q^2)\ldots | \emptyset \rangle  
.
\eeq

It is easy to check \eqref{Zconifoldvertex2} directly by explicit commutation. 
Indeed, from \eqref{GammaCR1} we have 
\beq
[A_-(x), A_+(y)]=\frac{1}{(1-x^{-1}y)^2} (1+q_1 x^{-1} y) (1+ q_1^{-1} x^{-1} y)\eeq
and therefore
\begin{align*}
\left[\prod_{m=1}^{\infty} A_-(q^{-m}), \prod_{n=0}^{\infty} A_+(q^n)\right]
&=
\prod_{m=1}^{\infty}  \prod_{n=0}^{\infty} \frac{1}{(1-q^{m+n})^2} (1+q_1 q^{m+n}) (1+ q_1^{-1} q^{m+n}) \\
&=\prod_{n=1}^{\infty} (1-q^n)^{-2n} (1+q^n Q)^{n}  (1+q^n Q^{-1})^{n}.
\end{align*}
This is precisely the BPS partition function of the conifold.

\begin{figure}[htbp]
\centering{\includegraphics[scale=0.15]{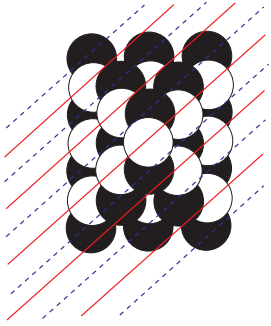}}
\caption[Slice of the conifold crystal by parallel planes.]{Slice of the conifold crystal by parallel planes. For a molten crystal, each of the slice gives a (complement of a) Young diagram.}
\label{conifoldslice1}
\end{figure}

Of course, the next natural question is where the expression \eqref{Zconifoldvertex} comes from. The answer is that it arises from a sequence of Young diagrams obtained as the slice the crystal.
%
As shown in Fig.\,\ref{conifoldslice1} we can slice the conifold crystal by an infinite number of parallel planes, and we have a set of Young diagrams $\{ \lambda(n+1/2) \}_{n\in \bZ}$, where we choose the label the planes (and therefore Young diagrams) by half-integers. By looking at the crystal structure carefully the melting rule \eqref{meltingruleold} is translated into the condition that Young diagrams satisfy the interlacing conditions
\beq
\ldots \lambda(-5/2) \ml \lambda(-3/2) \pl \lambda(-1/2) \ml \lambda(1/2) \pg \lambda(3/2) \mg 
\lambda(5/2) \pg \lambda(7/2)\ldots, 
\label{interlacing}
\eeq
where we used the notation
$
\lambda \pl \mu ~~(\lambda \ml \mu)
$
for 
$
\lambda \prec \mu ~~(\lambda^t \prec \mu^t)
$
, respectively. 
The vertex operators $\Gamma_{\pm}$ and $\Gamma'_{\pm}$ in \eqref{Zconifoldvertex} implement (in the language of free fermions) the interlacing conditions \eqref{interlacing}. 

It is also easy to generalize this analysis to other chambers. For example, in the chamber $C_2$ where the crystal takes the form shown in Fig.\,\ref{newgroundstate}, the analysis of the melting rule for the crystal suggests that we consider
\beq
\begin{split}
Z_{\rm BPS}(q_0,q_1)=\langle \emptyset | 
\ldots
&\Gamma_-(q^{-1})\Gamma^{'}_-(q^{-3} q_1) 
\Gamma_+(1) \Gamma^{'}_-(q^{-2} q_1) \\ 
\times 
&\Gamma_+(q) \Gamma^{'}_-(q^{-1} q_1) 
\Gamma_+(q^2) \Gamma^{'}_+( q_1) 
\Gamma_+(q^3) \Gamma^{'}_+(q q_1)
\ldots
| \emptyset \rangle ,
\end{split}
\label{Zconifoldvertex3}
\eeq
We can again easily verify explicitly that this coincides with \eqref{ZCnresult}. Note that the vertex operators appearing in \eqref{Zconifoldvertex3} are the same as in \eqref{Zconifoldvertex}, and we are just changing their orders!
This means that wall crossings are described by commuting vertex operators!
We will see in Section\,\ref{subsec.openvertex} that they hold more generally.

\section{Wall Crossing Formulas}\label{chap.WCF}

In the previous section we gave a derivation of the wall crossing phenomena by using the M-theory viewpoint and reformulating the BPS state counting problem as a counting problem of free spinning M2-branes. We have seen that when we cross a wall we obtain/lose one factor
$$
(1 -q^n Q^\beta)^{n N_{\beta}^n}.
$$

Historically, the M-theory argument was not the first method to derive these results. There is an independent derivation of the results, using the wall crossing formulas. 
The aim of this section is to explain the wall crossing formula, and to apply it to the examples we studied in the previous section.


\subsection{Overview: What Wall Crossing Formula Tells Us}

Let us first explain the meaning of the wall crossing formula before discussing the explicit form of it. 
Suppose that we already know the positions of walls of marginal stability (for example, by the central charges \eqref{wallposition}). Consider two chambers separated by a wall. Due to the wall crossing phenomena, the BPS index $\Omega(\gamma)$ in general jumps along the wall. The wall crossing formula relates the BPS indices on both sides of the wall. In other words, given an index on one side of the wall, wall crossing formula tells us the index on the other side of the wall.

This means that given the chamber structures and the BPS index in one chamber of the moduli space, wall crossing formula reproduces the BPS index in all the other chambers of the moduli space. Of course, we still need to know the BPS index in at least one chamber. In the example of the resolved conifold discussed in Section\,\ref{sec.conifold}, as well as in the examples of generalized conifolds in Section\,\ref{sec.gc}, there is a special chamber (core region) where the BPS partition function becomes trivial, and starting from this chamber we can reproduce the partition function in all other chambers.

In the literature there are several different versions of the wall crossing formula. The first one is the primitive/semi-primitive wall crossing formula, which
 is proposed by Denef and Moore in the context of supergravity and is discussed in detail in the next section. The second is the Kontsevich-Soibelman wall crossing formula \cite{KontsevichS}. There is also yet another formula by Joyce and Song \cite{JS,Joyce}. See also the recent paper \cite{SenWCF}. In this section we only discuss the semi-primitive wall crossing formula, and comment on the more general Kontsevich-Soibelman formula in \ref{app.KS}.

\subsection{Semi-Primitive Wall Crossing Formula}\label{sec.semiprimitive}

Let us begin with the primitive wall crossing formula. Suppose that
at the walls of marginal stability the decay pattern is
$\gamma\to \gamma_1+\gamma_2$, where $\gamma_1$ and $\gamma_2$ are primitive charges. Here primitive means that particle with that charge is stable at any value of the K\"ahler moduli. In this case, the primitive wall crossing formula says that the jump of $\Omega(\gamma)$ is given by
\beq
\Delta \Omega(\gamma)=(-1)^{\langle \gamma_1, \gamma_2 \rangle} \Big|\langle \gamma_1, \gamma_2 \rangle \Big|\Omega(\gamma_1) \Omega(\gamma_2),
\label{primitive}
\eeq
where $\langle \, , \, \rangle$ refers to the symplectic paring of the charge lattice $\Gamma$.
We do not need to specify the value of K\"ahler moduli for $\Omega(\gamma_1)$ and $\Omega(\gamma_2)$ since we assumed that $\gamma_1, \gamma_2$ are primitive.

Semi-primitive wall crossing formula deals with the decay pattern $\gamma\to \gamma_1+N_2 \gamma_2$. The formula says that 
\beq
\Omega(\gamma_1)+\sum_{N_2>0} \Delta \Omega(\gamma_1+N_2 \gamma_2) p^{N_2} =\Omega(\gamma_1) \prod_{k>0} \left(1-(-1)^{k \langle \gamma_1, \gamma_2 \rangle } p^k \right) ^{k \left|\langle \gamma_1, \gamma_2 \rangle\right| \Omega(k\gamma_2)},
\label{semiprimitive}
\eeq
where $p$ is a formal variable. Note that the term constant in $p$ gives a trivial relation, and the linear term gives the primitive wall crossing formula.

\bigskip

In general, the decay pattern is $\gamma\to N_1\gamma_1+N_2\gamma_2$  (see the comment below \eqref{wallposition} in Section\,\ref{sec.WC}), and we need a more general wall crossing formula, the one proposed by Kontsevich and Soibelman. However, the story simplifies considerably for a noncompact Calabi-Yau manifold with a single D6-brane. Since we are counting D6/D2/D0 bound states, $\gamma$ has one D6-brane charge. The fact that the Calabi-Yau manifold is non-compact with a single D6-brane means that it is impossible for a D6-brane to decay into an anti D6-brane and two D6-branes, for example; that will cost infinite energy. This means that either $\gamma_1$ or $\gamma_2$ has D6-brane charge one, while the other has charge zero.
Without losing generality we can assume that $\gamma_1$ contains a D6-brane, while $\gamma_2$ contains none. This means that the decay pattern is always $\gamma\to \gamma_1+N_2 \gamma_2$. Note that this is not the case if we have multiple D6-branes.

Two further simplifications occur in our setup. First, the wall of marginal stability is determined by $\gamma_2$ alone. As we have seen in \eqref{wallposition}, the wall of marginal stability is determined by
\beq
\textrm{Arg}\, Z(\gamma_1)=\textrm{Arg}\, Z(\gamma_2).
\eeq
However, since $\gamma_1$ contains D6-brane having an infinite mass, we have
\beq
\textrm{Arg}\,Z(\gamma_1)\sim \textrm{Arg}\,Z({\rm D6}),
\eeq
and the $\gamma_1$ dependence drops out from the story.

Second, the values of the Gopakumar-Vafa invariants given in Section\,\ref{sec.conifold} and \ref{sec.gc} (\eqref{conifoldGV} and \eqref{root}) says \footnote{In this section we are concentrating on toric examples, and do not discuss the non-toric example of Section\,\ref{subsec.nontoric}.} that Gopakumar-Vafa invariants vanish when we have more than one (anti) D2-branes. This means that when $\gamma_2$ contains (anti) D2, semi-primitive wall crossing formulas simplifies to 
\begin{equation}
\begin{split}
\Omega(\gamma_1)+&\sum_{N_2>0} \Delta \Omega(\gamma_1+N_2 \gamma_2) p^{N_2} \\ 
&=\Omega(\gamma_1)\prod_{k>0} \left(1-(-1)^{k\langle \gamma_1, \gamma_2 \rangle} p^k\right) ^{k|\langle \gamma_1,\gamma_2 \rangle | \Omega(k\gamma_2)} \\ 
&=\Omega(\gamma_1) \left(1-(-1)^{\langle \gamma_1, \gamma_2 \rangle} p\right) ^{|\langle \gamma_1,\gamma_2 \rangle | \Omega(\gamma_2)}.
\label{simpleWCF}
\end{split}
\end{equation}

\bigskip

Let us now use apply these formulas to the resolved conifold, whose chamber structure is summarized in Fig.\,\ref{conifoldchamber}.
We start with the simplest example, the wall between $\tilde{C}_0$ and $\tilde{C}_1$. At $\tilde{C}_0$ (core region), the degeneracies are simply
\beq
\Omega({\rm D6})=1, \quad \textrm{ all others zero},
\eeq
and the decay pattern at the wall of marginal stability is $\gamma\to \gamma_1+N_2 \gamma_2$ with 
\beq
\quad \gamma_1={\rm D6}, \quad \gamma_2={\rm D2+D0}, \quad \langle \gamma_1, \gamma_2 \rangle=1.
\eeq
This means, when combined with the expressions for the Gopakumar-Vafa invariants \eqref{conifoldGV} and the wall crossing formula \eqref{simpleWCF},
\beq
\Omega(\gamma_1+\gamma_2;\tilde{C}_1)=1, \quad
\Omega(\gamma_1+N_2 \gamma_2;\tilde{C}_1)=0 \, \textrm{ for all } \, N_2>1.
\eeq
and we have
\beq
Z_{\rm BPS}(q,Q;\tilde{C}_1)=1+qQ=1-(-q) Q.
\eeq

We can repeat the same argument for the next chamber $\tilde{C}_2$.
At $\tilde{C}_1$ , the degeneracies are
\beq
\Omega({\rm D6})=1, \quad \Omega({\rm D6+D2+D0})=1, \quad \textrm{ all others zero},
\eeq
and the decay pattern at the walls of marginal stability is $\gamma\to \gamma_1+N_2 \gamma_2$ with 
\beq
\quad \gamma_1=\textrm{ D6 or D6+(D2+D0)}, \quad \gamma_2={\rm D2+2D0}, \quad \langle \gamma_1, \gamma_2 \rangle=2.
\eeq
Therefore, \eqref{simpleWCF} and Gopakumar-Vafa invariants \eqref{conifoldGV} tell us that
\beq
\begin{split}
&\Omega(\gamma_1+\gamma_2;\tilde{C}_2)=-2, \quad
\Omega(\gamma_1+2\gamma_2;\tilde{C}_2)=1, \\
&\Omega(\gamma_1+N_2 \gamma_2;\tilde{C}_1)=0 \, \textrm{ for all } \, N_2>2,
\end{split}
\eeq
and we have
\beq
\begin{split}
&\Omega({\rm D6})=\Omega({\rm D6+D2+D0})=\Omega({\rm D6+2D2+4D0})=\Omega({\rm D6+3D2+5D0})=1 \\
&\Omega({\rm D6+D2+2D0})=\Omega({\rm D6+2D2+3D0})=-2.
\end{split}
\eeq
The partition function is then given by
\beq
Z_{\rm BPS}(q,Q;\tilde{C}_2)=\prod_{i=1,2} \left(1-(-q)^i Q \right)^i.
\eeq

In general, on the wall $\gamma\to \gamma_1+N_2 \gamma_2$ with $\gamma_2={\rm D2}+n \, {\rm D0}$, we have from \eqref{simpleWCF}
\beq
\sum_{\gamma_1} \Omega(\gamma_1) \bm{q}^{\gamma_1} 
+\Delta\sum_{N_2>0, \gamma_1} \Omega(\gamma_1+N_2\gamma_2) \bm{q}^{\gamma_1} p^{N_2}
=\sum_{\gamma_1} \bm{q}^{\gamma_1} \Omega(\gamma_1) \left(1-(-1)^n p\right)^n,
\eeq
where $\bm{q}=(q,Q)$ collectively refers to the D0-brane chemical potential $q$ and the D2-brane chemical potential $Q$.
If we set $p=\bm{q}^{\gamma_2}=q^n Q$ we have 
\beq
Z_{\rm BPS}(q,Q;\tilde{C}_{n})=Z_{\rm BPS}(q,Q;\tilde{C}_{n-1})\left(1-(-1)^n q^n Q\right)^n.
\eeq
This means in general
\beq
Z_{\rm BPS}(q,Q;\tilde{C}_n)=\prod_{j=1}^n \left(1-(-1)^j q^j Q\right)^j.
\eeq
and in particular 
\beq
Z_{\rm BPS}(q,Q;\tilde{C}_{\infty})=\prod_{j=1}^{\infty} \left(1-(-1)^j q^j Q\right)^j.
\eeq
The chamber between $\tilde{C}_{\infty}$ and $C_{\infty}$ is special, since this is the wall where the D6-brane makes a bound state with pure D0-branes without D2-branes, and therefore we should use the original formula \eqref{semiprimitive} instead of \eqref{simpleWCF}.
The wall crossing formula \eqref{semiprimitive} gives
\beq
Z_{\rm BPS}(q,Q;C_{\infty})=M(-q)^2\prod_{j=1}^{\infty} \left(1-(-1)^j q^j Q\right)^j.
\eeq
Repeating similar analysis, we have
\beq
Z_{\rm BPS}(q,Q;C_n)=M(-q)^2 \prod_{j=1}^{\infty} \left(1-(-1)^j q^j Q\right)^j \prod_{j>n}^{\infty} \left(1-(-1)^j q^j Q^{-1}\right)^j.
\eeq
and in particular
\beq
Z_{\rm BPS}(q,Q;C_0)=M(-q)^2 \prod_{j=1}^{\infty} \left(1-(-1)^j q^j Q\right)^j \prod_{j=1}^{\infty} \left(1-(-1)^j q^j Q^{-1}\right)^j.
\eeq
These results coincide with those of Section\,\ref{chap.M-theory} (see Section\,\ref{sec.conifold}). The same argument applies to the generalized conifolds discussed in Section\,\ref{sec.gc}.

\subsection{Derivation of the Wall Crossing Formulas} \label{sec.WCFderivation}

In the previous section we have seen that the semi-primitive wall crossing formula correctly reproduces the results obtained in Section\,\ref{chap.M-theory}.
However, we still do not see the physics behind the formula. In this section we give a derivation of the semi-primitive wall crossing formula following Ref.~\refcite{DenefM}. This will uncover tantalizing connections with the multi-centered black hole solutions described in \ref{app.multi}.

\begin{figure}[htbp]
\centering{\includegraphics[scale=0.2]{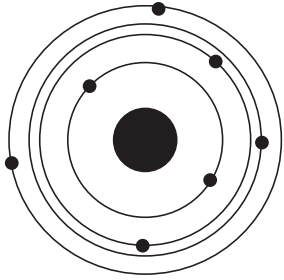}}
\caption[The core and the halo of a multi-centered black hole.]{The D6-brane (the core) sits in the center of the black hole, and is surrounded by D2/D0-brane charges (halos).}
\label{corehalo}
\end{figure}

In supergravity, a particle with primitive charge is described by a single-centered black hole. The BPS state with charge given by the sum of primitive charges, $\gamma=\gamma_1+N_2\gamma_2$, correspond to a multi-centered black hole. Since $\gamma_1$ contains the D6-brane and has an infinite mass, $\gamma_1$ is called the ``core'' of the black hole. The D0/D2 particles, which are sometimes called ``halos'', surround the core of the black hole (see Fig.\,\ref{corehalo}).

Let us begin with the primitive wall crossing formula.
Suppose that on one side of the wall we have a stable bound state with charge $\gamma=\gamma_1+\gamma_2$, where $\gamma_1, \gamma_2$ are primitive.
Suppose moreover that this particle become unstable on the other side of the wall. In supergravity, the particle corresponds to a 2-centered black hole, whose distance between the two centers given in \eqref{r12}. Since \eqref{r12} contains the central charge, the distance depends on the value of the K\"ahler moduli.
The distance between the two centers become infinite on the wall of marginal stability; this means that there is no longer a stable bound state, and the index jumps.

Since the two centers are far away from each other on the wall, the Hilbert space splits into direct products:
\beq
\Delta\scH(\gamma\to \gamma_1+\gamma_2)=[J_{12}]\otimes \scH(\gamma_1) \otimes \scH(\gamma_2), \label{Hprimitive}
\eeq
where we have included the rotational degrees of freedom coming from the angular momentum of the 2-centered black hole \eqref{J12}.
This means
\beq
\Delta\Omega(\gamma)=(-1)^{2J_{12}+1} (2J_{12}+1)\Omega(\gamma_1)\Omega(\gamma_2)
\eeq
which is the primitive wall crossing formula, when we use the expression for $J_{12}$ in \eqref{J12}.

We can also derive the semi-primitive wall crossing formula \eqref{semiprimitive} from supergravity viewpoint. The formula \eqref{semiprimitive} is written in the form of a generating function, and contains several terms when expanded with respect to $p$. For example, the quadratic term for $p$ contains two different contributions, one with two factors of $\Omega(\gamma_2)$, and one with a single $\Omega(2\gamma_2)$. What this means is simple: we can distribute the charge $2 \gamma_2$ into two particles each with charge $\gamma_2$, or to a single particle with $2\gamma_2$. Since the distance between the centers in \eqref{r12} depends on the charges, the corresponding halos have different radius.

Consider the more general situation $\gamma\to \gamma_1+N_2 \gamma_2$. We can again distribute the charge $N_2 \gamma_2$ to halos with different radius, each having $k_i \gamma_2$. Since each halo has mutually zero symplectic pairing, 
we can ignore the interactions and use the 2-centered black hole solution for each halo \footnote{The two assumptions in Section\,\ref{sec.idea} are the underlying assumptions behind this derivation.}.
This situation is concisely expressed if we introduce the generating function:
\beq\displaystyle
\bigoplus_{N_2}p^{N_2} \Delta\scH \Big|_{\gamma\to \gamma_1+N_2\gamma_2}
=\scH(\gamma_1)\bigotimes_k \scF \left(p^k [J_{\gamma_1,k\gamma_2}]\bigotimes \scH(\gamma_2) \right),
\eeq
where $\bigotimes\scF \left(p^k [J_{\gamma_1,k\gamma_2}]\bigotimes \scH(\gamma_2) \right)$ represents a Fock space spanned by halos with charge $k \gamma_2$, and $\bigoplus, \bigotimes$ represents a formal sum and a product, respectively.
This is a generalization of \eqref{Hprimitive}.

Now the each halo $[J_{\gamma_1,k\gamma_2}]\bigotimes \scH(\gamma_2)$ contributes a factor 
$$
\left(1-(-1)^{k\langle \gamma_1,\gamma_2 \rangle} p^k \right)
^{k \left|\langle \gamma_1, \gamma_2 \rangle\right| \Omega(k\gamma_2)}.
$$
The sign comes from the angular momentum $k (2J_{12}+1)=k \left|\langle \gamma_1, \gamma_2 \rangle\right|$, and the power is determined by the degeneracy $k (2J_{12}+1) \Omega(\gamma_2)$. This immediately gives the formula
\beq
\Omega(\gamma_1)+\sum_{N_2>0} \Delta \Omega(\gamma_1+N_2 \gamma_2) p^{N_2} =\Omega(\gamma_1) \prod_{k>0} \left(1-(-1)^{k \langle \gamma_1, \gamma_2 \rangle } p^k \right) ^{k \left|\langle \gamma_1, \gamma_2 \rangle\right| \Omega(k\gamma_2)},
\eeq
which is the semi-primitive wall crossing formula \eqref{semiprimitive}.

\bigskip
As this derivation shows, it is not a coincidence that the methods of Section\,\ref{chap.M-theory} and those of this section give the same answer. The derivation of the semi-primitive wall crossing formula uses the Fock space spanned by D2/D0 halos, whereas when lifted to M-theory we have a Fock space spanned by spinning M2-branes and the graviton. We should keep in mind, however, that the argument of Section\,\ref{chap.M-theory} does not require the validity of supergravity approximation.


\section{Open BPS Wall Crossing}\label{chap.open}

In this section, we generalize the analysis in previous sections to 
include ``open BPS invariants''. 
The BPS invariants in previous sections, which should be called ``closed BPS invariants'' more precisely, count D0/D2/D6-branes wrapping holomorphic 0/2/6-cycles.
In this section, we also consider D2-branes wrapping disks whose boundaries are on D4-branes wrapping non-compact Lagrangian 3-cycles. The open BPS invariants counts the degeneracy of such D-brane configurations. Since D4-branes break the supersymmetry to half and fill $\bR^{1,1}$ in $\bR^{3,1}$, from gauge theory viewpoint we are discussing the wall crossing in the two-dimensional $\scN=(2,2)$ theory.

The notion of open counting exists already in the topological string theory; of course, this is the open topological string theory. One of the most important results in the open topological string theory is that the partition function can be computed by gluing the topological vertex \cite{AKMV}, assigned to each trivalent vertex of the $(p,q)$-web.

Motivated by this result in the topological string theory,
we propose a generalization of the topological vertex, which we call  
the ``non-commutative topological vertex'' and which capture the open BPS invariants in the sense mentioned above. 
The vertex is defined combinatorially using the crystal melting model proposed 
recently, and depends on the value of closed string moduli at infinity.
The vertex in one special chamber gives the same answer as that computed by the ordinary topological vertex.

We prove an identify expressing the non-commutative topological vertex of a toric Calabi-Yau manifold $X$ as a specialization of the closed BPS partition function of an orbifold of $X$, thus giving a closed expression for our vertex.
We also clarify the action of the Weyl group of an affine $\hat{A}_{L-1}$
Lie algebra on chambers, and comment on the generalization of our results to the case of refined BPS invariants.

We also derive the wall crossing of open BPS invariants by the generalization of the M-theory argument in Section\,\ref{chap.M-theory}. This reproduces the results from the crystal melting model, and moreover predicts the wall crossing of open BPS invariants with respect to open as well as closed string moduli. We also comment on the consistency of this result with the work of Ref.~\refcite{CecottiVold}, which discuss the wall crossing in two-dimensional $\scN=(2,2)$ theories.

This section is based on the papers Refs.~\refcite{NY,AY}. See the paper Ref.~\refcite{Nagao3} for geometrical meaning of the open BPS invariants and the vertex operator representations. The vertex operator representation for closed BPS invariants is also discussed in Ref.~\refcite{Sulkowski}.

\subsection{Overview: Open BPS Invariants}\label{sec.intro}

We begin this section by reformulating the closed BPS invariants for generalized conifolds (Section\,\ref{sec.closed}). In this formulation,
the closed BPS partition function \footnote{The upper index $c$ stands for `closed'.}
$$
Z^c_{\textrm{BPS}, (\sigma',\theta')},
$$
which is defined in Section\,\ref{sec.WC} as the generating function of the degeneracy of D-brane BPS bound states,
depends on maps $\sigma',\theta'$ specifying a chamber in the K\"ahler moduli space \footnote{See \ref{app.Weyl} for details.}. The use of these maps is natural from the viewpoint of affine $\hat{A}_{L-1}$ algebra.
As we have seen in Section\,\ref{sec.derivation} (see the discussion around \eqref{DTChamber}), in one special chamber $\tilde{C}_{\rm top}$ of the K\"ahler moduli space, the BPS partition function is equivalent the topological string partition function \footnote{Actually, the topological string partition function depends on the choice of the resolution of the singular Calabi-Yau manifold $X$. This is related to the choice of the chamber, as will be explained in the main text.} (up to the change of variables, which we do not explicitly show here for simplicity):
\beq
Z^c_{\rm BPS} \Big|_{\tilde{C}_{\rm top}} =Z^c_{\rm top} \label{ZtoZtop}.
\eeq

It is natural to expect that similar story should exist for open BPS invariants as well. Namely, we expect to define
open version of the BPS partition function \footnote{The upper index $o$ stands for `open'.}
$$
Z^o_{\textrm{BPS},(\sigma,\theta)}
$$
depending on maps $\sigma, \theta$ specifying the chamber in the K\"ahler moduli space, such that the partition function reduces to the open topological string partition function in a special chamber $C_{\rm top}$:
\beq
Z^o_{\textrm{BPS}} \Big| _{C_{\rm top}}=Z^o_{\rm top}.
\eeq
The question is how to define open BPS degeneracies such that the generating function follows the conditions above.

As a guiding principle of our following argument, we use the crystal melting model developed in Section\,\ref{chap.crystal}. 
As we have seen, this crystal melting model generalizes the result of Ref.~\refcite{ORV} for $\bC^3$ to an arbitrary toric Calabi-Yau manifold. In the case of $\bC^3$, the crystal melting partition function with the boundary conditions specified by three Young diagrams $\lambda_1,\lambda_2,\lambda_3$ gives 
the topological vertex \cite{AKMV} $C_{\lambda_1,\lambda_2,\lambda_3}$. By using these vertices
as a basic building block, we can compute open topological string partition function with non-compact D-branes wrapping Lagrangian 3-cycles of the topology $\bR^2\times S^1$ included \cite{AV}.
In this story, generalization from closed to open topological string partition function corresponds to the change of the boundary condition for the crystal melting model for $\bC^3$.

Now the results presented in Section\,\ref{chap.crystal} show that the closed BPS partition function  discussed above can be written as a statistical mechanical partition function of the crystal model. This model applies to any toric Calabi-Yau manifold, and for $\bC^3$ the BPS partition function coincides with the topological string partition function. Similarly to the case of the topological string story mentioned in the previous paragraph, we hope to define the open version of the BPS invariants by changing the boundary condition for the crystal melting model. The invariants defined in this way will be defined in any chamber in the K\"ahler moduli space, and reduces to the ordinary topological vertex in a special chamber. We call such a generalization of the topological vertex ``the non-commutative topological vertex'' \footnote{The word `non-commutative' stems from the mathematical terminologies such as ``non-commutative crepant resolution'' \cite{VdB} and ``non-commutative Donaldson-Thomas invariant'' \cite{Szendroi}. The non-commutativity here refers to that of the path algebra of the quiver, and is different from the more familiar one (e.g. Refs.~\refcite{ConnesDS,SWNCG}) in string theory contexts. The quiver (together with a superpotential) determines a quiver quantum mechanics, which is the low-energy effective theory on the D0/D2-brane worldvolume (see Section\,\ref{chap.crystal}). See also Ref.~\refcite{BryanK} for ``closed topological vertex'', which correspond to $\bC^3/\bZ_2\times \bZ_2$. }, 
following ``the orbifold topological vertex'' named in Ref.~\refcite{BCY2}.

We will see that this expectation is indeed true. We adopt the
definition proposed in the mathematical literature \cite{Nagao2,Nagao3}, although we hope to have clarified the presentation by using the language of the crystal melting model. Our non-commutative topological vertex is defined for a Calabi-Yau manifold $X$ without compact 4-cycles, and a set of representations $\lambda$ assigned to external legs of the toric diagram. As in the case of topological vertex, $\lambda$ encodes the boundary condition of the D4-branes wrapping Lagrangian 3-cycles. We propose our vertex as the building block of open BPS invariants. 

We can provide several consistency checks of our proposal (see Section\,\ref{sec.check} for more details).
First, our vertex by definition reduces to the closed BPS invariant when all the representations $\lambda$ are trivial. Second, our vertex shows a wall crossing phenomena as we change the closed string K\"ahler moduli, and the vertex coincides with the topological vertex computation in the chamber where the closed BPS partition function reduces to the closed topological string partition function.
Third, the wall crossing factor is independent of the boundary conditions on D-branes, and is therefore the partition function factorizes into the closed string contribution and the open string contribution, as expected from Ref.~\refcite{OVknot} and the generalization of the arguments of Section\,\ref{chap.M-theory}.

Given a combinatorial definition of the new vertex, the next question is whether we can compute it, writing it in a closed expression. We show that the answer is affirmative, by proving the following statement.
For a Calabi-Yau manifold $X$, the non-commutative topological vertex $C_{\textrm{BPS},(\sigma,\theta;\lambda)}(X)$ is equivalent to the closed BPS partition function $Z^c_{\textrm{BPS},(\sigma',\theta')}(X')$ for an orbifold $X'$ of $X$
, under a suitable identification of variables explained in the following sections \footnote{More precisely, we need to specify the resolution of $X$ and $X'$. We also need to impose the condition that two of the representations $\lambda$ are trivial. See the discussions in the main text.}:
\beq
C_{\textrm{BPS},(\sigma,\theta;\lambda)}(X) =Z^c_{\textrm{BPS},(\sigma',\theta')}(X').
\label{eq.main}
\eeq
We will give an explicit algorithm to determine $X'$ and $\sigma', \theta'$, starting from the data on the open side.
Since an infinite-product expression for $Z^c_{\textrm{BPS},(\sigma',\theta')}(X')$ is already known \cite{Nagao1,AOVY}, this gives a closed infinite-product expression for our vertex.

The rest of this section is organized as follows. We begin in Section\,\ref{sec.closed} by slightly reformulating the results obtained in Section\,\ref{chap.M-theory} the closed BPS invariants, their wall crossings, and their relation with the topological string theory.
In Section\,\ref{sec.def} we define our new vertex using the crystal melting model. We also perform several consistency checks of our proposal. Section\,\ref{sec.closedexpression} contains our main result \eqref{eq.main}, which shows the 
equivalence of 
our new vertex with a closed BPS partition function under suitable parameter identifications. We also give an explicit algorithm for 
constructing the closed BPS partition function starting from our vertex.
In Section\,\ref{sec.openeg} we treat several examples in order to illustrate our general results. We also include Appendices \ref{app.Weyl}, \ref{app.core} and \ref{app.proof} for mathematical proofs and notations. 

\subsection{Closed BPS Invariants Revisited}\label{sec.closed}
Before discussing the open BPS invariants, we summarize in this section the definition and the properties of the closed BPS invariants. This section hopefully serves as a self-explanatory introduction to the closed BPS invariants, and prior reading of Section\,\ref{chap.M-theory} is not necessary for the first reading of this section.

Throughout this section, we concentrate on the case of 
the so-called generalized conifolds (see Section\,\ref{sec.gc}).
The reason for this is that wall crossing phenomena is understood 
well only in cases without compact 4-cycles (see Section\,\ref{chap.M-theory}), which means $X$ is either a generalized conifold or $\bC^3/(\bZ_2\times \bZ_2)$ \footnote{See Section\,\ref{sec.gc} for the proof of this statement.}. 

\begin{figure}[htpb]
\centering{\includegraphics[scale=0.2]{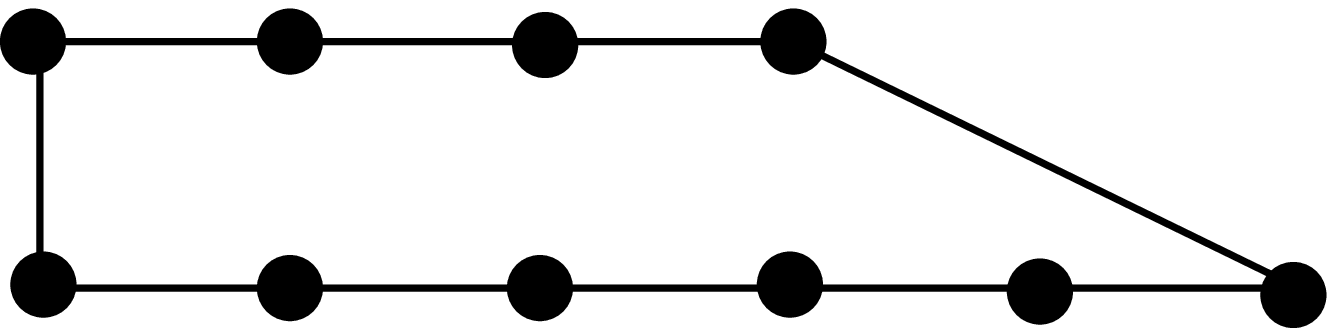}}
\caption[The toric diagram of a generalized conifold.]{The toric diagram of a generalized conifold, with $L_+=3, L_-=5$.}
\label{gctd}
\end{figure}

By a suitable $SL(2,\bZ)$ transformation, we can assume that the toric diagram of a generalized conifold is a trapezoid with height 1, with 
length $L_+$ edge at the top and $L_-$ at the bottom (see Fig.\,\ref{gctd}) \footnote{The Calabi-Yau manifold is determined by $L_+$ and $L_-$ as $xy=z^{L_+} w^{L_-}$.}. If we denote by $L=L_{+}+L_-$ the sum of the length of the edges on the top and the bottom of the trapezoid, this geometry has $L-1$ independent compact $\bP^1$'s. We label them by $\alpha_i$, borrowing the language of the root lattice of $\hat{A}_{L-1}$ algebra.

The language of the root lattice will be used extensively throughout this section \footnote{The root lattice of $\hat{A}_{L-1}$ is exploited in Refs.~\refcite{Nagao1,AOVY,Nagao2}. See also \ref{app.Weyl}.}.
We can also make more $\bP^1$'s by combining them. For example, combining all the $\bP^1$'s between $i$-th and $j$-th $\bP^1$ (assume $i<j$), we have another $\bP^1$ which we denote by 
\beq
\alpha_{i,j}:=\alpha_i+\ldots+ \alpha_j. \label{alphaij}
\eeq
This corresponds to a positive root of $\hat{A}_{L-1}$.

Suppose that we have a Calabi-Yau manifold $X$ without compact 4-cycles.
We also consider a single D6-brane filling the entire $X$ and D0/D2-branes
wrapping compact holomorphic 0/2-cycles specified by $n \in H_0(X;\bZ)$ and $\beta\in H_2(X;\bZ)$, respectively. We can then define the
BPS degeneracy $\Omega(n,\beta)$ counting BPS degeneracy of D-branes \footnote{More precisely, this BPS degeneracy is defined by the second helicity supertrace.}.
The closed BPS partition function is then defined by (see \ref{sec.WC})
\beq
Z^c_{\textrm{BPS}}(q,Q)=\sum_{n,\beta}\Omega(n,\beta) q^n Q^{\beta}.
\eeq

The closed BPS partition function for generalized conifolds is studied in Refs.~\refcite{Nagao1,AOVY}.
To describe the results, let us first specify the resolution (crepant resolution) of $X$ \footnote{This is not essential, since by varying the value of the K\"ahler moduli we can go to the geometry with other choices of resolutions. We just need to specify an arbitrary resolution in order to begin the discussion. See \ref{app.Weyl} for more on this.}.
Each of the $L-1$ $\bP^1$'s is either an $\scO(-1,-1)$-curve
or an $\scO(-2,0)$-curve.
In the language of the toric diagram, this choice is specified by the triangulation of the toric diagram. We specify this choice by a map
\beq
\sigma: \{ 1/2,3/2,\ldots, L-1/2 \} \to \{\pm 1 \}.
\eeq
In the following we sometimes write $\pm $ instead of $\pm 1$.
When $\sigma(i-1/2)=1$ ($\sigma(i-1/2)=-1$), the $i$-th triangle from the left has 
one of its edges on the top (bottom) edge of the trapezoid.
This means that the $i$-th $\bP^1$ is a $\scO(-1,-1)$-curve ($\scO(-2,0)$-curve) when $\sigma(i-1/2) = -\sigma(i+1/2)$ ($\sigma(i-1/2) = \sigma(i+1/2)$).
By definition, we have $\big| \sigma^{-1}(\pm 1)\big|=L_{\pm}$. 

\begin{figure}[htpb]
\centering{\includegraphics[scale=0.25]{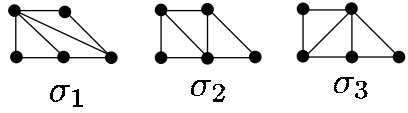}}
\caption[The choice of resolutions of a generalized conifold.]{The choice of resolutions of a generalized conifold ($L_+=1,L_-=2$).}\label{fig.sigmaex}
\end{figure}

For example, in the case of Suspended Pinched Point ($L_+=1, L_-=2$) whose toric diagram is shown in Fig.\,\ref{fig.sigmaex}, $L=3$ and there are 3 difference choice of resolutions. This is represented by
\begin{align}
\begin{split}
\sigma_1:\{ 1/2,3/2,5/2\}\to \{-,-,+ \}, \\
\sigma_2:\{ 1/2,3/2,5/2\}\to \{-,+,- \}, \\ 
\sigma_3:\{ 1/2,3/2,5/2\}\to \{+,-,- \}.
\end{split}
\label{SPPsigma}
\end{align}

Given $\sigma$, the topological string partition function is given by \cite{AKMV,IK}
\beq
Z_{\textrm{top}, \sigma}(q=e^{-g_s},Q=e^{-t})=\prod_{n=1}^{\infty}\prod_{\beta} (1-q^n Q)^{n N_{\beta}^{g=0}},
\eeq
where $N_{\beta}^{g=0}$ is the genus 0 Gopakumar-Vafa invariant \footnote{Higher genus Gopakumar-Vafa invariants vanish for generalized conifolds.}. For the 2-cycle $\beta=\alpha_i+\ldots +\alpha_j$, the explicit form of $N_{\beta}^0$ depends on $\sigma$ and is given by (see \eqref{root})
\begin{align}
\begin{split}
N_{\beta=\alpha_i+\ldots \alpha_j}^{g=0}&=(-1)^{1+\sharp \{k\mid i\le k\le j, \ \sigma(k-1/2)\ne \sigma(k+1/2) \}} \\
&=(-1)^{1+\sharp \{k\mid i\le k\le j, \ \alpha_k \textrm{ is a } \scO(-1,-1)-\textrm{curve} \}}.
\end{split}
\end{align}
By CPT invariance in five dimensions \cite{AOVY}, we have
$
N_{\beta=-(\alpha_i+\ldots \alpha_j)}^{g=0}
=N_{\beta=\alpha_i+\ldots \alpha_j}^{g=0}.
$
We also have $N_{\beta=0}^0=\chi(X)$, where the Euler character $\chi(X)$ for a toric Calabi-Yau manifold is the same as twice the area of the toric diagram.

As shown in Section\,\ref{chap.M-theory}, the closed BPS partition function is given by
\begin{eqnarray}
Z_{\rm BPS}(q,Q)&=&Z_{\rm top}(q,Q) Z_{\rm top}(q,Q^{-1}) \big|_{\rm chamber} \nonumber \\
&=& \prod_{(\beta,n): Z(\beta,n)>0} (1-q^n Q^{\beta})^{n N_{\beta}^0}, \label{Zchamber}
\end{eqnarray}
where the central charge $Z(\beta,n)$ is given in \eqref{Z}.
There $1/R$ denotes (up to proportionality constants) the central charge of the D0 brane, and following the discussion in Section\,\ref{sec.idea} we choose the complexified K\"ahler moduli to be real. 

Now suppose that $1/R$ is positive \footnote{Under this condition we are discussing only half of chambers of the K\"ahler moduli space, which lie between the Donaldson-Thomas chamber and the non-commutative Donaldson-Thomas chamber.
 The other half arises when $1/R$ is negative.}.
From \eqref{Zchamber} and \eqref{Z}, it follows that the wall crossing occurs when the integer part of the value of the B-field through the cycle change. For the cycle $\alpha_{i}+\ldots +\alpha_j$, this is given by
\beq
\left[ B(\alpha_i)+\ldots + B(\alpha_j) \right], 
\eeq
Since there are 
$L-1$ $\bP^1$'s in $X$, there are 
$L(L-1)/2$ such parameters.

We can take a special limit $B(\alpha_i)\to \infty$ for all $i$. Let us denote this special chamber by $\tilde{C}_{\rm top}$. As discussed at the end of Section\,\ref{sec.derivation}, in this limit the BPS partition function reduces to the closed topological string partition function:
\beq
Z^c_{(\sigma,\theta)}\Big|_{\tilde{C}_{\rm top}} =Z^c_{\rm top},
\eeq
just as advertised in \eqref{ZtoZtop}.

\medskip
For concreteness, let us discuss the example of the Suspended Pinched Point ($L=3$) using the triangulation $\sigma_1$ in \eqref{SPPsigma}. In this example, the topological string partition function is 
\beq
Z_{\textrm{top}, \sigma=\sigma_1}(q,Q)=M(q)^{3/2} \prod_{n=1}^{\infty} (1-q^n Q_1)^{-n}
\prod_{n=1}^{\infty} (1-q^n Q_2)^{n} \prod_{n=1}^{\infty} (1-q^n Q_1 Q_2)^{n},
\eeq
where $M(q)$ is the MacMahon function defined previously in \eqref{MacMahondef}.The BPS partition function is given by
\begin{align}
\begin{split}
Z_{\rm BPS}&(q,Q)=M(q)^{3} 
\prod_{n=1}^{\infty} (1-q^n Q_1)^{-n}
\prod_{n=1}^{\infty} (1-q^n Q_2)^{n} \prod_{n=1}^{\infty} (1-q^n Q_1 Q_2)^{n} \\%
&\times 
\prod_{n>\left[ B(\alpha_1) \right]}^{\infty} (1-q^n Q_1^{-1})^{-n}
\prod_{n>\left[ B(\alpha_2) \right]}^{\infty} (1-q^n Q_2^{-1})^{n} \prod_{n>\left[ B(\alpha_1+\alpha_2) \right]}^{\infty} (1-q^n (Q_1 Q_2)^{-1})^{n}.
\end{split}
\end{align}


\medskip

The parameters $\left[ B(\alpha_{i}+\ldots +\alpha_j) \right]$ specify the chamber, but as we can see from the definition they are not completely independent parameters. Since we only have 
$L-1$ real parameters $B_i$, it is likely that this parametrization is redundant. Indeed, as explained in the \ref{app.Weyl} 
we can specify the chamber by a map $\theta$, which is specified by 
$L$ half-integers, 
$\theta(1/2),\theta(3/2),\ldots, \theta(L-1/2)$, satisfying one constraint
\beq
\sum_{i=1}^L \theta\left(i-\frac{1}{2}\right)=\sum_{i=1}^L \left(i-\frac{1}{2}\right).
\eeq
This means we can indeed parametrize the chamber by 
$L-1$ independent (half-)integers, which is what we expected. As discussed in \ref{app.Weyl}, $\theta$ is an element of the Weyl group of 
$\hat{A}_{L-1}$.

\subsection{The Noncommutative Topological Vertex}\label{sec.def}

In this section we give a general definition of the non-commutative topological vertex using the crystal melting model. This definition is equivalent to the one given in Ref.~\refcite{Nagao2} using the dimer model \footnote{See \ref{app.PM}  for the equivalence between crystal melting model and the dimer model.}. 
See Ref.~\refcite{Nagao3} for more conceptual definition in terms of Bridgeland's stability conditions and moduli spaces. 

To define our vertex, we need the following set of data:

\begin{itemize}
\item A map 
$$
\sigma: \{ 1/2,\ldots, L-1/2\}\to \{\pm \}.
$$ 
As already explained in Section\,\ref{sec.closed}, this gives a triangulation of the toric diagram, or equivalently the choice of the resolution of the Calabi-Yau manifold. 

\item A map $\theta:\bZ_h \to \bZ_h$, where $\bZ_h$ is the set of half-integers. As explained in \ref{app.Weyl} in the case of closed BPS invariants, $\theta$ and $\sigma$ specify the chamber structure of the open BPS invariants. 

\item A set of Young diagrams $\lambda$, assigned to external legs of the $(p,q)$-web. This specifies the boundary condition of the non-compact D-branes ending on the $(p,q)$-web. We denote by 
$\lambda_1,\ldots,\lambda_{L}$ the Young diagrams for the top and the bottom edges of the trapezoid, and by $\lambda_+,\lambda_-$ the remaining two. We sometimes write $\lambda=(\mu,\nu)$, where 
$\mu=(\mu_1,\ldots, \mu_L)=(\lambda_1,\ldots,\lambda_L)$ and $\nu=(\lambda_+,\lambda_-)$.
In the example shown in Fig.\,\ref{lambdaconv}, there are 5+2 external legs and therefore 5+2 representations.

\begin{figure}[htbp]
\centering{\includegraphics[scale=0.3]{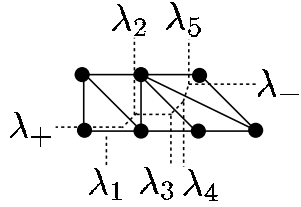}}
\caption[Representations assigned to external legs of the $(p,q)$-web.]{Representations assigned to external legs of the $(p,q)$-web. The dotted lines represent the $(p,q)$-web.}
\label{lambdaconv}
\end{figure}

For later purposes, we combine $\mu_1,\ldots, \mu_L$ into a single representation $\mu$ by
\beq
\mu(i-1/2+kL):=\mu_i(k-1/2). \label{combinedmu} 
\eeq
In other words, we choose $\mu$ such that L-quotients of $\mu$ give $\mu_1, \ldots, \mu_L$. By abuse of notation, we use the same symbol $\mu$ for a set of representations $\mu_1,\ldots, \mu_L$ as well as a single representation define above.
\end{itemize}

Given $\sigma, \theta$ and $\lambda$, we define the non-commutative topological vertex
$$
C_{\textrm{BPS},(\sigma,\theta;\lambda)}(q,Q).
$$
In the following we drop the subscript BPS for simplicity.

Before going into the general definition, we first illustrate our idea using simple example of the resolved conifold.

\subsubsection{Example: Resolved Conifold}\label{subsec.conifoldeg}

In this example there is only one $\bP^1$ and the BPS partition function depends on a single positive integer $N:=\left[ B(\alpha_1) \right]$. In the language of $\theta$,
\beq
\theta(1/2)=1/2-N, \quad \theta(3/2)=3/2+N. \label{thetaN}
\eeq
We fix $\sigma$ to be 
\beq
\sigma(1/2)=+, \quad \sigma(3/2)=-.
\eeq
Without losing generality we concentrate on $N\ge 0$, since $N<0$ corresponds to a flopped geometry, where $\sigma$ is replaced by $-\sigma$ (see \ref{app.Weyl}).

\begin{figure}[htbp]
\centering{\includegraphics[scale=0.15]{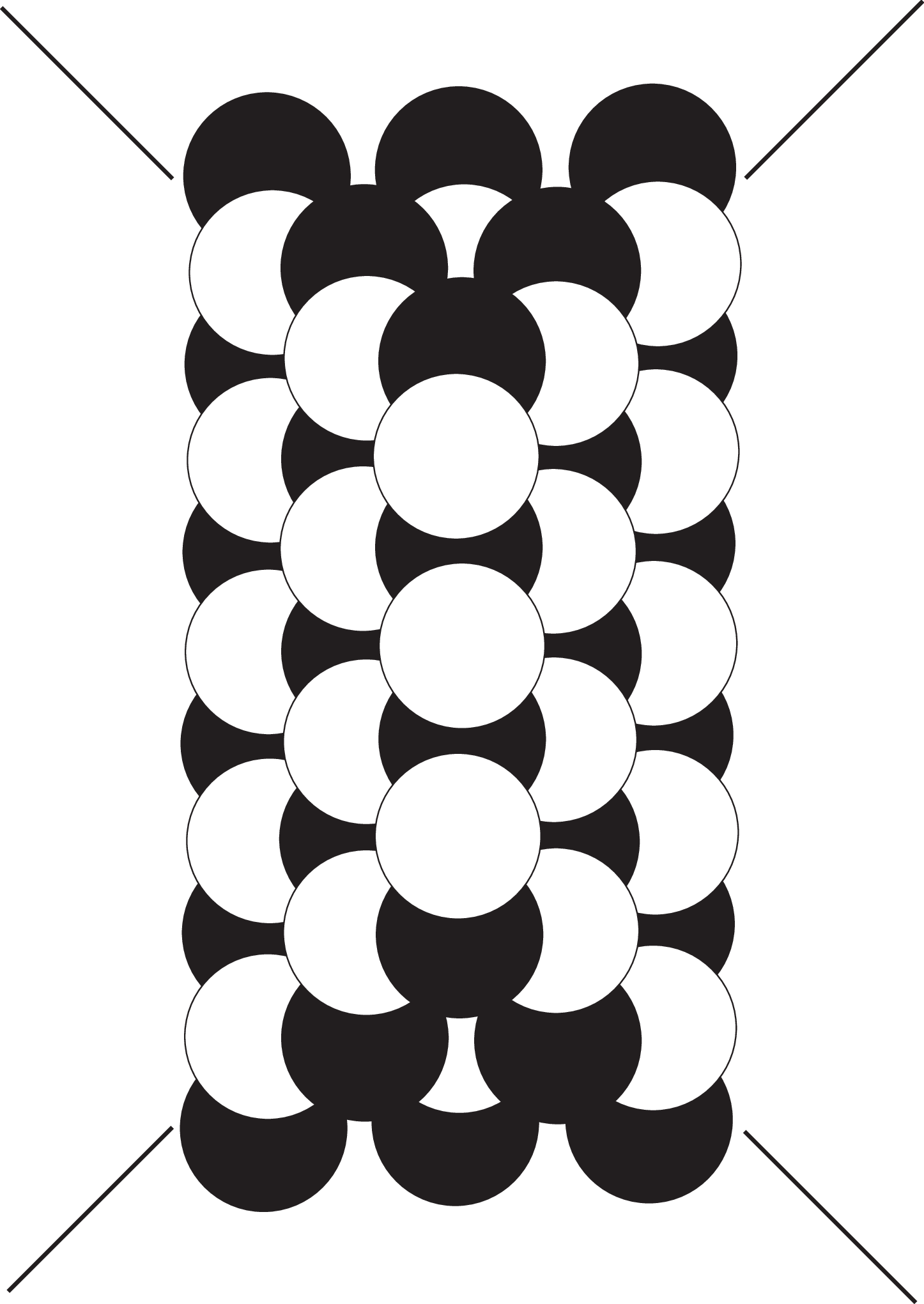}}
\caption[The ground state crystal for the resolved conifold.]{The ground state crystal for the resolved conifold for $N=2$. The crystal consists of an infinite number of layers, and only a finite number is shown here. The ridges of the pyramid are represented by four lines extending to infinity. This is the same figure as in Fig.\,\ref{newgroundstate}.}
\label{conifoldcrystal}
\end{figure}

The ground state crystal for $N=2$ is shown in Fig.\,\ref{conifoldcrystal} (this is the same figure as Fig.\,\ref{newgroundstate}, reproduced here for the convenience of the reader).
This crystal, sometimes called a pyramid, consists of infinite layers of atoms, the color alternating between black and white \cite{Szendroi,Young1}. 
In the $N$-th chamber there are $N+1$ atoms on the top.

The closed BPS partition function is defined by removing a finite set of atoms $\Omega$ from the crystal. When we do this, we follow the melting rule \eqref{meltingruleold} such that
whenever an atom is removed from the crystal, we remove all the atom above it.
We then define the partition function by summing over such $\Omega$:
\beq
Z=\sum_{\Omega} (q_0^{(N)})^{w_0(\Omega)} (q_1^{(N)})^{w_1(\Omega)},
\label{Zdefconifold}
\eeq
where $w_0(\Omega)$ and $w_1(\Omega)$ are the number of white and black atoms in $\Omega$, respectively.

\begin{figure}[htbp]
\centering{\includegraphics[scale=0.15]{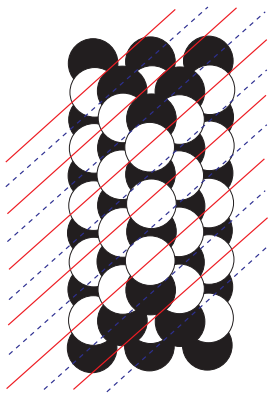}}
\caption[Slice of the conifold crystal.]{We can slice of the conifold crystal by an infinite number of parallel planes.}
\label{crystalslice}
\end{figure}

The weights $q_0^{(N)}$ ($q_1^{(N)}$) assigned to white (black) atoms in the $N$-th chamber are determined as follows. We can slice the crystal by the plane, 
and each slice is specified by an integer $i$ (see Fig.\,\ref{crystalslice}). We choose $i$ so that the slice $i=0$ cuts through the center of the crystal.

The weight $q_i^{(N)}$ depends on the chamber and is given by
\beq
 q_0^{(N)}=q_0^{-N}q_1^{-N+1}, 
\quad
 q_1^{(N)}=q_0^{N+1}q_1^{N},
\label{conifoldwt1}
\eeq
when $N$ is odd, and
\beq
 q_0^{(N)}=q_0^{N+1}q_1^{N},
\quad
 q_1^{(N)}=q_0^{-N}q_1^{-N+1},
\label{conifoldwt2}
\eeq
when $N$ is even. This coincides with the expression in \eqref{wtchg}, up to the exchange of two nodes of the quiver when $N$ is odd.
For example, $q_0^{(0)}=q_0, q_1^{(0)}=q_1$ when $N=0$, and 
$q_0^{(1)}=q_0^{-1} , q_1^{(1)}=q_0^2q_1$.
The change of variables arises from the Seiberg duality on the quiver quantum mechanics \cite{CJ}, geometrically mutations in the derived category of coherent sheaves \cite{CJ,NN} (see Section\,\ref{subsec.D6}), or in more combinatorial language the dimer shuffling \cite{Young1}. The parameters $q_0, q_1$ defined here are related to the D0/D2 chemical potentials introduced in Section\,\ref{sec.closed} by (see \eqref{paramchg}) \footnote{
The equation \eqref{conifoldwt2} is the same for $N$ odd and even if we suitably exchange the two nodes of the quiver diagram. 
The relation \eqref{chemicalpot} can also be written as
\beq
q=q_0^{(N)} q_1^{(N)}, \quad Q=(q_0^{(N)})^{N} (q_1^{(N)})^{N+1}, 
\eeq
when $N$ is even, and $q_0^{(N)}$ and $q_1^{(N)}$ exchanged when $N$ odd. This coincides with the expression in Ref.~\refcite{CJ}.
}
\beq
q=q_0 q_1, \quad Q=q_1.
\label{chemicalpot}
\eeq

Now let us discuss the open case. When non-trivial representations are assigned to each of the four external legs of the $(p,q)$-web, the only thing we need to do is to change the ground state of the crystal.

\begin{figure}[htbp]
\centering{\includegraphics[scale=0.15]{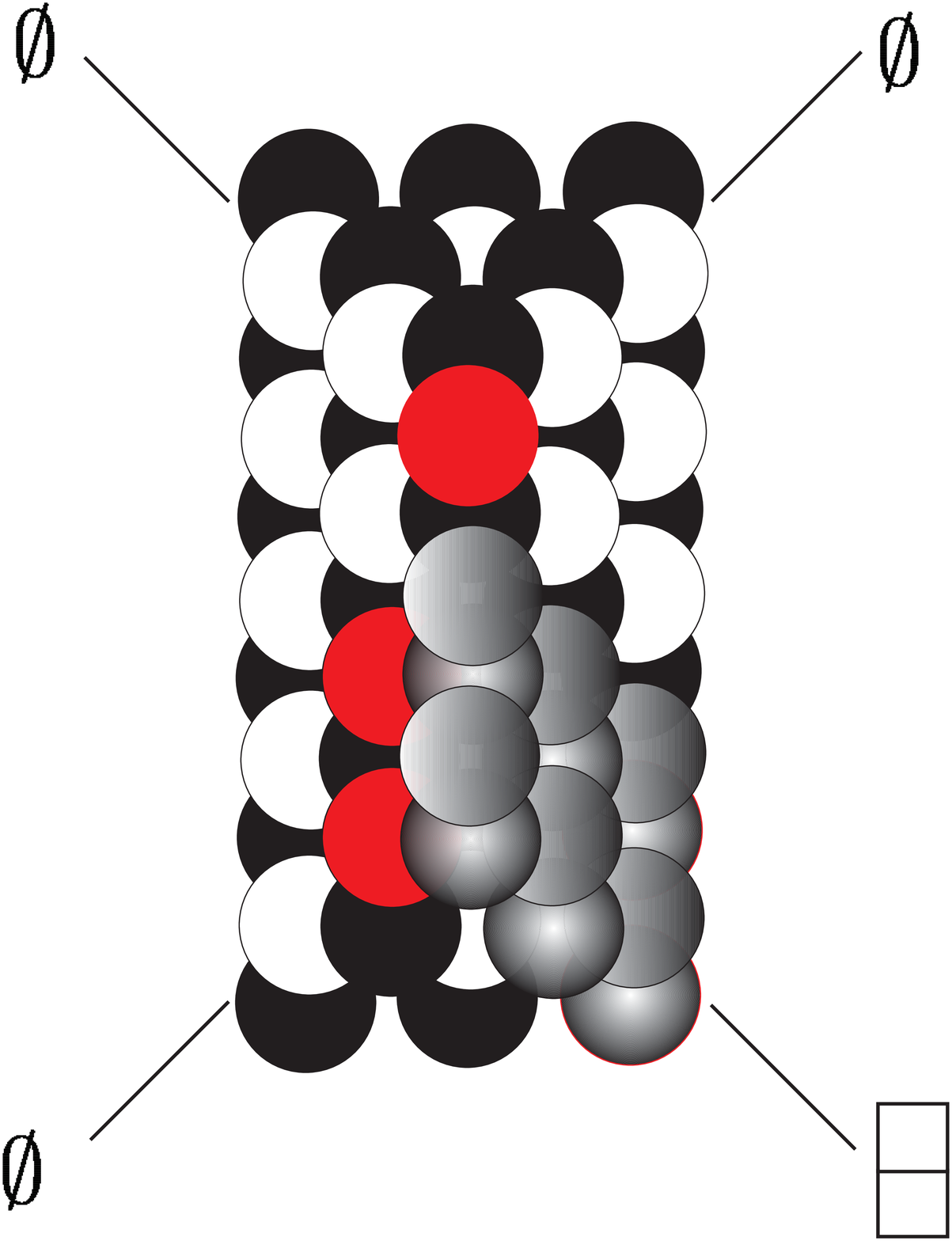}}
\caption[The pyramid for open BPS invariants.]{The pyramid for open BPS invariants. A non-trivial representation 
$(1,1)$ is placed on with one of the four external lines. As compared with the previous figure, atoms colored gray, corresponding to the Young diagram, are removed from the crystal. The red atoms have no atoms above them.}
\label{openconifoldeg}
\end{figure}

The crystal has four ridges, corresponding to four external legs of the $(p,q)$-web. When we assign a representation, we remove the atoms with the shape of the Young diagram. More precisely, we remove the atoms with the shape of the Young diagram in the asymptotic direction of the $(p,q)$-web, as well as all the atoms above them, so that the melting rule is satisfied. See Fig.\,\ref{openconifoldeg} for an example.

The partition function is defined in exactly the same way by \eqref{Zdefconifold}, and the result is denoted by $C_{(\sigma, \theta;\lambda)}$.

\bigskip

Several comments are now in order.

First, let us explain the origin of the name ``the non-commutative topological {\it vertex}''.
Recall that, in commutative case, topological vertex is defined for $\bC^3$. 
For a general affine toric Calabi-Yau manifold $X$, we divide the polygon into triangles and assign a topological vertex to each trivalent vertex of the dual graph. 
We can get the topological string partition function for the smooth toric Calabi-Yau manifold $Y$ (resolution of $X$) by gluing the topological vertices with propagators. 
Similarly, assume that a polygon is divided into trapezoids. 
Then we can assign a non-commutative topological vertex to each vertex of the dual graph and glue them by propagators. 
The BPS partition function defined in this way  is related to the 
topological string partition function 
via wall-crossing \footnote{
Given a division of a polygon into trapezoids, we get a partial resolution of $X$ and a non-commutative algebra $A$ over the partial resolution, which is derived equivalent to $Y$. The BPS partition function 
given by gluing non-commutative topological vertices counts torus invariants $A$-modules.}.
In Refs.~\refcite{BCY2,BCY1} they study the case when a polygon is divided into (not necessary minimal) triangles.

\begin{figure}[htbp]
\centering{\includegraphics[scale=0.3]{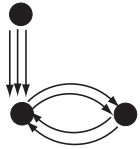}}
\caption[Quiver diagram for the open invariant with 
$\lambda=(1,1)$.]{Quiver diagram for the open invariant with 
$\lambda=(1,1)$. This is the Klebanov-Witten quiver \cite{KW} with an extra node and extra three arrows starting from it. The three arrows correspond to three red atoms in Fig.\,\ref{openconifoldeg}. Note that this is different from the quiver in Fig.\,\ref{fig.mutatedquiver}.}
\label{openquiver}
\end{figure}

Second, it is possible to give more geometric definition of the vertex 
(see Ref.~\refcite{Nagao3}).
For the closed BPS invariants, 
the crystal arises as a torus fixed point of the moduli space of the modules of the path algebra quiver (under suitable $\theta$-stability conditions). The moduli space is the vacuum moduli space of the quiver quantum mechanics arising as the low-energy effective theory of D-branes (Section\,\ref{chap.crystal}). The similar story exists in our case. Namely, the crystal is in one-to-one correspondence with the fixed point of the moduli space arising from a quiver diagram. For example, for conifold with $\lambda=${\tiny \yng(1,1)}, the quiver is given in Fig.\,\ref{openquiver}.

Third, in the case of $\bC^3$, our vertex reproduces the topological vertex for $\bC^3$
by definition.


\subsubsection{General Definition from Crystal Melting}

We next give a general definition of the vertex. Readers not interested in the details of the definition of the non-commutative topological vertex
can skip this section on first reading.

Given a boundary condition specified by $\sigma, \theta$ and $\lambda=(\mu,\nu)$, we would like to construct a ground state of the crystal, and
determine the weights assigned to the atoms of the crystal.

The basic idea is the same as in the conifold example. First, the closed string BPS partition function is equivalent to the statistical partition function of crystal melting. The ground state crystal can be sliced by an infinite number of parallel planes parametrized by integers $n\in \bZ$, just as in Fig.\,\ref{crystalslice}. On each slice, there are infinitely many atoms, labeled by integers $(x,y)\in \bZ_{\ge 0}^2$. Therefore, the atoms in the crystal are label by $(n,x,y)\in \bZ\times \bZ_{\ge 0}^2$.

\bigskip

\begin{figure}[htbp]
\centering{\includegraphics[scale=0.15]{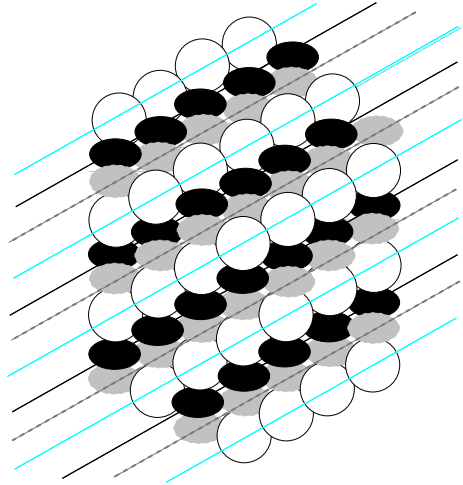}}
\caption[The crystal for the Suspended Pinched Point.]{The crystal for the Suspended Pinched Point. We can slice the crystal along planes represented by lines, which come with three different colors. This crystal appeared previously in Fig.\,\ref{fig.SPPatom}.}
\label{SPPcrystal}
\end{figure}

Let us show this in the example of the Suspended Pinched Point. The crystal in
Fig.\,\ref{SPPcrystal} clearly shows this structure.

\begin{figure}[htbp]
\centering{\includegraphics[scale=0.2]{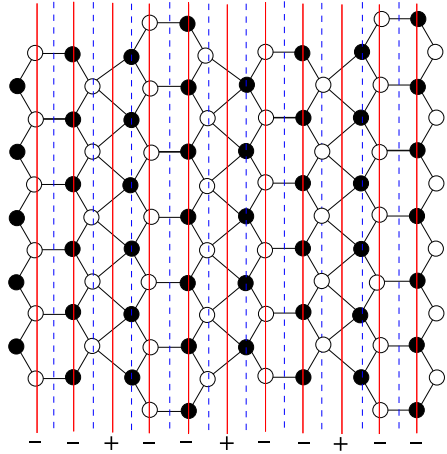}}
\caption[The bipartite graph for Suspended Pinched Point.]{The bipartite graph for Suspended Pinched Point. We here take $\sigma=\sigma_1$ and $\theta=id$. The red undotted (the blue dotted) lines have half-integer (integer) values of the coordinate along the horizontal axis. This figure is the universal cover the bipartite graph on $\bT^2$, discussed previously in Fig.\,\ref{fig.SPPbipartite} and in Fig.\ref{fig.SPPregion}.}
\label{SPPbipartite}
\end{figure}

Another way of explaining this is to 
construct a crystal starting from a bipartite graph on $\bR^2$, shown in Fig.\,\ref{SPPbipartite} \footnote{This is a universal cover of the bipartite graph on $\bT^2$, which appears in the study of four-dimensional $\scN=1$ quiver gauge theories. See Section\,\ref{sec.quiver}.}.
In this example, the bipartite graph consists of hexagons and squares, and 
periodically changes its shape along the horizontal directions.

The atoms of the crystal are located at the centers of the faces of the bipartite graph. and it thus follows we can slice the crystal along the horizontal axis. Each slice consist of an infinite number of atoms labeled by two integers $(x,y)\in (\bZ_{\ge 0})^2$, since there are two directions, the horizontal direction and the perpendicular direction to the paper \footnote{In general the bipartite graph is determined by $\sigma$. 
A hexagon (a square) corresponds to $\scO(-2,0)$-curve ($\scO(-1,-1)$-curve).
In other words, the $i$-th polygon is a hexagon (square) if 
$\sigma(i)=\sigma(i+1) (\sigma(i)\ne \sigma(i+1))$.}.

\bigskip

Now consider the open case. In this case, we construct a new ground state by removing atoms from the closed ground state. By the melting rule, the atoms 
removed from the $n$-th plane should be labeled by $(x,y)\in \scV(n)$, where
$\scV(n)$ is a Young diagram. Depending on the representations on external legs, $\scV(n)$ increases or decreases as we change $n$. Thus the ground state crystal for open BPS invariants are determined by such a sequence of Young diagrams $\{\scV(n)\}$, called transitions below. In the following we make this idea more rigorous.

Let us begin with some notations.
Let $\mu$ and $\mu'$ be two Young diagrams. 
We say $\mu\pg\mu'$ if the row lengths satisfy
\beq
\mu_1\geq\mu'_1\geq\mu_2\geq\mu'_2\geq\cdots,
\eeq
and $\mu\mg\mu'$ if the column lengths satisfy
\beq
\tenchi\mu_1\geq\tenchi\mu'_1\geq\tenchi\mu_2\geq\tenchi\mu'_2\geq\cdots.
\eeq
We define a {\it transition} $\V$ of Young diagrams of type $\type$ as a map 
from the set of integers $\bZ$ to 
the set of Young diagrams such that
\begin{itemize}
\item $\V(n)=\nu_-$ for $n\ll0$ and $\V(n)=\nu_+$ for $n\gg0$,
\item $\V(h-\mu\circ\theta(h)/2)\overset{\sigma\circ\theta(h)}{\succ}\V(h+\mu\circ\theta(h)/2)$ for $h$: half-integer.
\end{itemize}
Then as shown in Ref.~\refcite{Nagao2} there is a minimal transition $\Vmin$ of Young diagrams of type $\type$ such that for any transition $\V$ of Young diagrams of type $\type$  and for any $n\in\Z$ we have $\V(n)\supseteq \Vmin(n)$. 

For a transition $\V$ of Young diagram of type $\type$, the corresponding crystal configuration can be defined by
\beq
A(\V):=\{a(n,x,y)\mid n\in \bZ, x,y \in \bZ_{\ge 0}, \ (x,y)\notin \V(n)\}.
\eeq
where $a(n,x,y)$ denotes the atom at position $(n,x,y)$. In particular, $A(\Vmin)$ gives the ground state crystal.

Having defined the ground state crystal, the partition functions is defined again as the sum over a subset $\Omega$ of $A(\Vmin)$ satisfying the following two conditions: \footnote{It is straightforward to show that a subset $\Omega \subset A(\Vmin)$ satisfies the two conditions if and only if $A(\Vmin)\backslash \Omega=A(\V)$ for a transition $\V$ of Young diagram of type $\type$.}

\begin{itemize}
\item $\Omega$ is finite set, and 
\item $\Omega$ satisfies the melting rule \eqref{meltingruleold}. In other words, if 
$a'\in \Omega$ and $a'=a \alpha$ for an arrow $\alpha$, then $a\in \Omega$ \footnote{
We can also explicitly write down the melting rule using the coordinates $(n,x,y)$.
Let us write $a \sqsupset a'$ when there is a path (a composition of arrows) $\alpha$ such that $a'=a \alpha$. 
The partial order $\sqsupset$ is then generated by
$$
\begin{array}{lr}
a(h-1/2,x,y)\sqsupset a(h+1/2,x,y) & \mu\circ \theta(h)=+,\\
a(h+1/2,x,y)\sqsupset a(h-1/2,x,y) & \mu\circ \theta(h)=-,\\
a(h+1/2,x,y)\sqsupset a(h-1/2,x+1,y) & \mu\circ \theta(h)=+, \sigma\circ\theta(h)=-,\\
a(h+1/2,x,y)\sqsupset a(h-1/2,x,y+1) & \mu\circ \theta(h)=+, \sigma\circ\theta(h)=+,\\
a(h-1/2,x,y)\sqsupset a(h+1/2,x+1,y) & \mu\circ \theta(h)=-, \sigma\circ\theta(h)=-,\\
a(h-1/2,x,y)\sqsupset a(h+1/2,x,y+1) & \mu\circ \theta(h)=-, \sigma\circ\theta(h)=+,\\
a(n,x,y)\sqsupset a(n,x+1,y) & \sigma\circ\theta(n-1/2)=\sigma\circ\theta(n+1/2)=+,\\
a(n,x,y)\sqsupset a(n,x,y+1) & \sigma\circ\theta(n-1/2)=\sigma\circ\theta(n+1/2)=-.
\end{array}
$$
}. 
\end{itemize}

For a crystal $\Omega \subset A(\Vmin)$, we define the weight $w(\Omega)_i$ by the number of atoms with the color $i$ contained in $\Omega$:
\beq
w(\Omega)_i:=\sharp\{a(n,x,y)\in \Omega\mid n\equiv i\ (\mr{mod}\,L)\}.
\eeq
Also, for $\theta$, we put 
\beq
q^\theta_i:=q_{\theta^{-1}(i-1/2)+1/2}\cdot q_{\theta^{-1}(i-1/2)+3/2}\cdot\cdots\cdot q_{\theta^{-1}(i+1/2)-1/2}
\eeq
when $\theta^{-1}(i-1/2)<\theta^{-1}(i+1/2)$, and 
\beq
q^\theta_i:=
q_{\theta^{-1}(i-1/2)-1/2}^{-1}\cdot q_{\theta^{-1}(i-1/2)-3/2}^{-1}\cdot\cdots\cdot q_{\theta^{-1}(i+1/2)+1/2}^{-1} 
\eeq
when $\theta^{-1}(i-1/2)>\theta^{-1}(i+1/2)$.
Here we defined $q_i$ for $i\in \bZ$ periodically,
\beq
q_{i+L}=q_i.
\label{periodic}
\eeq
We then define the vertex by
\beq
C^{\rm ref}_{\type}(q_0,\ldots,q_{L-1}):=\sum_{\Omega} (q^{\theta}_0)^{w(\Omega)_0}\cdot\cdots\cdot (q^{\theta}_{L-1})^{w(\Omega)_{L-1}}.
\eeq
The parameters $q_0, \ldots, q_{L-1}$ defined here are related to the D0/D2 chemical potentials introduced in Section\,\ref{sec.closed} by
\beq
q=q_0\ldots q_{L-1}, \quad Q_i=q_i\quad (i=1,\ldots, L-1).
\label{chemicalweight}
\eeq


\subsubsection{Refinement}\label{subsec.refined}

We can also generalize the definition to include the open refined BPS invariants \footnote{See Refs.~\cite{KontsevichS,DG,BBS} for closed refined BPS invariants.}. 

Let us start by recalling the meaning of the refined BPS counting, first in the closed case \cite{GukovSV}.
As we have seen in Section\,\ref{chap.M-theory}, when the type IIA brane configuration is lifted to M-theory \cite{GV1} and when we use the 4d/5d correspondence \cite{GSY,DVVafa}, the D0/D2-branes are lifted to spinning M2-branes in $\bR^5$, which has spin under the little group in 5d, namely $SO(4)=SU(2)_L\times SU(2)_R$. The ordinary BPS invariant is defined as an index \eqref{Nindex}; it keeps only the $SU(2)_L$ spin, while taking an alternate sum over the $SU(2)_R$ spin. 
The refined closed BPS invariants is defined by taking both spins into account.

The situation changes slightly when we consider open refined BPS invariants.
The D4-branes wrapping Lagrangians, when included, are mapped to M5-branes on $\bR^3$. This means that $SO(4)$ is broken to $SO(2)$, and we have only one spin. However, there is an $SO(2)$ R-symmetry for $\scN=2$ supersymmetry in three dimensions, and in the definition of the ordinary open BPS invariants we keep only one linear combination of the two, while tracing out the other combination \cite{LMV}. The refined open BPS invariants studied here takes both of the two charges into account.

In the language of crystal melting used in this paper, the open refined BPS invariants are defined simply by modifying the definition of the weights. Here, we explain how to modify the weights in the case of $\mu=\emptyset$.

For an integer $n$ define
\beq
w(\Omega)_n:=\sharp\{ a(n,x,y)\in \Omega \}.
\eeq
We also define the weights by
\beq
\tilde{q}^\theta_n:=\tilde{q}_{\theta^{-1}(n-1/2)+1/2}\cdot \tilde{q}_{\theta^{-1}(n-1/2)+3/2}\cdot\cdots\cdot \tilde{q}_{\theta^{-1}(n+1/2)-1/2} 
\eeq
when $\theta^{-1}(n-1/2)<\theta^{-1}(n+1/2)$, and 
\beq
\tilde{q}^\theta_n:=\tilde{q}_{\theta^{-1}(n-1/2)-1/2}^{-1}\cdot \tilde{q}_{\theta^{-1}(n-1/2)-3/2}^{-1}\cdot\cdots\cdot \tilde{q}_{\theta^{-1}(n+1/2)+1/2}^{-1}\eeq
when $\theta^{-1}(n-1/2)>\theta^{-1}(n+1/2)$.
Here we defined
\beq
\tilde{q}_n:=q_n
\eeq
when $n \not\equiv 0 \ \textrm{mod} \ L$  (recall $q_n$ is understood to be periodic with respect to $n$ \eqref{periodic}), 
and
\beq
\tilde{q}_n:=
\begin{cases}
q_+ & n>0 \\
(q_+ q_-)^{1/2}&  n=0\\
q_- & n<0
\end{cases}
\eeq
when $n \equiv 0 \ \textrm{mod} \ L$.
We then define the refined vertex by
\beq
C_{(\sigma,\theta\sss;\sss \emptyset,\nu)}(q_+,q_-,q_1\ldots,q_{L-1}):=\sum_{\Omega} \prod_{n\in \bZ} (\tilde{q}^{\theta}_n)^{w(\Omega)_n}.
\eeq
By definition, the refined vertex reduces to the unrefined vertex by setting $q_+=q_-=q_0$.
The reader can refer to Ref.~\refcite{Nagao2} for the definition of weights in general cases $\mu\ne \emptyset$.

\subsubsection{Consistency Checks of Our Proposal}\label{sec.check}

In \ref{app.proof} we gave a purely combinatorial definition of the non-commutative topological vertex. We now claim that is captures open BPS invariants in the following sense.

Consider a generalized conifold (a toric Calabi-Yau manifold without compact 4-cycles) with representations assigned to each leg of the $(p,q)$-web. Each representation specifies a boundary condition on the non-compact D4-brane wrapping Lagrangian 3-cycle of topology $\bR^2\times S^1$ \cite{AV}.

In the absence of D4-branes, we are counting particles of D0/D2-branes wrapping 0/2-cycles, which makes a bound state with a single D6-brane filling the entire Calabi-Yau manifold. When the D4-branes are included, D2-branes can wrap disks ending on the worldvolume of D4-branes. The degeneracy of such D-brane configurations is what we mean by the open BPS degeneracies. Note that supersymmetry is broken by half due to the inclusion of D4-branes; our counting of BPS particles makes sense because we are counting BPS states in lower dimensions, where the minimal amount of supersymmetry is lower.

We can provide several consistency checks of our proposal.
First, our vertex by definition reduces to the closed BPS invariant when all the representations $\lambda$ are trivial:
\beq
C_{\textrm{BPS}, (\sigma,\theta;\lambda=\emptyset)}=Z^c_{\textrm{BPS},(\sigma,\theta)}.
\eeq
The second consistency check comes from the wall crossing phenomena. As shown in Ref.~\refcite{Nagao2}, the vertex goes through a series of wall crossings as we move around the closed string moduli space (K\"ahler moduli of the Calabi-Yau manifold), just as in the case of closed invariants. 
It was also shown in Ref.~\refcite{Nagao2} that in the chamber $C_{\rm top}$ where the closed BPS partition function reduces to the closed topological string partition function, our vertex gives the same answer as that computed from the topological vertex (in the standard framing):
\beq
C_{\textrm{BPS}, (\sigma,\theta;\lambda)} \Big|_{C_{\rm top}}=C_{\textrm{topological vertex}, \lambda}.
\eeq
The third consistency check comes from the fact that the wall crossing factor is independent of representations. In other words,
\beq
\bar{C}_{\type}(q_0,\ldots,q_{L-1}):=
\frac{C_{\type}(q_0^\theta,\ldots,q_{L-1}^\theta)}{C_{(\sigma,\theta\sss;\sss\emptyset,\emptyset)}(q_0^\theta,\ldots,q_{L-1}^\theta)}.
\label{independent}
\eeq
does not depend on $\theta$ \cite{Nagao2,Nagao3} (see also
Section\,\ref{subsec.openvertex} for an explanation)
\footnote{\label{normalization}To be exact, we have to normalize the
generating function by a monomial. See Ref.~\refcite{Nagao3}, Corollary 3.21, for the precise statement.}. This means that the open BPS partition function, which is defined by the sum over representations, takes a factorized form 
\beq
Z^{o}_{\rm BPS} =\frac{Z^{o}_{\rm BPS}}{Z^{o}_{\rm top}} Z^o_{\rm top}
= \frac{Z^{c}_{\rm BPS}}{Z^{c}_{\rm top}} Z^o_{\rm top}.
\label{Zoproduct}
\eeq
Since $Z^{c}_{\rm BPS}/Z^{c}_{\rm top}$ takes an infinite product form as explained in Section\,\ref{sec.closed} and $Z^o_{\rm top}$ also takes the infinite product form \cite{OVknot}, $Z^{o+c}_{\rm BPS}$ itself should take an infinite product form,
which is consistent with a suitable generalization of the argument of Section\,\ref{chap.M-theory}. We will see discuss this in more detail in Section\,\ref{sec.openMtheory}.

Using \eqref{Zoproduct}, we can compute our vertex using the ordinary topological vertex formalism.
In the next section, we give a yet another way of computing the non-commutative topological vertex. The advantage of our approach is that the final expression manifestly takes a simple infinite product form, and we do not have to worry about the summation of Schur functions.

\subsection{The Closed Expression for the Vertex}\label{sec.closedexpression}

In this section we give a closed expression for our non-commutative topological vertex. We do this by proving a curious identity stating the equivalence of our vertex for a toric Calabi-Yau manifold $X$ with a closed BPS partition function for an orbifold of $X$ \footnote{This is reminiscent of story of Ref.~\refcite{GO1}, where the `bubbling geometry' $X'$ is constructed for given a toric Calabi-Yau manifold $X$ such that the open+closed topological string partition function on $X$ is equivalent to closed topological string partition function on $X'$. However, our story is different in that the vertex computes only a part of the full open BPS partition function; the partition function itself is given by summation of our vertices over representations.}. We also comment on another method via vertex operators.

Start from a non-commutative topological vertex for a generalized conifold $X$, which has $L-1$ compact $\bP^1$'s. As we discussed above, for the definition of the vertex we need 
(1) $\sigma$ for a choice of the crepant resolution of $X$,
(2) a map $\theta$ specifying the chamber together with $\sigma$
(3) a set of representations $\lambda=(\mu,\nu)$,
and the resulting vertex is denoted by $C_{\sigma,\theta;\lambda}(q,Q)$. In the following we consider the special case $\nu=(\lambda_+,\lambda_-)=(\emptyset, \emptyset)$.

Choose an integer $M$ and consider the $\bZ_M$ orbifold $X'$ of $X$. 
We choose the orbifold action such that when the toric diagram of $X$ is a trapezoid with a top and the bottom edge of length $L_+$ and $L_-$ respectively, $X'$ has length $M L_+$ and $M L_-$.
We also choose map
\beq
\sigma': \{1/2,3/2,\ldots, ML-1/2\}\to \{\pm 1\}.
\eeq
and
\beq
\theta': \bZ_h\to \bZ_h, \quad \theta'(h+ML)=\theta'(h)+ML
\eeq
such that
\beq
\sigma\circ \theta=\sigma' \circ\theta', \quad \mu\circ \theta=\emptyset\circ \theta'.
\label{assumptionmain}
\eeq

Then
\beq
C_{(\sigma,\theta\sss;\sss \emptyset,\mu)}(q,Q)=
C_{(\sigma',\theta'\sss;\sss \emptyset, \emptyset)}(q',Q')|_{q^{\theta}_i=q'{}^{\theta'}_i=q'{}^{\theta'}_{i+L}=\cdots q'{}^{\theta'}_{i+(M-1)L}},
\label{main2}
\eeq
where $i=0,\ldots, L-1$.

See \ref{app.proof} for an explicit method for choosing such $M,\sigma', \theta'$ satisfying \eqref{assumptionmain} as well as generalization of \eqref{main2} to the case of refined BPS invariants.

Since an infinite-product expression for closed BPS partition function for a generalized conifold is already known (Section\,\ref{sec.closed}), we have a closed expression of our vertex when $\nu=\emptyset$.

\subsubsection{Examples}\label{sec.openeg}
Let us illustrate the above procedure by several examples.

\bigskip
\noindent {\bf  Example: $\mathbb{C}^3$}\\
First, we begin with the non-commutative topological vertex for $\mathbb{C}^3$. Since there is no wall crossing phenomena involved in this case \footnote{There is one wall between $R>0$ and $R<0$, however we do not discuss such a wall since we are specializing to the case $R>0$.}, the vertex should coincide with the ordinary topological vertex, thus providing a useful consistency check of our proposal.
For $\bC^3$, we have $L=1, \sigma(1/2)=-1, \theta=id$.

Take $\mathbb{C}^3$ with representation $\lambda=(\mu,\nu=\emptyset)$ with $\mu=(N,N-1,\ldots, 1)$ at one leg.
The method in \ref{app.proof} gives
$M=2$ (therefore $X'=\bC^2/\bZ_2\times \bC$), and 
\beq
\theta'(1/2)= 1/2-N,\quad  \theta'(3/2)=3/2+N, \quad  \sigma'=-1,
\eeq
and thus
\beq
\left[ B(\alpha_1) \right]=N.
\eeq
The weight is given by \eqref{conifoldwt1} or \eqref{conifoldwt2}.
By solving for 
\beq
q'{}_0^{\theta'}=q'{}_1^{\theta'}=q,
\eeq
we have 
\beq
q'{}_0=q^{-2N+1}, \quad q'{}_1=q^{2N+1}
\eeq
in the case of $N$ odd.
Substituting this into the closed BPS partition function
\beq
\prod_{n>0}(1-q'{}_0^n q'{}_1^n)^{-2n} 
\prod_{n>0}(1-q'{}_0^n q'{}_1^{n+1})^{-n} \prod_{n>N}(1-q'{}_0^n q'{}_1^{n-1})^{-n},
\eeq
we have
\beq
M(q)/ \prod_{n=1}^N (1-q^{2n-1})^{(N+1)-n}.
\label{egresult1}
\eeq
This coincides with the know expression for the topological vertex \cite{AKMV} \footnote{In the normalization of Ref.~\refcite{AKMV}, \eqref{egresult1} coincides with 
$M(q)\, q^{-\| \mu^{\tenchi} \| /2 }\, C_{\mu,\emptyset,\emptyset}$}. The case of $N$ even in similar.

\bigskip
\noindent {\bf Example: Resolved Conifold}\\
Now let us discuss the next simplest example, the resolved conifold.

Consider the representation $\mu=({\rm \tiny \yng(2,1)},{\rm \tiny \yng(1)})$, with 
\beq
\theta=id, \quad \sigma(1/2)=+, \quad \sigma(3/2)=-.
\eeq
In this case, the method in \ref{app.proof} gives 
\beq
\theta'(1/2)=-7/2,\quad \theta'(3/2)=-1/2,\quad \theta'(5/2)=11/2,\quad 
\theta'(7/2)=13/2
\eeq
with
\beq
\sigma'(1/2)=+,\quad \sigma'(3/2)=-,\quad \sigma'(5/2)=-,\quad \sigma'(7/2)=+.
\eeq
Then we have
\beq
\begin{split}\label{eq_Bforconifold}
& [B(\alpha_1)]=[B(\alpha_3)]=0,\quad 
[B(\alpha_2)]=[B(\alpha_2+\alpha_3)]=1,\\
&[B(\alpha_1+\alpha_2)]=[B(\alpha_1+\alpha_2+\alpha_3)]=2,  
\end{split}
\eeq
and 
\beq
\begin{split}\label{eq_qforconifold}
q'_0&=q_0^{-3}q_1^{-3}=q^{-3}, \quad
q'_1=q_0 q_1^{2}=q Q , \\
q'_2&=q_0^{3}q_1^{3}=q^3, \quad
q'_3=q_0=q Q^{-1}.
\end{split}
\eeq
The closed BPS partition function corresponding to the B-field \eqref{eq_Bforconifold} is the following:
\begin{align}
\prod_{n>0} (1-q'^n)^{-4n} 
\prod_{n>0} (1-q'^n Q'_1)^n \prod_{n>0} (1-q'^n Q'{}_1^{-1})^n
\prod_{n>0} (1-q'^n Q'_2)^{-n} \prod_{n>1} (1-q'^n Q'{}_2^{-1})^{-n}\notag\\
%
\prod_{n>0} (1-q'^n Q'_3)^n \prod_{n>0} (1-q'^n Q'{}_3^{-1})^n
\prod_{n>0} (1-q'^n Q'_1 Q'_2)^n \prod_{n>2} (1-q'^n Q'{}_1^{-1} Q'{}_2^{-1})^n
\notag\\
%
\prod_{n>0} (1-q'^n Q'_2 Q'_3)^n \prod_{n>1} (1-q'^n Q'{}_2^{-1} Q'{}_3^{-1})^n \notag\\
%
\prod_{n>0} (1-q'^n Q'_1 Q'_2 Q'_3)^{-n} \prod_{n>2} (1-q'^n Q'{}_1^{-1} Q'{}_2^{-1} Q'{}_3^{-1})^{-n},\label{eq_closedBPS}
\end{align}
where as in \eqref{chemicalweight} we defined
\beq
q'=q'_0 q'_1 q'_2 q'_3, ~\quad Q'_i=q'_i~ \quad (i=1,2,3).
\eeq

Substituting \eqref{eq_qforconifold} for \eqref{eq_closedBPS} under the identification \eqref{chemicalweight}, we obtain the open BPS partition function:
\beq
Z^c_{\mathrm{NCDT}}(q,Q)\cdot (1-q)^{-3}(1-q^3)^{-1}(1-Q)^2(1-q^2 Q)(1-q^2Q^{-1}),
\eeq
where 
\begin{align}
Z^c_{\mathrm{NCDT}}(q,Q):&=
\prod_{n>0} (1-q^n)^{-2n} 
\prod_{n>0} (1-q^n Q)^n 
\prod_{n>0} (1-q^n Q^{-1})^n \\
&=Z^c_{\mathrm{top}}(q,Q) Z^c_{\mathrm{top}}(q,Q^{-1}) .
\end{align}
This coincides with the expression computed from the result of Ref.~\refcite{IK}
\beq
M(q)^2 \prod_{n>0} (1-q^n Q) \frac{1}{(1-q)^3 (1-q^3)} (1-Q)^2 (1-Q q^2)(1-Q q^{-2})
\eeq
up to the wall crossing factor and the normalization by a monomial as remarked in footnote \ref{normalization}.

\subsubsection{Expression via Vertex Operators}\label{subsec.openvertex}

As we have seen the computation of our vertex from the relation \eqref{main2} is always possible but time-consuming in general. One efficient method is to use the vertex operator formalism, explained in Section\,\ref{sec.WCvertex}. By straightforward generalization of the argument there we can write down a rather general formula for the vertex, for all chambers and for all toric Calabi-Yau manifold without compact 4-cycles. 
Indeed, from the definition of the crystal it follows that \footnote{To be precise, we have to divide this result by an overall monomial when either $\lambda_+$ or $\lambda_-$ is non-trivial. See Proposition 3.15 of Ref.~\refcite{Nagao3}.}
\beq
C_{\type}(q_0,\ldots, q_{L-1})=\langle \lambda_+ | \ldots O(-1/2) O(1/2) O(3/2)\ldots
| \lambda_- \rangle,
\eeq
where 
$$
O(h):=\Gamma^{\sigma(\theta^{-1}(h))}_{\lambda(\theta^{-1}(h))} (q_{\theta^{-1}(h)}).
$$
The fact that \eqref{independent} is independent of the closed string moduli (i.e. the map $\theta$) can be verified from this expression. This result shows that wall crossing can be understood as the result of commuting vertex operators. We do not discuss the details here and interested readers are referred to Ref.~\refcite{Nagao3}.


\subsection{Open BPS Wall Crossing and M-theory}\label{sec.openMtheory}

In previous sections we discussed wall crossings of open BPS invariants with respect to the closed string moduli. However, we expect that there should be wall crossing associated with the open string moduli as well. Therefore, the question arises: At which values of the open string moduli is our vertex defined? How does the vertex change as we change the open string moduli? 
The answers to these question are discussed in the recent paper \cite{AY} \footnote{Related results on open BPS invariants have been obtained independently in Ref.~\refcite{DSV}. 
We thank R. Dijkgraaf, P. Sulkowski and C. Vafa for sharing this information with us.}, by generalizing the argument of Ref.~\refcite{AOVY} discussed in Section\,\ref{chap.M-theory} \footnote{Chamber dependence of open BPS states, in the context of surface operators of ${\cal N}=2$, $d=4$ gauge theories, was discussed recently in Ref.~\refcite{GaiottoSurface}. See also Refs.~\refcite{LeeYi,Krefl} for related discussions.}.
To explain this is the topic of this section.

Our main result in this section is stated as follows. Let us denote by  $Z_{\rm BPS}$ the BPS partition function including all BPS particles, open and closed. Then M-theory lift predict that the chamber dependence
of this is captured by the open+closed  partition function as
\beq\label{wc}
Z^{o+c}_{\rm BPS}(q,Q,v)=Z^{o+c}_{\rm top}(q,Q,v) Z^{o+c.*}_{\rm top}(q,Q^{-1},v^{-1}) \Big|_{\rm chamber},
\eeq
where $Z^{o+c.*}_{\rm top}(q,Q^{-1},v^{-1})$, whose closed string part coincides with $Z^c_{\rm top}(q,Q^{-1})$, will be defined shortly. 

We first begin in Section\,\ref{subsec.opentop} with the review the open topological string partition function, focusing on the integrality \cite{OVknot,LMV} and the connection with counting M2-branes ending on M5-branes. We also generalize the work of Ref.~\refcite{DVVafa} to open topological string and open BPS bound states.
Based on these results we derive in section \ref{sec.openwc} the wall crossing of the open BPS partition function. In the next section (Section\,\ref{sec.CV}) we discuss the relation to the work of Cecotti and Vafa.
\\

As before, consider adding $M$ D4 branes wrapping a special Lagrangian submanifold
$L$ of $X$. The D4 branes fill $\bR^{1,1}$ subspace of the flat space, and break half the supersymmetry of the Calabi-Yau. The theory on the branes thus has ${\cal N}=(2,2)$ supersymmetry in
two dimensions. We now get new kinds of BPS particles, corresponding to D2 branes wrapping disks and ending on the D4 brane \cite{OVknot}. These particles can form BPS bound states with the D6 brane and the closed D2 branes and the D0 branes we had before, and pin them to the D4 branes. These are the open BPS invariants that we would like to count \footnote{The counting of open BPS states was
also studied in Ref.~\refcite{ANV}. In that context, the instead of the D6 branes wrapping $X$ the authors had $N$ D4 branes on a divisor in $X$.}.

A D2 brane ending on a D4 brane is magnetically charged under 5d gauge field $A$ on the D4 brane, and electrically charged under the corresponding dual two-form $B$, $dA =*dB$. Suppose $b_1(L) = r$. We get $r$ $1-$cycles on $L$ that are contractible in $X$ and fill in to holomorphic disks in $X$. We will assume, for simplicity that $r=1$, tough the generalization to arbitrary $r$ is manifest. Integrating the B-field on the $S^1$ we get a  $U(1)^M$ magnetic gauge group generated by $\int_{S^1} B$. The D2 branes wrapping the holomorphic disks and ending on the D4 brane
are BPS particles charges under the magnetic $U(1)^M$. The $U(1)^M$ gauge fields in two dimensions sit in the twisted chiral multiplets $\Sigma_i$, $i=1, \ldots M$, whose lowest components $u_i$ enter the BPS masses of the particles charged under them\footnote{In fact, as we will need later, $u = \int_{\rm disk} k + i \int_{S^1} A$, where $A$ is the electric $U(1)$ on the D4 brane. We can write this, equivalently as $u = \int_{\rm disk} (k + iB_{NS}),$ since on the D4 brane $B_{NS} - dA$ is the gauge invariant combination.}. Since we are not interested in the gauge dynamics on the D4 branes, we will view the gauge symmetry as a global symmetry, and $\Sigma_i$ as background multiplets. The charges of a D2 brane under $U(1)^M$ keep track of which D4 branes the D2 brane has boundaries on, and how many times it wraps the corresponding $S^1$.
More precisely, since the $M$ D4 branes are identical, we have $S_M$ permutation invariance so the particles are representations $R$ of $U(1)^M/S_M$ (which can also be viewed as representations of $U(M)$).

When we lift this to M-theory on $X \times {\rm Taub} \hbox{-} {\rm NUT} \times \bR$, we now get an M5 brane wrapping the Lagrangian $L$
and filling $R^{2,1}$, where open D2 branes become M2 branes ending on the M5 brane. On the M5 branes, there is a $U(1)^M$ gauge theory, with ${\cal N}=2$ supersymmetry, where the open M2 branes are BPS particles.
Under the same assumptions as in the previous section (we will explain below why these are justified), the computation of BPS bound states in IIA reduces to computation of BPS degeneracies of a gas of free particles in M-theory, now living on the flat $R^{2,1}$ world volume of the M5 brane.
Below, we will show,  following  Ref.~\refcite{OVknot}, that the degeneracies of the spinning open M2 branes on $R^{2,1}$ are computed by the open topological
string, corresponding to A-model on $X$ with a Lagrangian $L$. To extract the degeneracies of the BPS bound state of one D6 brane with the open and closed D2 branes and D0 branes, we have to restrict the Fock space to those states preserving the same supersymmetry.

\subsubsection{Open M2 Branes and the Topological String}\label{subsec.opentop}

The BPS particles are now labeled by their $U(M)$ representation $R$, bulk class $\beta$, spin $s$ and $R$-charge $r$.
The presence of the M5 branes breaks the symmetries: $SO(4) = SU(2)_L\times SU(2)_R$ is now broken to $SO(2)_L\times SO(2)_R$: these now correspond to the little group of a particle in three dimensions and the R-symmetry of the ${\cal N}=2$ theory. More precisely $s_L = s+r, s_R = s-r$ where $s$ is the spin and $r$ is the R-charge. $s_R$ permutes the fields within the same multiplet, and $s_L$ annihilates the state.

We denote the number of M2-brane particles with $intrinsic$ spin $(s_L, s_R)$ and wrapping $\beta$ and representation $R$ by $N_{\beta, R}^{(s_L,s_R)}$, and define the index
$$
N_{\beta, R}^{s_L}=\sum_{s_R} (-1)^{s_R} N_{\beta, R}^{(s_L,s_R)}.
$$
Now, again, each such 3d particle gives rise to a field $\Phi$, and excitations of this field on $R^2$ are the particles we want to count
$$\Phi(z) = \sum_{n} \alpha_n z^n.
$$
Here we have set $z=z_1$, since M5 brane wraps the and $z_2=0$ subspace of $R^{4,1}$.

The partition function of these particles can be written as follows.  Introduce chemical potentials $q$, corresponding to the spin $s$, $Q$ corresponding to the charge $\beta$ and $v_i$
corresponding to the charges of the M2 branes under the $U(1)^M$.  For each field in representation $R, \beta, s$, we get a
contribution
$$
\prod_{{\vec k}_R}\prod_{n=1}^{\infty} (1- q^{s+n} Q^\beta  v^{\vec k_R})^{m_{{\vec k}_R}},
$$
where the product is over the weight vectors  ${\vec k}_R$ are of the representation $R$, ${m_{{\vec k}_R}}$ are the corresponding multiplicities and
$$v^{\vec k_R}=\prod_{i=1}^M v_i^{k_{R_i}}.$$
The full Fock-space partition function is
\beq
Z^c_{\rm Fock}=\prod_{\beta, s, R} \prod_{\vec{k}_R} \prod_{n=1}^{\infty}
(1-q^{s+n}v^{\vec k_R} Q^{\beta})^{{m_{{\vec k}_R}} N_{\beta,R}^s}.
\eeq
Note that, as in the closed string case, this should include M2 branes with both orientations. Flipping the orientation corresponds to sending $\beta$ to $-\beta$, $s$ to $-s$ and $R$ to ${\bar R}$, simultaneously, where CPT ensures
\beq
N_{\beta,R}^s = N_{-\beta,{\bar R}}^{-s}\, .\label{anti}
\eeq

As explained in Ref.~\refcite{OVknot}, the degeneracies of open M2 branes ending on an M5 brane wrapping $L$ are computed by the open topological string partition function in the presence of a D4 brane wrapping a Lagrangian $L$.
It was shown in Ref.~\refcite{OVknot} that the open topological string partition function has the following simple expansion, similar to the Gopakumar-Vafa expansion \cite{GV1,GV2} in the closed case:
\beq
Z^o_{\rm top}=\exp\left(
\sum_{d=1}^{\infty} \sum_R f_R(q^d,Q^d) \textrm{Tr}_R \frac{V^d}{d}
\right),
\label{OVexpansion}
\eeq
where
$$
f_R(q,Q)=\sum_{s,\beta>0} \frac{N_{R,\beta,s}}{q^{1/2}-q^{-1/2}} Q^{\beta} q^{s-{1\over 2}},
$$
and where  $N_{R,\beta,s}$ is an integer counting M2-brane particles with $R,\beta$ and $s$\footnote{We have shifted the definition of spin by $1/2$ relative to Ref.~\refcite{LMV}. In that paper, spins were taken to be half integral, while here we take them to be integral.}.
The symbol $\textrm{Tr}_R V$ denotes the holonomy of the gauge field on the D4-brane. Here $V$ captures the BPS masses of the open M2 branes, namely
$$
V=\textrm{diag}(v_1,v_2,\ldots,v_M).
$$
We can write
\beq
\textrm{Tr}_R V^d=\sum_{\vec{k}_R} {m_{{\vec k}_R}} \prod_i v_i^{d k_i^R},
\label{TrRV}
\eeq
where $k_i^R$ are the weights of the representation $R$ and $m_{{\vec k}_R}$ their multiplicities.
Then expression \eqref{OVexpansion} can be rewritten as the Fock space trace:
\beq
Z^o_{\rm top}=\prod_{n=0}^{\infty}\prod_{R,s,\beta>0} \prod_{\vec{k}_R} \left(1-q^{n+s} Q^{\beta} v^{\vec{k}^R} \right)^{{m_{{\vec k}_R}}N_{R,\beta,s}},
\eeq
where $v^{\vec{k}_R}:=\prod_i v_i^{\vec{k}^R_i}$.
This naturally corresponds to the half of the Fock space corresponding to M2 branes wrapping holomorphic curves only. In other words, we have
\beq
Z^o_{\rm Fock}=
Z^o_{\rm top}(q,Q,v) Z^{o,*}_{\rm top}(q,Q^{-1},v^{-1}),
\label{openZsquare}
\eeq
where $Z^{o,*}_{\rm top}$ contains the contributions of the anti-M2 branes,
\beq
Z^{o,*}_{\rm top}(q,Q^{-1},v^{-1})=\prod_{n=0}^{\infty}\prod_{R,s,\beta>0} \prod_{\vec{k}_R} \left(1-q^{n-s} Q^{-\beta} v^{-\vec{k}^R} \right)^{{m_{{\vec k}_R}}N_{R,\beta,s}},
\eeq
where we used the \eqref{anti} and the fact that $\textrm{Tr}_{{\bar R}}V = \textrm{Tr}_{R} V^{-1}$. Note that
$Z^{o,*}_{\rm top}(q,Q,v)$ is almost the same as $Z^{o}_{\rm top}(q,Q,v)$, the only difference being that the spin $s$ in the power of $q$ is replaced by $-s$.

\subsubsection{Open BPS Wall Crossing and M-theory}\label{sec.openwc}

Up to this point we discuss all M2-brane particles. Consider the central charges of the particles, which determine which ones are mutually BPS.
The central charge of the open D2 brane wrapping a holomorphic disk in class $k_R$, bulk class $\beta$ and D0 brane charge $n$
is given by\footnote{As in Ref.~\refcite{AOVY}, there is a complex proportionality constant between $R$ in the formulas here and the radius of the Taub-NUT; this is irrelevant for the discussion of wall crossing.}
\beq\label{ccnew}
Z(D2)=\left(u(k_R)+t(\beta) +n\right)/R,
\eeq
where
$$
t = \int_{S^2} i k + B,
$$
and
$$
u= \int_{\rm disk} ik + B.
$$
Here, $B$ is the NSNS B-field, and $k$ is the K\"ahler class and $R$ is the radius of the Taub-NUT defined above. The central charge receives contributions from the background twisted chiral multiplets where gauge fields and their superpartners reside (see, for example Ref.~\refcite{HananyHori}). The central charge is of the form $\sigma(q) = q \cdot \sigma$ where $q$ is the charge of the state under the gauge symmetry, and $\sigma$ the lowest component of the twisted chiral multiplet. The first term in \eqref{ccnew} comes from the $U(1)^M$ gauge symmetry on the D4 branes. The second two come from the bulk gauge fields, reduced to two dimensions\footnote{The 4d ${\cal N}=2$ bulk vector multiplet splits into two twisted chiral multiplets. These determines the 2d ${\cal N}=(2,2)$ central charges of the 4d electrically and magnetically charged states.  Namely, the twisted chiral multiplet coupling to electrons contains the the $d=4$ complex scalar, and the longitudinal components $A_{0,1}$ of the $d=4$ vector. The twisted chiral multiplet coupling to monopoles is slightly more complicated. Start with the chiral multiplet containing two transverse components of the gauge fields $A_{2,3}$. The dual twisted chiral multiplet determines the $d=2$ central charge of the magnetically charged particles, in particular of the D6 brane. We will need this in a later section.}.
To satisfy our Assumption 1 (page \pageref{Assumptions}), that all the central charges of D2 and D0 branes align, we had set the area of all the disks to zero and 2-cycles to be zero. This is possible, since $t$ and $u$ are  the moduli of our our theory in two dimensions, which we can dial at will.
The central charge is then given by
$$
Z(D2)=\left(B(k_R)+ B(\beta)+{n}\right)/R,
$$
in IIA, or equivalently
$$
Z(M2)=C(k_R)+C(\beta)+{n/R},
$$
in M-theory.
The BPS particles are particles with
$$
Z(D2)>0.
$$
This means that the complete open+closed BPS partition function is
$$
Z^{o+c}_{\rm BPS}(q,Q,v)=Z^{o+c}_{\rm Fock}(q,Q,v) \Big|_{Z(D2)>0},
$$
where the Fock-space partition function is computed by the open+closed topological string partition function,
$$
Z^{o+c}_{\rm BPS}(q,Q,v)=Z^{o+c}_{\rm top}(q,Q,v) Z^{*,o+c}_{\rm top}(q,Q^{-1},v^{-1})  \Big|_{Z(D2)>0}.
$$


\subsubsection{Examples}\label{sec.examplesNCTV}

In Section\,\ref{sec.def}, we discussed  a crystal melting model for the open BPS invariants, generalizing the previous results for the closed invariants explained in Section\,\ref{chap.crystal}. There the jump of the open BPS invariants under the change of closed string moduli was studied. However, it was not clear there at which value of the open string moduli the crystal computation corresponded to, or what happens as open string moduli are varied.
The partition function computed there had two properties: a) in the limit large
values of closed K\"ahler moduli it reduced to a single copy of open topological string, and b)
as the closed string moduli are varied, only the degeneracies of closed BPS states jumped.

From our results, property a) implies that BPS invariants studied in Ref.~\refcite{NY}
are in the special chambers where
$$R>0, \qquad u^i \to\infty,$$
where $u^i$ are as defined in section 3. This is because then, fixing $R$, and $u^i$ and taking $\textrm{Re}(t)$ to infinity one indeed recovers the topological string amplitude.
For example, when $0<\textrm{Re}(t)<1$, we have
$$
Z_{\rm BPS}=Z^{o+c}_{\rm top}(q,Q,v) Z^c(q,Q^{-1}).
$$
and when $\textrm{Re}(t)\to\infty$, we have
$$
Z_{\rm BPS}=Z^{o+c}_{\rm top}(q,Q,v).
$$
In this case, taking $\textrm{Re}(t)$ to infinity in addition, only M2 branes ever contribute, and no anti-M2 branes. This means that $Z_{\rm BPS}$ has the same open part as the topological string theory.

We can consider another extreme $0<\textrm{Re}(u^i)<1$. In this case, open part has contributions both from M2-branes and anti M2-branes. For example if $0<\textrm{Re}(t)< 1$
$$
Z_{\rm BPS}(q,Q,v)=Z^{o+c}_{\rm top}(q,Q,v) Z^{o+c,*}_{\rm top}(q,Q^{-1},v^{-1} ).
$$
We can also consider $R<0$. For example, there is a chamber $R<0, \textrm{Re}(u^i)\to\infty, \textrm{Re}(t)\to \infty$, where the partition function simply becomes one.


\subsection{Relation to the Work of Cecotti and Vafa}\label{sec.CV}

The wall crossing of the closed BPS invariants discussed in Chaps.\,\ref{chap.M-theory}, \ref{chap.WCF} is a special case of a more general problem, the wall crossing of BPS bound states in four dimensional theories with ${\cal N}=2$ supersymmetry.
In this context, Kontsevich and Soibelman \cite{KontsevichS} recently conjectured that the degeneracies on two sides of the wall are related by commuting certain symplectomorphisms of complex tori. In the special case of ${\cal N}=2$ gauge theories in four dimensions, this remarkable structure was explained from several different perspectives \cite{GMN1,GMN2,CecottiV}. In particular, in Ref.~\refcite{CecottiV} both the appearance, and the particular choice of symplectomorphism, was beautifully illuminated. In some cases, the statements that follow from Ref.~\refcite{KontsevichS} are particularly simple, and were predicted in physics literature \cite{DenefM} a while back. As we saw in Section\,\ref{chap.WCF}, when the state of charge $\gamma$
decays into two primitive states $\gamma_1$ and $\gamma_2 = \gamma-\gamma_1$,
$$
\gamma\; \rightarrow \; \gamma_1 + \; \gamma_2,
$$
the degeneracies of single particle states $\Omega(\gamma)$ jump\footnote{In what follows, we will be cavalier about whether we gain or lose the states in the jump, i.e. about the sign of $\Delta\Omega$, since that depends on the direction of crossing.} as \cite{KontsevichS,DenefM}
$$
\Omega(\gamma) \rightarrow \Omega(\gamma) + \langle\gamma_1, \gamma_2\rangle \;\Omega(\gamma_1)\;\Omega(\gamma_2).
$$
%
More generally, when the central charges of $\gamma_1$ and $\gamma_2$ align, they do so for any multiple of $\gamma_1$ and $\gamma_2$ as well, so one can a-priori lose an infinite number of states. While there are no known such simple formulas for the general case, Refs.~\refcite{KontsevichS,DenefM} predicted that in the semi-primitive case, where only one of the charges is primitive, the degeneracies $\Omega(\gamma_1+n \gamma_2)$
jump as
%
%
$$
Z(q) \rightarrow  Z(q)       \;\prod_{n=1}^{\infty}(1-q^n)^{-n \langle\gamma_1, \gamma_2\rangle\Omega(n\gamma_2)},
$$
where we defined
\beq\label{zq}
Z(q)=\sum_{n=0}^{\infty} \Omega(\gamma_1+n \gamma_2) q^n.
\eeq
As noted in Section\,\ref{chap.M-theory}, M-theory provides a derivation of this formula when $\gamma_2$ carries
D0 and D2 brane charges, and $\gamma_1$ in addition carries one unit of D6 brane charge. Note that the jumps are governed by a Mac-Mahon function type formula, $\prod_{n=1}^{\infty}(1-q^n)^{-n \omega(n)}$.

In the present context, we count open BPS states. We have seen that
M-theory predicts that the BPS states jump as
\beq\label{jump}
\Omega(\gamma) \rightarrow \Omega(\gamma) + \Omega(\gamma_1)\;\Omega(\gamma_2).
\eeq
in the primitive case, and more generally as,
\beq\label{jumptwo}
Z(q) \;\;\rightarrow \;\;Z(q) \; \prod_{n=1}^{\infty}(1-q^n)^{-\Omega(n\gamma_2)},
\eeq
in the semi-primitive case, where $Z(q)$ is as in \eqref{zq}. In other words, M-theory predicts that the open BPS states jumps are governed by eta-function $\prod_{n=1}^{\infty}(1-q^n)^{-1}$ type formulas.

In fact, this could have been anticipated. The open BPS states are BPS states in the ${\cal N}=(2,2)$ abelian gauge theory in two dimensions on the world-volume of the D4 branes. The open D2 branes (bound to the closed D0, D2 branes and the D6 brane), as discussed above, are charged particles in this two-dimensional theory.
The particles in two dimensions are solitons. This is so even for fundamental matter, simply because a particle is a codimension one object, so the vacuum of the theory can change from one side to the other. Soliton spectra of the massive ${\cal N}=(2,2)$ theories were studied in Refs.~\refcite{CFIV,CecottiVold}. The fact that the spectrum of BPS states can jump was in fact discovered in this context. We will now first review the essential results from Ref.~\refcite{CecottiVold}, and then explain how to apply them in the present context.

\subsubsection{Review of the Results by Cecotti and Vafa}
Consider a massive ${\cal N}=(2,2)$ theory.
BPS particles are solitons $\Delta_{ik}$ interpolating between the vacua $i$ and $k$ at spatial infinities. Let
$$
\mu_{ik}={\rm Tr}_{ik}  (-1)^F {\rm F}
$$
be the ``number''
of such solitons, 
or more precisely the index  that weights the solitons with their fermion number charge $F$. Only the BPS solitons contribute to the index. These live in short, two dimensional representations of the ${\cal N}=(2,2)$ supersymmetry, and the index is the same as counting the multiplets weighted by $(-1)^F$, where $F$ is the fermion number of the lowest component in the multiplet \cite{CecottiVold}.

The central charge, $Z(\Delta_{ik})$ of the soliton depends only on the vacua $i$ and $k$,
so we can define $W_{i}$, so that
$$Z(\Delta_{ik}) = W_i - W_k.
$$
If the theory at hand is a Landau-Ginsburg theory, then $W_i$ is the value of the superpotential $W$ in the $i$-th vacuum.
If, as we cross the line of
    marginal stability, $W_j$ crosses the line in the $W$-plane interpolating between $W_i$ and $W_k$, the soliton $\Delta_{ik}$ decays as
$$\Delta_{ik}\rightarrow \Delta_{ij}+\Delta_{jk},$$
and moreover the number of solitons jumps as
%
\beq\label{cvone}
\mu_{ik} \rightarrow \mu_{ik} + \mu_{ij}\mu_{jk}.
\eeq
More generally, vacua $j_1, j_2, \ldots j_N$ can cross the straight line connecting $W_i$ and $W_j$, where the order is set by the order of interception points. The state $\Delta_{ik}$ can have different decay channels, corresponding to all the different ways of getting from $i$ to $k$ while passing through the intermediate vacua.
The index in the $ik$ sector jumps as
\beq\label{cvtwo}
\mu_{ik} \rightarrow \mu_{ik} + \sum_{\stackrel{1\leq n \leq N}{1\:\leq s_1 <s_2 <\ldots< s_n \leq N}}
\mu_{ij_{s_1}}\mu_{j_{s_1},j_{s_2}}\ldots \mu_{j_{s_n}, k}.
\eeq
This has a simple interpretation. The jump corresponds to counting all the chains of BPS solitons interpolating from vacuum $i$ to vacuum $k$ via the intermediate vacua.
More precisely, we are counting the lowest components of these multiplets, weighted by $(-1)^F$. Since the net fermion number is simply the sum of the fermion numbers, the result is simply the product of the BPS degeneracies in the individual sectors of the chain.

\subsubsection{Counting of Open BPS States}

In the case at hand, we are counting massive BPS particles, which carry charges under the $U(1)$ gauge fields from the D4 brane and the bulk.
%
%
The corresponding central charge that enters the ${\cal N}=(2,2)$ supersymmetry algebra is
determined by the lowest components $\sigma_{\alpha}$ of the
twisted chiral multiplets $\Sigma_{\alpha}$\footnote{In this section, for simplicity of the notation, we use $\sigma$ to denote both open and closed twisted masses of the particles on the D4 branes. These include the central charges of D0, D2 and D6 branes, bound to open D2 branes. In addition, since we are not interested in the dynamics of the gauge theory on the D4 branes, but only in the BPS particle content, we have been viewing $\Sigma_{\alpha}$ as non-dynamical, so that the charges $q^{\alpha}$ are simply the global symmetry charges.}. 
For a particle $\gamma$
of charge $q^{\alpha}$, the central charge is
\beq\label{cc2}
Z(\gamma) = \sum_{\alpha} q^{\alpha}  \sigma_{\alpha}.
\eeq
Per definition then, the solitons of the theory are the fundamental
particles. Consider a vacuum of the theory $|i\rangle$. Adding a BPS
particle $\gamma$ to it changes the vacuum of the theory from
$|i\rangle$ on one side to $|j\rangle$ on the other where, again per definition, 
$$
W_j - W_i = Z(\gamma).
$$
We can think of the vacua as {\it labeled by the charges of the particles needed to create it}, from an arbitrary but fixed vacuum. Note that we have extra structure here, that is not present in an arbitrary massive ${\cal N}=(2,2)$ theory.  Namely, given any vacuum $|i'\rangle$, the BPS particle $\gamma$ takes it to a vacuum $|j'\rangle$ where $W_{j'}$ and $W_{i'}$ differ by the BPS mass of $\gamma$,  $W_{j'} - W_{i'} = Z(\gamma)$.

\begin{figure}[htbp]
\centering{\includegraphics[scale=0.35]{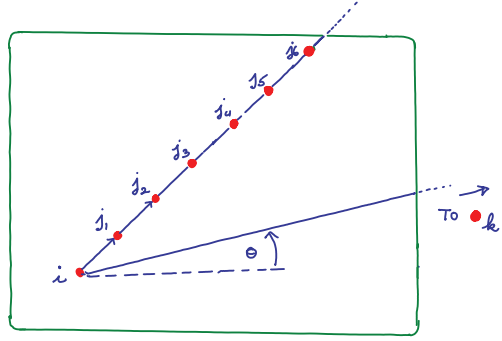}}
\caption[The value of $W$, shown in the complex plane.]{In the $W$-plane, all the $W_{j_i}$'s are on a single line emanating from $W_i$. $W_k$ is infinitely large, and we cut off the $W$-plane at large radius $\Lambda$, shown here in green.}
\label{W-plane}
\end{figure}

Among the particles in the theory is a D6 brane wrapping the whole non-compact Calabi-Yau. The focus of our paper has been understanding the wall crossing of particles carrying one unit of D6 brane charge. However, the corresponding central charge is strictly speaking infinite. If $|k\rangle$ is a vacuum with one unit of D6 brane charge, then $W_k$ is strictly at infinity in the $W$ plane.
To deal with this infinity we need a regulator. The correct way to do this, it turns
out, is to cut off the W-plane at some large but finite radius $\Lambda$. $\Delta_{ik}$ is then a non-compact cycle with a boundary. The relevant data that affects the problem is only the angle $\theta$ that $W_k-W_i$ makes with the real axis in the $W$-plane\footnote{This parallels the treatment in Ref.~\refcite{JM} where the wall crossing of closed BPS bound states corresponding to D6 branes on non-compact Calabi-Yau was studied.} (see Fig. \ref{W-plane}). Moreover, it is easy to see that $\theta$ is independent of any additional D2 or D0 brane charges that the D6 brane may carry, since the central charges of these are finite. So, any soliton $\Delta_{jk}$
where $W_j$ is in $|W| \leq \Lambda$, corresponds to a state with the same value of $\theta$, but whose central charge $Z(\Delta_{jk})$ differs from $Z(\Delta_{ik})$ by a finite amount $W_j-W_i$.

Fix now $\gamma = \Delta_{ik}$, corresponding to one unit of D6 brane charge, and some D2 brane and D0 brane charges. The degeneracy $\mu_{ik}$ of domain walls interpolating between the vacua $i$ and $k$ is simply the degeneracy of the 1-particle states with charge $\gamma$,
$$\mu_{ik} = \Omega(\gamma).$$
As we vary the moduli, suppose vacuum $j$ inside the cutoff $W$-plane crosses the straight line between $i$ and $k$. This then implies that corresponding state $\gamma_2$ is a single particle state, with some D0 and D2 brane charges, and no D6 brane charge. So, $\Delta_{ij} = \gamma_2$, and $\Delta_{jk} = \gamma-\gamma_2=\gamma_1$ carries one unit of D6 brane charge. The soliton degeneracies are simply the degeneracies of the corresponding one-particle states, so
$$\mu_{ij} = \Omega(\gamma_2),\qquad\mu_{jk} = \Omega(\gamma_1),$$
and the jump \eqref{cvtwo} in the degeneracies of $\gamma$ predicted from Ref.~\refcite{CecottiVold} agrees with \eqref{jump}.

Moreover, if the vacuum $|j\rangle$ corresponding to $\gamma_2$ crosses the $ik$ line in the $W$ plane, a large number of other vacua will cross as well.
These are vacua which can be obtained from $|i\rangle$ by adding $n$ $\gamma_2$ particles. They are all collinear, located at points $W_{j_n} = W_i + n Z(\gamma_2)$. This holds for any $n$, as long as $W_{j_n}$ is inside the cut-off W-plane. These additional critical points lead to more complicated decays.
Let's fix $\gamma_1=\Delta_{j_n k}$, the decay product with one unit of D6 brane charge, and consider all the different ways that we can obtain it from $\gamma = \Delta_{ik}$. There is is a two-soliton decay,
$$
\Delta_{ik} \rightarrow \Delta_{i j_n}+\Delta_{j_n k}.
$$
where $\gamma=\gamma_1+n \gamma_2$
decays into $\gamma_1$ and $n \gamma_2 =\Delta_{i j_n}$
This, according to Ref.~\refcite{CecottiVold} corresponds to a change in the degeneracies
$$
\mu_{ik} \rightarrow \mu_{ik} + \mu_{ij_n}\mu_{j_nk},
$$
or
$$
\Omega(\gamma)  \rightarrow \Omega(\gamma)+ \Omega(\gamma_1) \Omega(n\gamma_2).
$$
This is exactly as predicted from M-theory.

There are also channels where $n\gamma_2$ is split into more particles, each carrying some multiple of $\gamma_2$ charge. Here we need to be careful. Much of the discussion of Ref.~\refcite{CecottiVold} assumes that no three vacua are collinear in the $W$-plane\footnote{If collinearity is coincidental, one can remove it without loss of generalities, simply by displacing the vacua slightly. However in the present case it is physical, and there is no small deformation of the theory that can remove it.}. Here, we are in the opposite regime, of a large number of collinear vacua. In this regime, a very naive application of \eqref{cvtwo} overcounts the change in the soliton number.

Recall that, from \eqref{cvtwo}, the number of solitons gained or lost, is $\Omega(\gamma_1)$ times the number of ways of assembling a chain of solitons of total charge $n\gamma_2$ out of the available $\Omega(k \gamma_2)$ one particle BPS states particles of charge $k\gamma_2$, where $k=1,2 \ldots $. In counting the number of possibilities, the order of in which we string the $\gamma_2$-solitons in a chain would matter. When the vacua are not collinear, the order indeed matters. However here this manner of counting would lead to a contradiction.
Consider example, in a decay channel where $\gamma$ splits into $\gamma_1$ and $n$ copies of $\gamma_2$,
$$
\Delta_{ik} \rightarrow  \Delta_{ij_1}+\Delta_{j_1 j_2}+\ldots +\Delta_{j_nk},
$$
The $n$ $\gamma_2$ particles are all mutually BPS, so in counting the number of such $n$ particle states, we can simply use the free Fock-space, generated by $1$ particle states. The dimension of the corresponding $n$-particle Hilbert space is not $\Omega(\gamma_2)^n$, as the naive application of Ref.~\refcite{CecottiVold} would suggest, but $\Omega(\gamma_2) (\Omega(\gamma_2)+1)\ldots
(\Omega(\gamma_2)+n-1)/n!.$ The number of states we gain or lose is just this times $\Omega(\gamma_1)$. One should in principle be able to verify this by a careful computation of $n$-soliton contributions to the index in Ref.~\refcite{CFIV,CecottiVold}. More generally,
we can consider more complicated splits, where $n$ units of $\gamma_2$ charge are split among $m_{\ell}$ particles of charge $\ell \gamma_2$, with $\sum_{\ell} m_{\ell} \ell = n$. The number of states is just the product of the dimensions of $m_{\ell}$ particle Hilbert spaces. 
Of course, this is precisely the counting predicted from M-theory.

.

\section{Discussion and Future Directions}\label{chap.discussion}

In this paper we have seen that many problems in the BPS state counting and wall crossing phenomena have natural and beautiful solutions, which are obtained by rapid developments in this field over the last a few years. However, there are many interesting and important topics which require further exploration. Of course, many of the problems are interrelated, and the solution of one problem is likely to shed light on other problems as well.

\bigskip

\begin{list}{\labelitemi}{\leftmargin=2em}

\item The formula we proved in Section\,\ref{chap.M-theory}
\beq
Z_{\rm BPS}=Z_{\rm top}^2 \Big| _{\rm chamber},
\label{Z^2}
\eeq
is highly reminiscent of the Ooguri-Strominger-Vafa relation \cite{OSV}
\beq
Z_{\rm BPS}=|Z_{\rm top}|^2.
\eeq
It would be interesting to clarify the relation between them. Of course, one should keep in mind that there are important differences between the two. First, the former relation is verified only for the Calabi-Yau manifold without compact 4-cycles,
 whereas the latter is believed to hold for any Calabi-Yau. Second, the first relation applies to any chamber, while the second holds only in a special chamber of the moduli space. In particular, as emphasized in Ref.~\refcite{DenefM}, we need to specify the chamber of $Z_{\rm BPS}$ in order to even formulate OSV conjecture. Third, the arguments of $Z_{\rm top}$ are different in the two relations.
Finally, we are typically considering different set of D-brane charges; OSV is mostly for D4/D2/D0, whereas \eqref{Z^2} is currently only for D6/D2/D0.

\item It would be interesting to extend our analysis to a Calabi-Yau manifold with compact 4-cycles and with multiple D6-branes. As we have seen, it seems to be rather non-trivial to extend the arguments in Section\,\ref{chap.M-theory} and Section\,\ref{chap.WCF} to these situations. This is an important problem since Calabi-Yau geometry with compact 4-cycles frequently appears in geometric engineering and in black hole physics, and comparison with semiclassical entropy formula of black holes (e.g. BMPV black hole \cite{BMPV}) often requires large D6-brane charges. See Refs.~\cite{TodaD6,Stoppa,CDP} for recent discussion of multiple D6-brane charges.

\hspace{3mm}
One should remember that in simple examples, the answer is given by using the non-commutative topological vertex in Section\,\ref{sec.def}. 
For example, if we have a canonical bundle over $\bP^1\times \bP^1$ and if the B-field through one of the $\bP^1$'s is very large, we can divide the toric diagram into two along the $\bP^1$ and glue the two non-commutative vertices by a propagator as in the usual topological vertex. However, this simply gives the topological string partition function of $\bP^1\times \bP^1$ times the wall crossing factor for $\bC^3/\bZ_2$. 
The interesting question is what happens when the B-fields through both $\bP^1$'s become small.

\hspace{3mm}
One should also remember that crystal melting model we defined in Section\,\ref{chap.crystal} can be defined for toric Calabi-Yau manifolds with compact 4-cycles. The only question is how that partition function is related to the topological string theory, where we have fewer parameters. 

\hspace{3mm}
Further generalization is to consider compact Calabi-Yau manifolds. This will lead us back to the original motivation in the introduction. This is a notoriously difficult problem in general, but the recent results on the computation of topological string partition function on a some compact example up to high genus \cite{HKQ} will be helpful for this purpose.

\item It is an interesting problem to generalize our crystal melting model to string theory compactifications other than Calabi-Yau 3-folds. In principle, when we compactify string theory on a non-compact manifold preserving $\scN\ge 2$ supersymmetry and we have BPS particle of D-branes, the computation of BPS index should reduce to the computation of the Witten index of the supersymmetric quantum mechanics arising as the low energy effective field theory of D-branes. The possible drawback is that in general the connection with topological string theory will be lost. Moreover, identifying the precise form of the quantum mechanics, especially the superpotential, is a difficult problem in general. However, a technique generalizing the brane tiling method in Section\,\ref{sec.quiver} is available for M2-branes probing Calabi-Yau 4-fold singularities \footnote{See Refs.~\refcite{HananyM2-1,HananyM2-2,HananyM2-3,MartelliS1,MartelliS2,UYM2,IKcrystal} for some early papers. Ref.~\refcite{Aganagic} also discuss $G_2$ holonomy manifolds.}, giving rise to a 3d Chern-Simons-matter theories generalizing the ABJM theory \cite{ABJM}. We should also note that there is a quiver description for compact examples, such as quintic \cite{StromingerQuintic}.

\item 
It would be an important problem to check our prediction for non-toric examples, for example the one in Section\,\ref{subsec.nontoric}. Recently the paper Ref.~\refcite{Gholampour} studies examples $\bC^2/\gamma\times \bC$, where $\gamma$ is a discrete subgroup of $SU(2)$. The subgroup $\gamma$ has a ADE classification, and even for D and E type singularities they proved the relation
\beq
Z_{\rm NCDT}(q,Q)=Z_{\rm top}(q,Q) Z_{\rm top}(q,Q^{-1}),
\eeq
which is consistent with the results of Section\,\ref{chap.M-theory}.

\item In our discussion of wall crossing from wall crossing formulas in Section\,\ref{chap.WCF}, semi-primitive wall crossing formula is sufficient since the Calabi-Yau manifold is non-compact and we have only a single D6-brane.
However, if we want to discuss multiple D6-branes on compact Calabi-Yau manifold, we need the general Kontsevich-Soibelman wall crossing formula. In these case the simple relation like \eqref{Z^2} would presumably not hold, however it is interesting to see if there is still a connection of the BPS state counting problem with topological string theory. The work of Ref.~\refcite{CecottiV} is interesting from this viewpoint. It is also important to physically understand the Kontsevich-Soibelman formula, see Refs.~\cite{GMN1,GMN2}.

\item It would be interesting to see if there is a generalization of Gopakumar-Vafa large $N$ duality \cite{GVdual} including the background dependence, and whether there is a Chern-Simons interpretation of the wall crossing phenomena (see the recent paper Ref.~\refcite{CecottiV} for related ideas). This will potentially lead to new persepctives in knot theory.

\item Similarities and differences between wall crossings in $\scN=2$ and $\scN=4$ theories are interesting problems to pursue. For example in $\scN=4$ dyon counting in type II compactification on $K3\times \bT^2$, there is a universal formula for the degeneracy \cite{DVVerlinde,SSY}, stating the BPS degeneracy $\Omega(P,Q)$ is given by 
\beq
(-1)^{P\cdot Q+1}\Omega(P,Q)=\oint dM \frac{e^{\pi i (P^2\rho+Q^2 \sigma-2P\cdot Q \nu)}}{\Phi(M)}, \quad 
  M=\left(
\begin{array}{cc}
        \sigma & \nu\\
	\nu & \rho
\end{array}
\right).
\eeq
Here $P,Q$ are magnetic and electric charges, $\Phi(M)$ is unique weight 10 Igusa cusp form of $Sp(2,\bZ)$.
Moreover, the wall crossing phenomena is simply interpreted as the change of the choice of the contour \cite{SenWMS,ChengVdying}.
 It is an interesting problem to see if we can write the $\scN=2$ BPS partition function in a similar form, namely a universal expression with the moduli dependent choice of contour. The cusp form $1/\Phi$ can be written as a square of the character of the Borcherds-Kac-Moody algebra, which
is highly reminiscent of our expression $Z_{\rm top}^2$ in \eqref{Z^2}.

\item A related question is to identify the BPS state algebra \cite{HarveyM1,HarveyM2} for the $\scN=2$ theories discussed in this paper. In the $\scN=4$ case this is given by a Borcherds-Kac-Moody algebra \cite{ChengV,CD}, and for ALE spaces corresponding to ADE affine Dynkin diagrams, by an affine Kac-Moody algebra \cite{Nakajima}. Mathematically, the BPS state algebra is believed to be captured by the the motivic Hall algebra studied in Ref.~\refcite{KontsevichS}. See Ref.~\refcite{KontsevichS3} for more along these lines.

\item The D6/D2/D0 systems studied in this paper is T-dual to the D5/D1/D(-1) system, discussed in a series of papers \cite{Uranga1,Uranga2,Collinucci09}.
Euclidean D-brane instantons (D1/D(-1)-instantons in this example) produce non-perturbative terms, which are 
of phenomenological interest since they can generate perturbatively forbidden couplings \cite{BlumenhagenCW,IbanezU,Florea} and have applications to moduli stabilization \cite{KKLT}. It was noticed  in particular in Ref.~\refcite{Collinucci09} that topological string theory lies behind the continuity of the non-perturbative term across walls of marginal stability, which is similar in spirit to the work of Ref.~\refcite{AOVY} explained in Section\,\ref{chap.M-theory}. Since there has recently been significant developments in the understanding of D-brane instantons, these works can provide a different perspective on the problem discussed in this paper, as well as phenomenological motivations.

\item 
Crystals similar to the one discussed in this paper appear in the context of counting gauge invariant operators in four-dimensional $\scN=1$ supersymmetric quiver gauge theories (see for example Refs.~\refcite{FHZ,ButtiCounting}) \footnote{This counting problem is also related to the index of 4d $\scN=1$ superconformal field theories \cite{4dindex}. }. The difference is that they only consider gauge invariant operators, whereas in our construction of the crystal we used a path algebra, an algebra consisting of all possible operators (regardless of whether gauge invariant or not). This means that our crystal contains the crystal of Ref.~\refcite{FHZ} as a subset (sublayers). Also, in the definition of the partition function we are using a set of atoms (satisfying the melting rule) , whereas Ref.~\refcite{FHZ} counts the atoms themselves. This is attributed to the difference between the Euler character and the Hilbert series of the moduli space.

\hspace{3mm}
Despite these differences, there are some similarities and it would be interesting to understand the relation better. For example, the paper Ref.~\refcite{FHZ} observes that their crystal changes as they change the value of the B-field and go through the flop transition. This is part of the chamber structure of the K\"ahler moduli space. Also, the counting of gauge invariants operators in toric examples reduces to counting of the lattice points in the convex polytope (image of the moment map), which is in the same spirit of the discretization of space we saw in Section\,\ref{chap.thermodynamic}.

\item As mentioned in Section\,\ref{chap.crystal}, the atomic structure we have discovered would have possible implications for quantum aspects of black hole physics.  Such discretized structures are common in geometric quantizations, and is reminiscent of the bubbling geometry \cite{LLM} and Fuzzball conjecture \cite{Mathur}

\item Fluctuation of the dimer model from the thermodynamic limit is an interesting problem. In Section\,\ref{chap.thermodynamic}, we have seen that the thermodynamic limit of the crystal partition function gives a genus $0$ topological string partition function, and the dependence on the wall crossing disappears in the thermodynamic limit. By studying the fluctuation of the crystal from the limit shape, we should be able to see the higher genus topological string partition function as well as chamber dependence of BPS invariants \footnote{See Ref.~\refcite{OkounkovNCG} for a recent attempt.}. This should be a doable problem since the dimer model is a kind of an exactly solvable model whose correlation function is known \cite{OR,KOS} in the mathematics literature.
The higher genus contributions are also captured by the matrix model (``remodeling'' the B-model) \cite{EynardO,BKMP}, and we expect a more direct connection between the dimer fluctuations and the matrix model. See Refs.~\refcite{OSY,Sulkowski2} for recent discussion.


\item In this paper we only discussed the classical version of the crystal melting model, or equivalently the dimer model. In the literature, there is a quantum version of the dimer model \cite{RK}, which originally arises from the resonating valence bond models for superconductivity and is known to have an interesting phase structure. There is a speculation that quantum dimer model is related to the topological M-theory \cite{topologicalM}. See Ref.~\refcite{DOR} for recent discussions.

\item Many of the four-dimensional $\scN=2$ theories arise from six dimensions, by wrapping M5-brane on a punctured Riemann surface. These theories have recently attracted considerable attention due to the work of Ref.~\refcite{GaiottoN2}, which generalizes the result of Ref.~\refcite{WittenSolutions} and proposes a new family of interacting 4d conformal field theories. Furthermore, a new connection between 4d gauge theories and 2d Liouville theory on the Riemann surface was proposed \cite{AGT}. It would be interesting to explore the implications of these results to the results presented in this paper. See the works Refs.~\refcite{GMN2,GaiottoSurface} for discussion along these lines.

\end{list}

The list can go on, but let us stop here. We hope that future developments will solve some of these problems, and will ultimately lead to a better understanding of string theory as a theory of quantum gravity.


\thispagestyle{empty}

\section*{Acknowledgments}



This research project would not have been possible without the help, support and inspiration I obtained from many people.

First and foremost, it is a pleasure to express my gratitude to Prof.\ Hirosi Ooguri. He was my adviser during the last two years of my Ph.D. course at IPMU and Caltech, and this paper owes much to the discussions and collaborations with him. I learned a lot from him, and he offered me invaluable assistance, guidance and support.

I owe my deepest gratitude to my adviser Prof.\ Tsutomu Yanagida.
During my stay at the University of Tokyo I often enjoyed the discussion with him about physics, which was always inspiring and suggestive to me.
I also had a wonderful opportunity to collaborate with him.

I am also deeply indebted to Prof.\ Tohru Eguchi, who is my mentor since my Master course study at the University of Tokyo. The research on brane tilings and quiver gauge theories, which I have undertaken under his supervision, provides the  basic ground which lead me to the study described in this paper.

Special thank go to my collaborators: Mina Aganagic, Kentaro Nagao, Cumrun Vafa. It was a wonderful and an exciting experience to collaborate with you. I also obtained valuable feedback from the discussion with Alexei Borodin, Tudor Dimofte, Lotte Hollands, Sergei Gukov, Daniel Krefl, Andrei Okounkov, Kazushi Ueda, Jaewon Song, Piotr Su{\l}kowski, Yukinobu Toda.

I would also like to express my gratitude to the Ph.D. paper committee, especially Yoichi Kazama, for criticism and comments on the manuscript which is partly incorporated into this final version.

I am indebted to my colleagues at the University of Tokyo, both at the Department of Physics and the Institute for the Physics and Mathematics of the Universe. I would also like to thank the high energy theory group at Caltech for hospitality, where I enjoyed my one year stay in total. I would also like to thank the secretaries, Mami Hara at Tokyo and Carol Silberstein at Caltech.

The contents in this paper are presented at numerous universities and workshops I attended. I convey thanks to all the participants for their interest, questions and feedback which greatly help to improve this paper.

During my Ph.D study I have been supported by the JSPS fellowships for Young Scientists, by the Global COE Program for Physical Sciences Frontier at 
the University of Tokyo funded by MEXT of Japan, by DOE grant DE-FG03-92-ER40701 and by the World Premier International Research Center Initiative of MEXT of Japan. I would like thank these funding agencies for support.

Last but not least, I wish to thank my parents for their understanding,
love and support.

\bigskip
{\it Note added for the journal version}: I would like to thank the anonymous referee for careful reading of the manuscript and 
 for pointing out numerous typos and mistakes in the original version of this paper.

\newpage



\addtocontents{toc}{\vspace{2em}} 

\appendix

\section{Appendices}\label{chap.appendix}

\subsection{Perfect Matchings}\label{app.PM}
In this appendix we are going to explain the one-to-one correspondence between a molten crystal discussed in Section\,\ref{chap.crystal} and a perfect matching of the bipartite graph. 
This means that the problem of counting BPS states can also be reformulated as a problem of counting  perfect matchings of the bipartite graph \footnote{A bipartite graph is a graph with vertices colored either black or white and edges connecting vertices of different colors. A perfect matching of a bipartite graph is a subset of edges of the bipartite graph such that each vertex is contained exactly once. See Ref.~\refcite{Kenyon} for an introduction to the dimer model.}.
Such a description is useful for the discussion of thermodynamic limit in Section\,\ref{chap.thermodynamic}.
The contents of this appendix is basically a recapitulation of Ref.~\refcite{MR}.

In Section\,\ref{crystal.subsec} we considered a quiver $\tilde{Q}=(\tilde{Q}_0,\tilde{Q}_1)$, which is a universal cover of the quiver $Q$ on $\bT^2$. The dual graph of $\tilde{Q}$, which we denote by $\tilde{\gamma}$, can be made bipartite using 
orientation of arrows of $\tilde{Q}$, and is a universal cover of 
the bipartite graph $\gamma$ on $\bT^2$ described in Section\,\ref{tiling.sec}.
What we are going to do is to give an explicit correspondence between a perfect matching of the bipartite graph 
$\tilde{\gamma}$ and a configuration of molten crystal.

\begin{figure}[htbp]
\includegraphics[scale=0.14]{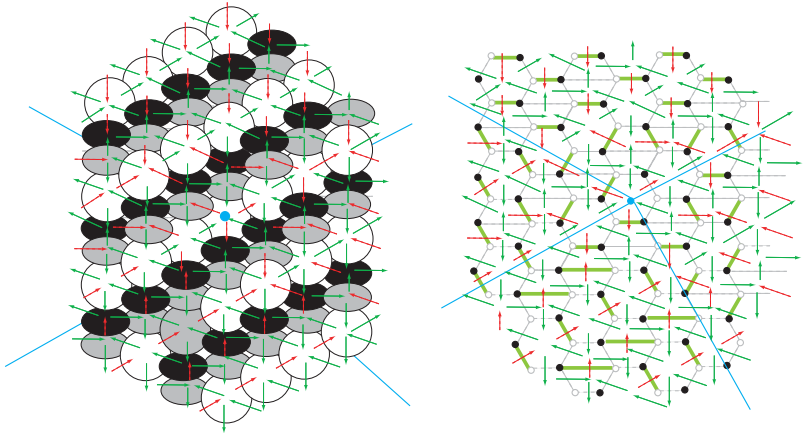}
\caption[A molten crystal is equivalent to a perfect matching.]{Given a configuration of a molten crystal, we can construct a perfect matching of the bipartite graph. Each arrow is colored green if it is along the surface of the crystal, and red otherwise. The set of dual of arrows colored red gives a perfect matching of the bipartite graph.}
\label{fig.SPPmodule}
\end{figure}

We first construct a perfect matching from a molten crystal.
Given a molten crystal as shown in Fig.\,\ref{fig.SPPmodule}, choose all the arrows of $\tilde{Q}$ which are along the surface of the crystal. In the example of Fig.\,\ref{fig.SPPmodule}, 
such arrows are colored green in Fig.\,\ref{fig.SPPmodule}, 
while the remaining arrows are colored red. 
Take the set of the dual of edges colored red. 
It is proven by Ref.~\refcite{MR} that such a subset of edges of $\tilde{\gamma}$ is a perfect matching. This is the perfect matching we wanted to construct.

In the case when no atoms are removed from the crystal, the perfect matching obtained by this method is called the canonical perfect matching, which we denote by $m_0$. Since 
only a finite number of atoms are removed from the crystal, 
the perfect matching obtained from a molten crystal by the above method coincides with $m_0$ when sufficiently away from the reference point $i_0$.

\begin{figure}[htbp]
\centering{\includegraphics[scale=0.14]{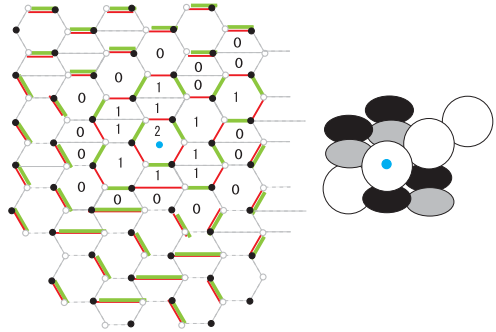}}
\caption[The height function for a perfect matching.]{By superimposing a perfect matching of Fig.\,\ref{fig.SPPmodule} with the canonical perfect matching shown later in Fig.\,\ref{fig.SPPempty}, we have a  set of loops, which defines a height function $h_m$. From this function we can recover a molten crystal.}
\label{fig.SPPheight}
\end{figure}

Conversely, given a perfect matching $m$ which coincides with $m_0$ when sufficiently away from the reference point $i_0$, we can reproduce a molten crystal.
Let us superimpose $m$ with $m_0$, and we have a finite number of loops, as shown in Fig.\,\ref{fig.SPPheight} in the case of Suspended Pinched Point. 
Define a height function $h_m$ such that 

\medskip
(1) $h_m(j)=0$ when sufficiently away from $i_0$.

\smallskip
(2) $h_m$ increases by one whenever we cross the loop and go inside it.
\medskip

\noindent The example of $h_m$ for the case of Suspended Pinched Point is shown in Fig.\,\ref{fig.SPPheight}.

By removing $h_m(j)$ atoms from each $j\in \tilde{Q}_0$, we can construct a molten crystal. 
It was proven in Ref.~\refcite{MR} that the set of atoms removed from the crystal so defined satisfies the melting rule of Section\,\ref{molten.subsec}. 
This establishes the one-to-one correspondence between a molten crystal and a perfect matching of the bipartite graph, meaning that BPS states can also be counted by perfect matchings of the bipartite graph $\tilde{\gamma}$.

\begin{figure}[htbp]
\includegraphics[scale=0.15]{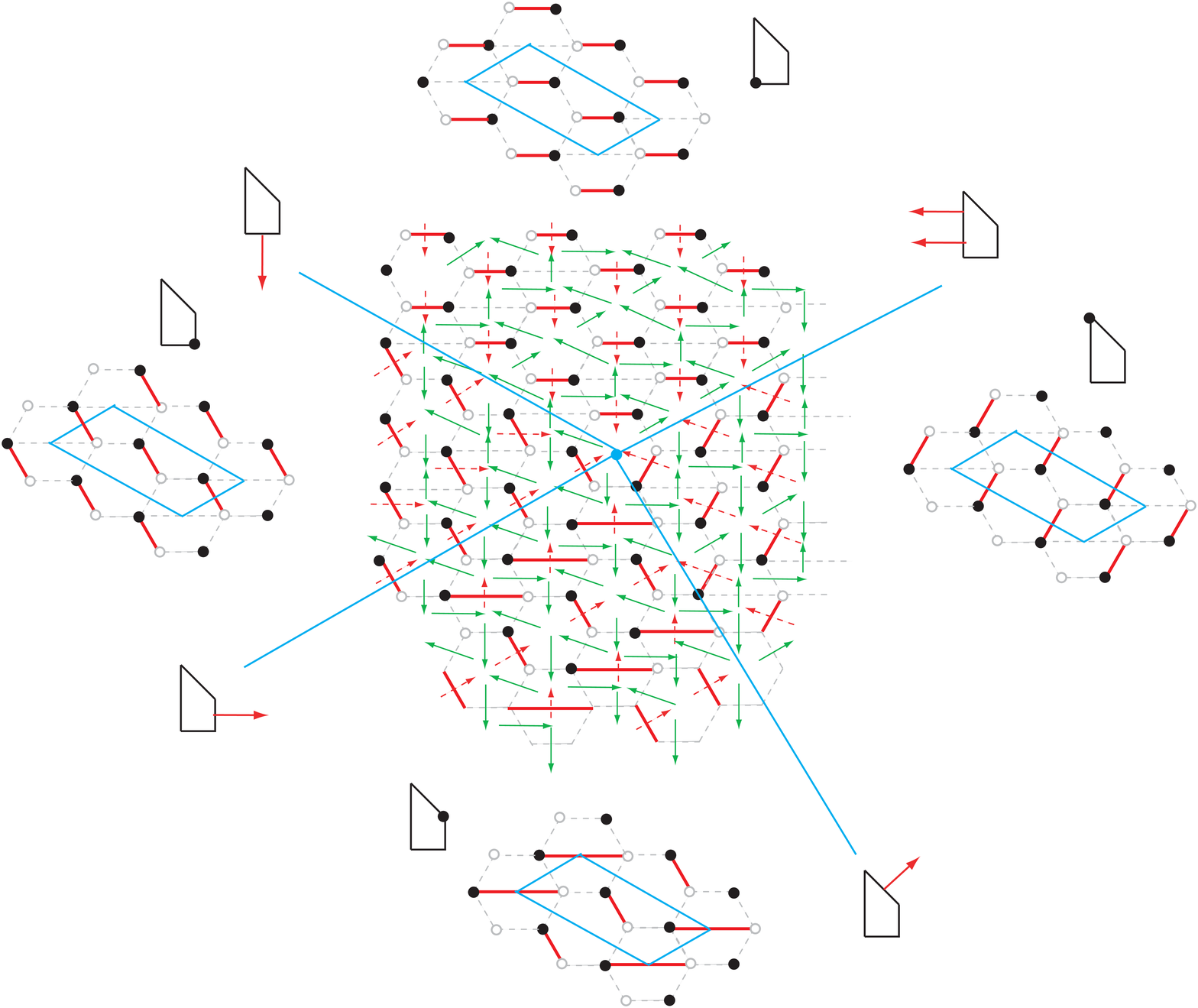}
\caption[An example of the canonical perfect matching.]{The canonical perfect matching of the bipartite graph for
the Suspended Pinched Point singularity. Asymptotically, 
the perfect matching corresponds to one of the
 four perfect matching of the bipartite graph corresponding 
to vertices of the toric diagram. The blue borders between 
different choices of perfect matchings
represent the $(p,q)$-web.}
\label{fig.SPPempty}
\end{figure}

Finally, let us finish this appendix by pointing out  an interesting connection
 of the canonical perfect matching $m_0$  with toric geometry.
The example of 
 canonical perfect matching $m_0$ for the Suspended Pinched Point is shown in Fig.\,\ref{fig.SPPempty}. 
In this example, the asymptotic form of the bipartite graph 
has four different patterns. 
Each of four patterns is periodic and therefore an be thought of as a perfect matching of the bipartite graph on $\bT^2$.
 In the brane tiling literature, 
a perfect matching on the bipartite graph on $\bT^2$ is 
known to correspond to one of the lattice points of the 
toric diagram \cite{BT1,FV} \footnote{For this correspondence, 
we consider superimposition of perfect matchings and 
define a $\bZ_2$-valued height function, which is similar
to the height function $h_m$ defined previously.}. 
We recognize that
the four perfect matchings are identified with the four corners
of the toric diagram in Fig.\,\ref{fig.SPPtiling}-(a) and that 
the borders between different patterns are identified with 
the blue lines in Fig.\,\ref{fig.SPPtiling}-(b), which makes 
the $(p,q)$-web of the diagram.

In general, for an arbitrary toric Calabi-Yau manifold, we can use the same 
pattern to construct a perfect matching.
Divide the universal covering of the bipartite graph
into segments separated by the $(p,q)$-web of the toric 
diagram \footnote{We choose the diagram that corresponds to 
the most singular Calabi-Yau manifold.}.
The perfect matching in each segment is periodic and 
is identified with one of the perfect matchings of bipartite 
graphs on $\bT^2$, which corresponds to one of the lattice points of the 
toric diagram and 
the lattice point in question is precisely the vertex 
surrounded by the two $(p,q)$-webs on $\bT^2$ \footnote{In a consistent quiver gauge theory, 
it is believed that the multiplicity of perfect matchings 
at the vertices of the toric diagram is one \cite{BT4}.}.
This determines a perfect matching. In particular,
this means that the ridges of the crystal line up with the
$(p,q)$-web of the toric diagram. 

This pattern holds for all toric Calabi-Yau manifolds without compact 4-cycles, and we conjecture that
the perfect matching constructed in this way is canonical. 
We would like to stress again that this conjecture is not
needed to construct the crystal melting model. 
Here we are simply pointing out that, in the examples
we have studied, the crystalline structures fit beautifully 
with the corresponding toric geometries.

\subsection{Multi-Centered Black Holes}\label{app.multi}
In this appendix we briefly summarize the multi-centered black hole solutions
 discovered in Refs.~\refcite{DenefFlow,DenefCorrespondence,DGR,BatesD}.

Why do we need to bother about these black holes? In many examples in the literature, including the Strominger-Vafa paper \cite{StromingerVafa}, the D-brane configurations correspond to single centered black holes. This is not always the case \cite{DenefFlow}, however, and the BPS index for the D-branes contain contribution from multi-centered black holes in general \cite{DenefMessay}. In these cases, we need to discuss multi-centered black holes in order to correctly account for the BPS index of the D-branes.
Indeed, most of the BPS states discussed in this paper correspond to multi-centered black holes. Moreover, the supergravity viewpoint provides an intuitive description of BPS bound states and their wall crossing phenomena, and even yields the (semi-)primitive wall crossing formula as discussed in Section\,\ref{sec.WCFderivation}. We should also keep in mind, however, that there is a limitation to the supergravity approach, since some of the D-brane configuration discussed in this paper does not have large black holes and therefore validity of supergravity approximation is not guaranteed.

Multi-centered black holes are specified by a set of charges $\gamma_i$ at the 
``centers'' $\vec{x}_i$ of the black hole .
The Ansatz for the metric is of the form
\beq
ds^2=-e^{2 U} (dt+\omega)^2+e^{-2U} d\vec{x}^2,
\eeq
where the function $U(\vec{x})$ is a warp factor and $\omega$ is a one-form. They are determined by BPS equations of motion. In this appendix we neglect higher derivative corrections to the equations of motion.

The first BPS equation of motion states 
\beq
2 e^{-U} \textrm{Im} (e^{-i \alpha} \Omega)=-H,
\label{BPSeq1}
\eeq
where $\alpha=\textrm{Arg}\,\left(\sum_i Z(\gamma_i) \right)$ and the $\Omega\sim e^{-it}$ is the period vector with suitable normalization \cite{DenefM}.
Another equation is 
\beq
*_3 d\omega=\langle dH,H \rangle,
\label{BPSeq2}
\eeq
where $*_3$ is the Hodge star in $\bR^3$, $H$ is a function with values in the even cohomology (specifying D-brane charges) of the Calabi-Yau manifold, and $\langle \, , \,\rangle$ is the symplectic paring of the even cohomology, as in the main text.
The function $H$, which determines $\omega$ through \eqref{BPSeq2}, is a harmonic function with poles at the position $\vec{x}_i$ of the charges. For $N$-centered black holes it is given by
\beq
H(\vec{x})=\sum_{i=1}^n \frac{\gamma_i}{|\vec{x}-\vec{x}_i|}-2 \textrm{Im} (e^{-i\alpha} \Omega ) \Big|_{r=\infty}.
\label{eq.H}
\eeq
The equation \eqref{BPSeq2} has a nonsingular solution if and only if the following integrability condition is satisfied:
\beq
\sum_{j=1,\, j\ne i}^N \frac{\langle \gamma_i,\gamma_j \rangle }{|\vec{x}_i-\vec{x}_j|}
=2\textrm{Im}(e^{-i\alpha}Z(\gamma_i))\Big|_{r=\infty},
\eeq
where we acted $d*_3$ on both sides of \eqref{BPSeq2} and used the relation
\beq
Z(\gamma_i)=\langle \gamma_i,\Omega \rangle.
\eeq
In particular, for 2-centered black holes this simplifies to 
\beq
r_{12}:=|\vec{x}_1-\vec{x}_2|=\left.\frac{\langle \gamma_1,\gamma_2 \rangle}{2\textrm{Im}(e^{-i\alpha} Z(\gamma_1))} \right|_{r=\infty}
=\left.\frac{\langle \gamma_1,\gamma_2 \rangle}{2}\frac{|Z(\gamma_1)+Z(\gamma_2)|}
{\textrm{Im}(Z(\gamma_1) \overline{Z(\gamma_2)})} \right|_{r=\infty}.
\label{r12}
\eeq
Since the distance between the center should clearly be positive, we have the condition 
\beq
\langle \gamma_1,\gamma_2\rangle \textrm{Im} \left( Z(\gamma_1) \overline{Z(\gamma_2)} \right)>0.
\label{Denefstability}
\eeq

\noindent
which is sometimes called the Denef stability condition \footnote{In general multi-centered black holes, the integrability is not enough to ensure the existence of physically interesting solutions. For example, we have to check that the discriminant is positive, and that K\"ahler moduli remain in the physical moduli space. In general, it was conjectured in Ref.~\refcite{DenefM} that the components of the moduli space of multi-centered black hole solutions is classified by attractor flow trees.}. Wall crossing occurs when $r_{12}$ goes to infinity. This is the case when
\beq
Z(\gamma_1)=c Z(\gamma_2), \quad c\in \bR_+.
\eeq
This coincides with the result of Section\,\ref{sec.WC}.

Finally, the black hole carries the angular momentum
\beq
J_{12}= \frac{1}{2} \sum_{i\ne j}\langle \gamma_i,\gamma_j \rangle \frac{x_i-x_j}{|x_i-x_j|}.
\eeq
In particular, for a two-centered configuration, 
\beq
J=\frac{1}{2} {| \langle \gamma_1,\gamma_2 \rangle| } .
\eeq
This is a classical result, but there is a quantum correction \cite{DenefQuantum} which modifies the result to
\beq
J_{12}=\frac{1}{2} {\left(|\langle  \gamma_1,\gamma_2 \rangle | -1\right)}. \label{J12}
\eeq
This angular momentum plays crucial roles in the derivation of the primitive wall crossing formula in Section\,\ref{sec.WCFderivation}.

\subsection{Kontsevich-Soibelman Formula}\label{app.KS}

In this appendix we briefly comment on the more general Kontsevich-Soibelman formula.
Since the formula itself is not needed in the rest of this paper, the discussion is brief and the interested readers are referred to the original papers \cite{KontsevichS,KontsevichS2} for more information.

Let us start from the charge lattice $\Gamma$, which is the even coholomogy of the Calabi-Yau manifold. We have a natural symplectic pairing $\langle \, , \, \rangle$ on $\Gamma$. Consider a Lie algebra with generators $e_{\gamma}, \gamma\in \Gamma$, satisfying the commutation relations
\beq
[e_{\gamma_1},e_{\gamma_2}]=(-1)^{\langle \gamma_1, \gamma_2 \rangle }
\langle \gamma_1, \gamma_2 \rangle e_{\gamma_1+\gamma_2}.
\label{eCR}
\eeq
We also define a multiplication by
\beq
e_{\gamma_1} e_{\gamma_2}=e_{\gamma_1+\gamma_2}.
\eeq
We define a group element
\beq
U_{\gamma}:=\exp\sum_{n=1}^{\infty} \frac{1}{n^2}e_{n\gamma}.
\eeq
which acts on $e_{\gamma'}$ by an adjoint action. Using the commutation relation \eqref{eCR}, we have
\beq
U_{\gamma}: e_{\gamma'} \mapsto (1-(-1)^{\langle \gamma,\gamma'\rangle }e_{\gamma})^{\langle \gamma,\gamma'\rangle} e_{\gamma'}.
\label{Uaction}
\eeq

As we have seen in Section\,\ref{sec.WC},  on the generic walls of marginal stability the decay pattern is $\gamma\to N_1\gamma_1+N_2\gamma_2$ with 
$Z(\gamma_1)/Z(\gamma_2)\in \bR_+$. 
On both sides of the walls of marginal stability consider the product
\beq
A:=\prod_{\gamma=N_1\gamma_1+N_2 \gamma_2, \, N_1>0, N_2>0}^{\curvearrowright} U_{\gamma}^{\Omega(\gamma;t)},
\label{Au}
\eeq
where the ordering of the product is given by the clockwise ordering of the central charge $Z(\gamma)$. When we cross the wall of marginal stability, the ordering of the central charges changes, and also the BPS invariants $\Omega(\gamma;t)$ jumps. However, the wall crossing formula states that the whole product \eqref{Au} stays invariant:
\beq
A(t_+)=A(t_-).
\eeq
where $t_{\pm}$ is the value of the K\"ahler moduli on different sides of the wall.

\bigskip

Let us derive the primitive wall crossing formula from the Kontsevich-Soibelman formula. For a decay $\gamma\to \gamma_1+\gamma_2$ with $\gamma_1, \gamma_2$ primitive, the wall crossing formula says
\beq
U_{\gamma_1}^{\Omega(\gamma_1;t_+)} U_{\gamma_1+\gamma_2}^{\Omega(\gamma_1+\gamma_2;t_+)}
U_{\gamma_2}^{\Omega(\gamma_2;t_+)} =
U_{\gamma_2}^{\Omega(\gamma_2;t_-)} U_{\gamma_1+\gamma_2}^{\Omega(\gamma_1+\gamma_2;t_-)}
U_{\gamma_1}^{\Omega(\gamma_1;t_-)}.
\eeq
Since what matters here is only the part generated by $e_{\gamma_1}, e_{\gamma_2}$ and $e_{\gamma_1+\gamma_2}$, we consider a quotient of the Lie algebra with all other generators. It is easy to see that this is consistent with the commutation relation \eqref{eCR}.
We have $\Omega(\gamma_i;t_+)=\Omega(\gamma_i;t_-)$ for $i=1,2$ because $\gamma_{1,2}$ are primitive. We simply use the notation $\Omega(\gamma_i)$ in the following. We also note that $\gamma_1+\gamma_2$ is a central element of the quotient Lie algebra. This means that 
$$
U_{\gamma_1+\gamma_2}^{\Delta\Omega(\gamma_1+\gamma_2;t_+)} 
 =
U_{\gamma_2}^{\Omega(\gamma_2)} 
U_{\gamma_1}^{\Omega(\gamma_1)} 
U_{\gamma_2}^{-\Omega(\gamma_2)}
U_{\gamma_1}^{-\Omega(\gamma_1)}.
$$
Using the commutation relation \eqref{eCR} again we have
\beq
\Delta\Omega(\gamma_1+\gamma_2)=(-1)^{\langle \gamma_1,\gamma_2 \rangle}\langle \gamma_1,\gamma_2 \rangle \Omega(\gamma_1)\Omega(\gamma_2).
\eeq
This is (up to an overall sign, which depends on the definition of $\Delta\Omega$) the primitive wall crossing formula \eqref{primitive}.

It is also straightforward to derive the semi-primitive wall crossing formula from the Kontsevich-Soibelman formula \footnote{The following arguments are known to experts, but as of this writing there is no published account to the best of the author's knowledge.}.
\beq
\begin{split}
U_{\gamma_1}^{\Omega(\gamma_1;t_+)}& U_{\gamma_1+\gamma_2}^{\Omega(\gamma_1+\gamma_2;t_+)} U_{\gamma_1+2\gamma_2}^{\Omega(\gamma_1+2\gamma_2;t_+)} \ldots
U_{\gamma_2}^{\Omega(\gamma_2;t_+)} U_{2\gamma_2}^{\Omega(2\gamma_2;t_+)}\ldots \\
=
&U_{\gamma_2}^{\Omega(\gamma_2;t_-)} U_{\gamma_2}^{\Omega(\gamma_2;t_-)}\ldots \ldots U_{\gamma_1+2\gamma_2}^{\Omega(\gamma_1+2\gamma_2;t_-)}
U_{\gamma_1+\gamma_2}^{\Omega(\gamma_1+\gamma_2;t_-)}
U_{\gamma_1}^{\Omega(\gamma_1;t_-)}.
\end{split}
\eeq
In this case, we consider the quotient algebra by $e_{N_1\gamma_1+N_2\gamma_2}$ with $N_1>1$. This means that $e_{\gamma_1+N_2\gamma_2}$'s commute each other in the quotient algebra. Also, since $\gamma_2$ is primitive $\Omega(N_2 \gamma_2)$ does not have moduli dependence (at general value of K\"ahler moduli) .
This means that 
\beq
\begin{split}
&U_{\gamma_1}^{\Omega(\gamma_1;t_+)} U_{\gamma_1+\gamma_2}^{\Omega(\gamma_1+\gamma_2;t_+)} U_{\gamma_1+2\gamma_2}^{\Omega(\gamma_1+2\gamma_2;t_+)} \ldots \\
=
& \left( U_{\gamma_2}^{\Omega(\gamma_2)} U_{2\gamma_2}^{\Omega(2\gamma_2)}\ldots
\right)
U_{\gamma_1}^{\Omega(\gamma_1;t_-)} 
U_{\gamma_1+\gamma_2}^{\Omega(\gamma_1+\gamma_2;t_-)}
\ldots
\left( U_{\gamma_2}^{\Omega(\gamma_2)} U_{\gamma_2}^{\Omega(\gamma_2)}\ldots
\right)^{-1}.
\end{split}
\label{temporaryeq}
\eeq
Since 
\beq
U_{\gamma_1}^{\Omega(\gamma_1;t_{\pm})} U_{\gamma_1+\gamma_2}^{\Omega(\gamma_1+\gamma_2;t_{\pm})}  \ldots 
=\exp\left(\sum_{k\ge 0} \Omega(\gamma_1+k \gamma_2) e_{\gamma_1+k\gamma_2}\right)
\eeq
in the quotient algebra, \eqref{temporaryeq} with \eqref{Uaction} gives the relation
\beq
\sum_{k\ge 0} \Omega(\gamma_1+k \gamma_2;t_+) e_{\gamma_1+k\gamma_2}
=\prod_{l>0} \left(1-(-1)^{l\langle \gamma_1,\gamma_2 \rangle } e_{l\gamma_2}\right)^{l\langle \gamma_1,\gamma_2 \rangle \Omega(l\gamma_2)} \sum_{k\ge 0} \Omega(\gamma_1+k \gamma_2;t_-) e_{\gamma_1+k\gamma_2}.
\eeq
If we remove the common $e_{\gamma_1}$ and replace $e_{\gamma_2}$ by a formal variable $p$, we have 
\beq
\sum_{k\ge 0} \Omega(\gamma_1+k \gamma_2;t_+) p^k
=\prod_{l>0} \left(1-(-1)^{l\langle \gamma_1,\gamma_2 \rangle } p^l\right)^{l\langle \gamma_1,\gamma_2 \rangle \Omega(l\gamma_2)} \sum_{k\ge 0} \Omega(\gamma_1+k \gamma_2;t_-) p^k.
\eeq
Now recall that $\Omega(\gamma_1+k\gamma_2)$ with $k>0$ is zero on one side of the wall. This means $\Omega(\gamma_1+k \gamma_2;t_-)=0$ for $k>0$ \footnote{
This should be replaced by $\Omega(\gamma_1+k \gamma_2;t_+)=0$ depending on the sign of $\langle \gamma_1,\gamma_2\rangle$. The Denef stability condition \eqref{Denefstability} tells us on which side of the wall BPS bound states are stable. } and we have
\beq
\Omega(\gamma_1)+\sum_{k> 0} \Delta\Omega(\gamma_1+k \gamma_2;t_+) p^k
=\prod_{l>0} \left(1-(-1)^{l\langle \gamma_1,\gamma_2 \rangle } p^l\right)^{l\langle \gamma_1,\gamma_2 \rangle \Omega(l\gamma_2)} \Omega(\gamma_1) ,
\eeq
which coincides with the semi-primitive wall crossing formula \eqref{semiprimitive}. This is what we wanted to show.

\subsection{The Parametrization of Chambers by the Weyl Group} \label{app.Weyl}
In this section we explain the parametrization of chambers of closed BPS invariants by maps $\sigma$ and $\theta$, as claimed in the main text.

The map $\theta$ is defined to be a map from the set of half-integers $\bZ_h$ to itself
$$
\theta: \bZ_h\to \bZ_h,
$$
satisfying the following two conditions.
First,
\beq
\theta(h+L)=\theta(h)+L \label{theta+L}
\eeq
for any $h\in\bZ_h$. In other words, $\theta$ is periodic with period $L$.
Second,
\beq
\sum_{i=1}^L \theta\left(i-\frac{1}{2}\right)=\sum_{i=1}^L \left(i-\frac{1}{2}\right).
\label{thetasum}
\eeq
Therefore $\theta$ is specified by 
$L-1$ (half-)integers, namely 
$L$ half-integers 
$$
\theta(1/2),\ldots, \theta(L-1/2),
$$
subject to one constraint \eqref{thetasum}. 
Let us assume for the moment that $\theta$ satisfies the condition 
\beq
\theta\left(\frac{1}{2}\right)< \theta\left(\frac{3}{2}\right) < \ldots 
< \theta\left(L-\frac{1}{2}\right).
\label{thetaincrease}
\eeq
We will discuss other cases later.

Given $\sigma$, we have a specific choice of resolution having 
$L-1$ $\bP^1$'s.
Moreover, given a map $\theta$ we can determine the corresponding value of the B-field $B_{\theta}$ by
\beq
\left[ B_{\theta}(\alpha_i+\ldots +\alpha_j) \right]=
\sharp \{m\in\Z\mid \theta(i-1/2)<mL<\theta(j+1/2)\}.  \label{thetaB}
\eeq
It is easy to see that this gives well-defined values of the integer parts of the B-field, which parametrize the chamber as explained in the main text.
Conversely, it can also be proven that given any B-field, we can find a corresponding $\theta$ uniquely \footnote{This comes from the fact that the action of the affine Weyl group on the space of B-fields is faithful.}.

When the condition \eqref{thetaincrease} is not satisfied, we need to change the choice of the crepant resolution.
Choose a permutation $\Sigma$ of $\{ 1/2,\ldots, L-1/2\}$ such that
\beq
\theta\left(\Sigma\left(\frac{1}{2}\right)\right)< \theta\left(\Sigma\left(\frac{3}{2}\right)\right) < \ldots 
< \theta\left(\Sigma\left(L-\frac{1}{2}\right)\right)
\label{permdef}
\eeq
holds. Then we replace $\theta$ by 
$\theta':=\theta\circ \Sigma$ 
and we choose $\sigma'$ so that $\sigma\circ\theta=\sigma'\circ\theta'$ \footnote{In fact, we can take $\sigma\circ\underline{\theta}\circ\Sigma^{-1}\circ\underline{\theta}^{-1}$ as $\sigma$, where $\underline{\theta}$ is the permutation induced by $\theta$.}.
Note that the combination $\sigma\circ\theta$, which appears in the definition of the vertex in Section\,\ref{sec.def}, remain invariant under this process.
This means that sometimes different $\sigma,\theta$ and $\sigma', \theta'$ corresponds to the same chamber. We can either fix $\sigma$ and change $\theta$ to parametrize chambers, or change both $\sigma$ and $\theta$ for convenience. The latter parametrization is redundant, but sometimes useful.

\bigskip

In the above discussion $\theta$ appears somewhat artificially, but $\theta$ is often used in the mathematical literature.
The reason is that the maps $\theta$ makes a group, which is the Weyl group of $\hat{A}_{L-1}$.
As is well-known, the Weyl group of $A_{L-1}$ is the $L$-th symmetric group, which is a set of isomorphism $\{1/2,\ldots, L-1/2 \}\to \{1/2,\ldots, L-1/2 \}$.
The map $\theta$ gives a generalization to the affine case.
The affine Lie algebra $\hat{A}_{L-1}$ appears in the formula for the BPS partition function \cite{Nagao1,AOVY}, and as we have seen specifies a chamber structure. This is reminiscent of the appearance of the Weyl group of the Borcherds-Kac-Moody algebra in $\mathcal{N}=4$ wall crossing \cite{ChengV,CD}. It would be interesting to explore this point further (see Section\,\ref{chap.discussion} for more on this).

\bigskip
Finally, let us illustrate this parametrization with examples. Consider the resolved conifold ($L=2$) with the resolution $\sigma(1/2)=+, \sigma(3/2)=-$.
Due to the condition \eqref{thetasum}, we can write
\beq
\theta(1/2)=1/2-N,\quad \theta(3/2)=3/2+N,
\label{theta1}
\eeq
and this integer $N$ parametrize the chambers. This integer $N$ is the same integer $N$ appearing in \eqref{thetaN}.
When $N\ge 0$, condition \eqref{thetaincrease} is satisfied and we are in one resolution $\sigma$. When $N<0$, \eqref{thetaincrease} is not satisfied and by a flop transition we are in a different crepant resolution specified by $\sigma'(1/2)=-, \sigma'(3/2)=+$.
If we define $\theta'$ by
$$
\theta'(1/2)=1/2+N, \quad \theta'(3/2)=3/2-N,
$$
then $\sigma,\theta$ for $N>0$ and $\sigma',\theta'$ for $N<0$ parametrize the same chamber. In this sense, we can either choose $\sigma$ to be fixed and change $\theta$, or change both $\sigma$ and $\theta$, although the latter is a redundant parametrization.
The chamber corresponding to topological string theory for one resolution $\sigma$ is given by $N\to\infty$ in \eqref{theta1}, and another resolution $\sigma'$ given by $N\to -\infty$.

As a next example suppose $L=3$.
When 
$$
\theta(1/2)=-5/2, \quad \theta(3/2)=3/2, \quad \theta(5/2)=11/2,
\label{theta3-1}
$$
we have
$$
\left[B_{\theta}(\alpha_1)\right]=\left[B_{\theta}(\alpha_2)\right]=1, \quad \left[B_{\theta}(\alpha_1+\alpha_2)\right]=2.
$$
This is also given by
$$
\theta(1/2)=11/2, \quad \theta(3/2)=3/2, \quad \theta(5/2)=-5/2.
\label{theta3-2}
$$
However, they parametrize different chambers in general \footnote{Sometimes they give the same chamber. This happens, for example for $\bC^2/\bZ_2\times \bC$, where there is a unique choice of crepant resolution.}. This is because the two $\theta$'s are related by a permutation $\Sigma: \{1/2.3/2,5/2\}\mapsto \{5/2,3/2,1/2 \}$, and correspondingly we have to change the choice of crepant resolution $\sigma$ by $\sigma\circ \Sigma$ as mentioned around \eqref{permdef}.

More generally, if we have 
$$
\theta\left(\frac{1}{2}\right)= \frac{1}{2}-2 L_1- L_2, \quad
\theta\left(\frac{3}{2}\right)= \frac{3}{2}+ L_1-L_2, \quad
\theta\left(\frac{5}{2}\right)= \frac{5}{2} +L_1 +2L_2.
$$
then
$$
\left[B_{\theta}(\alpha_1)\right]=L_1, \quad \left[B_{\theta}(\alpha_2)\right]=L_2, \quad \left[B_{\theta}(\alpha_1+\alpha_2)\right]=L_1+L_2,
$$
and if we have
$$
\theta\left(\frac{1}{2}\right)= -\frac{1}{2}-2 L_1- L_2, \quad
\theta\left(\frac{3}{2}\right)= \frac{3}{2}+ L_1-L_2, \quad
\theta\left(\frac{5}{2}\right)= \frac{7
}{2} +L_1 +2L_2.
$$
then
$$
\left[B_{\theta}(\alpha_1)\right]=L_1, \quad \left[B_{\theta}(\alpha_2)\right]=L_2, \quad \left[B_{\theta}(\alpha_1+\alpha_2)\right]=L_1+L_2+1.
$$


\subsection{Young Diagrams and Maya Diagrams} \label{app.core}

A partition (Young diagram) $\lambda$ is a set of non-increasing positive integers 
$\lambda=(\lambda_1,\lambda_2,\ldots )$, $\lambda_1\ge \lambda_2 \ge \ldots \ge 0$. Given a partition $\lambda$, its transpose $\lambda^t$ is defined by
$\lambda^t{}_i:= \# \{ j | \lambda_j \ge i \}$, and its length $|\lambda|$ by $|\lambda|:=\sum_i \lambda_i$. 
For example, the transpose of $(4,2,1)$ is given by $(3,2,1,1)$, and both have length $7$.

\begin{figure}[htbp]
\centering{\includegraphics[scale=0.35]{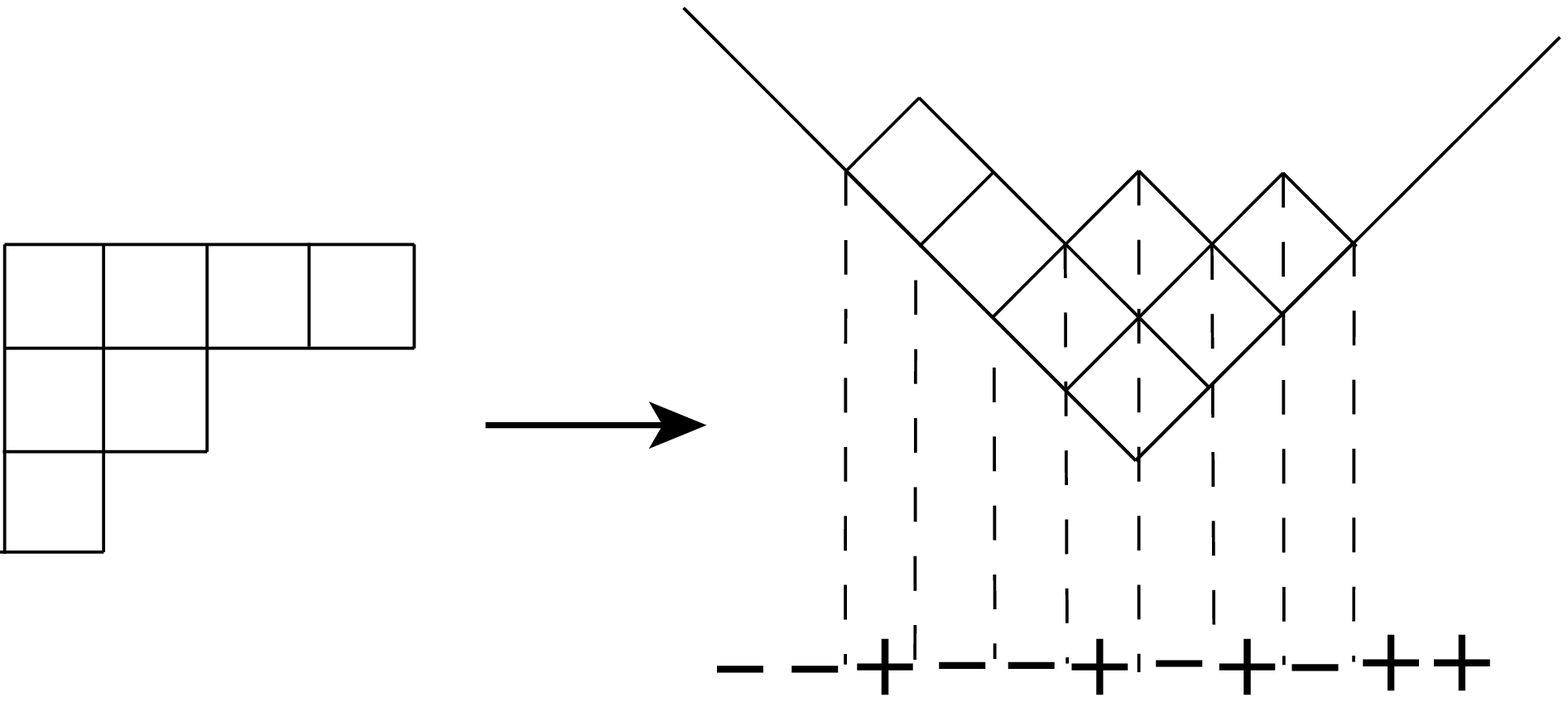}}
\caption{The convention of the Maya diagram in this paper.}
\label{Mayaconv}
\end{figure}

As is well-known, this is equivalently 
represented by a Maya diagram, namely a map
$$
\lambda: \bZ_h\to \{\pm 1\}
$$
such that $\lambda(h)=\pm 1$ for $\pm h \gg  1$.
We sometimes represent $\lambda$ by
$$
\lambda=\cdots \quad \lambda(-5/2) \quad \lambda(-3/2) \quad \lambda(-1/2) \quad \Big| \quad \lambda(1/2) \quad \lambda(3/2) \quad \lambda(5/2)\quad \cdots,
$$
where the symbol $\Big|$ represents the position of the origin.
For notational simplicity, we use the same symbol $\lambda$ for a Maya diagram.
Our convention is shown in Fig.\,\ref{Mayaconv}.

For example, 
\begin{eqnarray}
{\rm \tiny \yng(2,1,1)}&=\cdots----+-\Big|++-+++ \cdots,\\
{\rm \tiny \yng(3,1)}&=\cdots ---+--\Big|+-++++ \cdots. \label{yng31}
\end{eqnarray}

For a Young diagram and a positive integer $M$, define the quotients 
$\lambda_i (i=1,\ldots,M)$ by
\beq
\lambda_i (h)=\lambda\left(i-1/2+(h-1/2)M\right) \quad \textrm{for}\quad h\in \bZ_h.
\eeq

As an example, let us consider $\lambda={\tiny \yng(3,1)}$ shown in \eqref{yng31}. If you take $M=2$, then $(M=2)$--quotients are given by
\begin{align*}
\lambda_1&=\cdots------\Big| ++++++\cdots, \\
\lambda_2&=\cdots----+-\Big| -+++++\cdots.
\end{align*}

Suppose that $M$ is chosen such that the representation $\lambda_i$ is trivial for all $i=1,\ldots, M$.
This means that $\lambda_i$ 
can be written as 
\beq
\lambda_i (h)=\emptyset( h+ M N(i)).
\eeq
The integers $N(j)$ are called $M$-cores of $\lambda$, and satisfies
$$
\sum_{i=1}^M N(i)=0.
$$
For example, if we take $M=3$ for \eqref{yng31}, 
\begin{align*}
\lambda_1&=\cdots-----+\Big| ++++++\cdots, \\
\lambda_2&=\cdots------\Big| -+++++\cdots, \\
\lambda_3&=\cdots------\Big| ++++++\cdots,
\end{align*}
and 
\begin{align*}
N(1)=-1, \quad N(2)=1, \quad N(3)=0.
\end{align*}

\subsection{Proof of the Closed Expression for the Vertex}\label{app.proof}

In this appendix we give a proof of \eqref{main2}.
First, the following is clear from the definition:

\medskip
\noindent{\bf Proposition:}
Let $\sigma, \theta$ to be maps specifying the chamber for a Calabi-Yau manifold $X$.
Choose an integer $M$ and $\sigma', \theta'$ for $X'=X/\bZ_M$
such that the following condition holds:
\begin{equation*}
\sigma\circ\theta=\sigma'\circ\theta',\quad \mu\circ\theta=\mu'\circ\theta'.
\end{equation*}
Then we have
\[
C_{(\sigma',\theta'\sss;\sss\mu',\nu')}(q',Q')|_{q^{\theta}_i=q'{}^{\theta'}_i=q'{}^{\theta'}_{i+L}=\cdots q'{}^{\theta'}_{i+(M-1)L}}
=C_{\type}(q,Q).
\label{propassumption}
\]


\medskip
This is because the both sides of the equation are defined by the same crystal with the same weights. The crystal for an orbifold is the same as the original crystal, the only difference being the difference of the weights; the crystal for the orbifold has more colors (variables). 
 However, in the above equation we are specializing the variables so the weights are also the same.

\medskip
\noindent{\bf Lemma:}
For any $\sigma, \theta, \lambda$, we can take $M, \sigma', \theta'$ such that
\begin{equation}
\sigma\circ\theta=\sigma'\circ\theta',\quad \mu\circ\theta=\emptyset\circ\theta'.
\label{lemmaeq}
\end{equation}
%
\begin{proof}
We choose an integer $M$ such that all the $M$-quotients \footnote{See \ref{app.core} for the definition of $M$-quotients.} of $\mu_i$'s become trivial, i.e. $ML$-quotients of the combined representation $\mu$ (see \eqref{combinedmu}) is trivial.
For example, this is satisfied if we define
\[
h_-:=\mathrm{min}\{h\in\Zh \mid \mu\circ \theta(h)=+\},\quad 
h_+:=\mathrm{max}\{h\in\Zh \mid \mu\circ \theta(h)=-\},
\]
and take $M$ so that $ML>h_+-h_-$ \footnote{The choice of $M$ is not unique. 
But the final result is independent of the choice of $M$. For practical computation it is useful to take the minimum $M$.}.
This means that for any half-integer $1/2\leq h\leq ML-1/2$ we can take $N(h)\in \Z$ such that 
\begin{align*}
\mu\circ \theta(h+NML)&=
\begin{cases}
- & (N<N(h)),\\
+ & (N\geq N(h))
\end{cases} \\
&=\emptyset \left(h+(N-N(h))ML\right).
\end{align*}
In other words, $N(j)$ is the $ML$-core \footnote{See \ref{app.core} for the definition of $M$-coress.} of $\mu$.

Therefore the second condition of \eqref{lemmaeq} holds 
if we define $\theta'\colon\Zh\to\Zh$ by 
\[
\theta'(h)=h-N(h)ML\]
for $h\in \bZ_h$ \footnote{For practical computations, it is useful to further perform a permutation to $\theta$ such that \eqref{thetaincrease} holds. See \ref{app.Weyl}.}.
It is clear that the first condition of \eqref{lemmaeq} determines $\sigma'$ uniquely.
\end{proof}

Our theorem follows from the Proposition and the Lemma given above.
\medskip
\noindent{\bf Theorem:}
For $\sigma, \theta, \lambda$, take $M, \sigma', \theta'$ as above. 
Then we have
\[
C_{(\sigma,\theta\sss;\sss \emptyset,\mu)}(q,Q)=
C_{(\sigma',\theta'\sss;\sss \emptyset, \emptyset)}(q',Q')|_{q^{\theta}_i=q'{}^{\theta'}_i=q'{}^{\theta'}_{i+L}=\cdots q'{}^{\theta'}_{i+(M-1)L}}.
\]
%

It is straightforward to generalize this theorem to the case of refined BPS invariants discussed in Section\,\ref{subsec.refined}.


\providecommand{\href}[2]{#2}\begingroup\raggedright\endgroup

\end{document}